\documentclass[iop]{emulateapj-rtx4}
\shortauthors{Sekanina \& Kracht}
\shorttitle{Disintegration of Comet C/2012 S1 (ISON)}
\slugcomment{Version May 8, 2014; substantially updated August 11, 2014}
\newcommand{\Rsun}{$R_{\mbox{\scriptsize \boldmath $\odot$}}$}

\newcommand{\lapeq}{\mbox{\raisebox{0.3ex}{$<$}\hspace{-0.28cm}\raisebox{-0.75ex}{$\sim$}\,}}
\newcommand{\tend}{$t_{_{\hspace{-1pt}\mbox{\footnotesize \boldmath $\star$}}}$\hspace{-1pt}}
\newcommand{\ttend}{\scriptsize $t_{_{\hspace{-1.5pt}\mbox{\scriptsize \boldmath $\star$}}}$\hspace{-2pt}}
\newcommand{\rend}{$r_{_{\hspace{-1.2pt}\mbox{\footnotesize \boldmath $\star$}}}$}
\newcommand{\rrend}{\scriptsize $r_{_{\hspace{-1.7pt}\mbox{\scriptsize \boldmath $\star$}}}$\hspace{-1pt}}
\newcommand{\lpts}{\llap{.}\,.\rlap{\,.}}
\begin{document}
\title{Disintegration of Comet C/2012 S1 (ISON) Shortly Before
Perihelion:\\ Evidence from Independent Data Sets}
\author{Zdenek Sekanina$^1$ and Rainer Kracht$^2$}
\affil{$^1$Jet Propulsion Laboratory, California Institute of
Technology, 4800 Oak Grove Drive, Pasadena, CA 91109, U.S.A.\\
$^2$Ostlandring 53, D-25335 Elmshorn, Schleswig-Holstein, Germany}
\email{Zdenek.Sekanina@jpl.nasa.gov, R.Kracht@t-online.de}
\begin{abstract}
As an Oort Cloud object with a record small perihelion distance of 2.7~{\Rsun}
and discovered more~than a year before its encounter with the Sun, comet
C/2012~S1 is a subject of considerable scientific interest.  Its activity
along the orbit's inbound leg evolved through a series of cycles.  Two
remarkable events preserved in SOHO's and/or STEREO's near-perihelion images
of its tail were an early massive production of gravel at heliocentric
distances of up to $\sim$100~AU(!), evidently by the annealing of amorphous
water ice on and near the nucleus' surface; and, about a week before perihelion,
a rapid series of powerful explosions, from the comet's interior, of water
vapor with dust at extremely high rates, causing precipitous fragmentation of
the nucleus, shattering it into a vast number of sublimating boulders, and
ending up, a few days later, with a major, sudden drop in gas emission.
The disintegration of the comet was completed by about 3.5 hours before
perihelion, at a heliocentric distance of 5.2~{\Rsun}, when C/2012 S1 ceased
to exist.  The orbital motion in this period of time was subjected to
progressively increasing outgassing-driven perturbations.  A comprehensive
orbital analysis results in successfully fitting the comet's observed
motion from 2011 to $\sim$7 hours before perihelion.
\end{abstract}
\keywords{comets: general --- comets: individual (C/1959 Y1, C/1962 C1,
 C/1999 S4, C/2003 A2, C/2011 W3, C/2012~S1) --- methods: data
analysis{\vspace{-0.15cm}}}
\section{Introduction}
There are several reasons for an unusually intense scientific interest in
comet C/2012 S1.  Perhaps the most compelling one is its record small
perihelion distance, merely 2.7 solar radii (1 solar radius = 1 {\Rsun} =
0.0046548 AU), among known dynamically new comets, i.e., those arriving
from the Oort Cloud.  This perihelion distance beats the previous record,
held by comet C/1962 C1 (Seki-Lines) by more than 4 \Rsun.  Also beneficial
are the early discovery of C/2012 S1 by Nevski \& Novichonok (2012) and
subsequent detections of pre-discovery images of the comet when it was as
far as 9.4~AU from the Sun.  The early discovery allowed a comprehensive
monitoring of the comet's activity on its way to perihelion.

Anticipated with particular interest was the comet's behavior near
perihelion and chances of its survival.  An optimistic side of the
controversy was argued e.g.\ by Knight \& Welsh (2013), while the most
skeptical view was Ferr\'{\i}n's (2013, 2014), who nearly two months before
perihelion predicted the comet's impending demise. Much effort in the
present investigation is expended to examine extensive evidence on the
comet's physical state as a function of time, especially in the last
weeks before perihelion when the brightness, the coma and tail
morphology, and the orbital motion were subject to rapid and profound
changes.

\section{Light Curve, Water Production, and Mass Loss of the Nucleus}

The first of the examined data sets are the comet's observed light curve
and a representative H$_2$O production-rate curve.  The light curve is
defined as variations in the total brightness expressed in magnitudes,
normalized to a distance $\Delta$ of 1 AU from the observer (the Earth
for ground-based observations, a spacecraft for spaceborne observations)
by employing the usual correction term \mbox{$5 \log \Delta$} and to a
zero phase angle (backscatter) by applying a standard correction based
on a modified Henyey-Greenstein phase law introduced by Marcus (2007).
Both the brightness and H$_2$O production rate variations, the latter
measuring a fraction of the comet's total mass loss, are plotted
against time and heliocentric distance.

\subsection{The Light Curve}
Two types of brightness observations are employed in the following:\
those made by ground-based observers, up to 2013 November~22; and
those measured in the comet's images from space, between 2013 November
20 and 30.  The employed magnitudes by selected ground-based observers
are referred to the visual spectral region, even though some of them
were obtained using CCD detectors.\footnote{Three primary sources of
the brightness data have been, respectively, the web site of the {\it
International Comet Quarterly\/} (ICQ), {\tt
http://www.icq.eps.harvard.edu/CometMags.html};~the\,\mbox{reports}\,by
observers, {\tt http://groups.yahoo.com/neo/groups/CometObs/info}; and
selected sets of {\it total\/} magnitudes (marked T) from a synopsis of
astrometric observations in the database maintained by the {\it Minor
Planet Center\/}, {\tt http://www.minorplanetcenter.net/db\_search}.}
Each observer measures the brightness of a comet in his own photometric
system; the heterogeneity introduced by combining data from the various
observers is minimized by implementing corrections to convert the data
to a standard photometric system.  The details related to C/2012 S1 have
been described in \mbox{Sekanina} (2013a); an estimated uncertainty of
the normalized magnitudes is about $\pm$0.3~magnitude.
\begin{figure*}[ht]
\vspace{-2.6cm} 
\hspace{-0.2cm}
\centerline{
\scalebox{0.53}{ 
\includegraphics{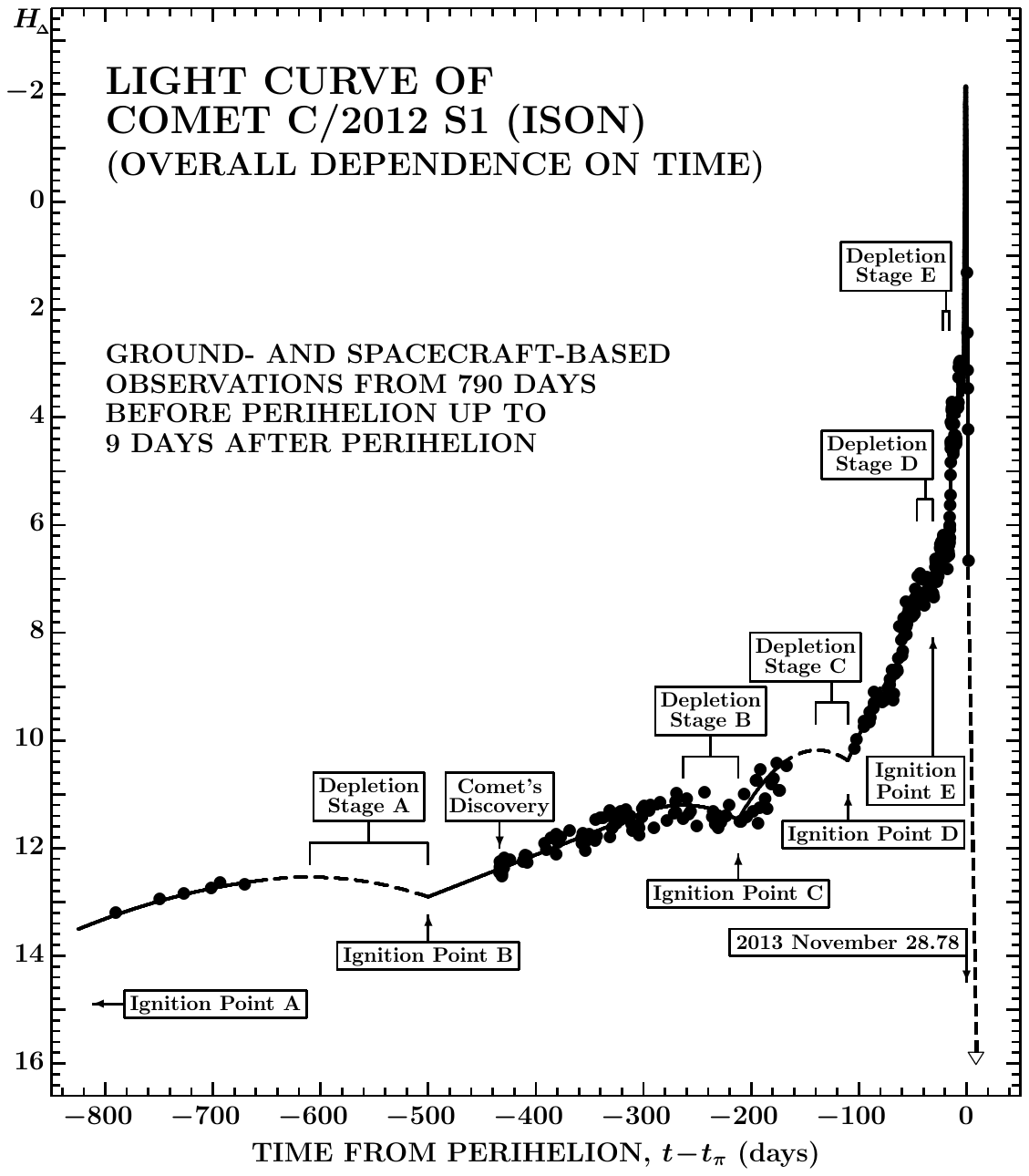}}} 
\vspace{-0.65cm}
\caption{Overall light curve of comet C/2012 S1, based on about 300 selected
ground-based observations (footnote 1) shown as solid circles and on
spaceborne photometry:\ the HI1 imager on board STEREO A (Hui, footnote~2)
and the C3 coronagraph on board SOHO (Knight \& Battams 2014; Nakano 2013a;
Ye, footnote 1).  The results based on Hui's and Knight \& Battams'
preperihelion data are shown as thick lines near the peak, the results
based on Nakano's and Ye's post-perihelion data as solid circles.  A
negative observation by Sako et al.\ (2013) from December 7 is depicted
with a triangle.  The magnitude $H_\Delta$, plotted against time $t$ reckoned
from perihelion, $t_\pi$, is normalized to 1~AU from the observer and phase
corrected using the Marcus (2007) law.  The plotted data cover the period
from 2011 September 30 (790 days before perihelion, the first pre-discovery
observation) to 2013 December 7 (9 days after perihelion).  Up to the time
of the first major outburst, about two weeks before perihelion, the comet's
light curve had been evolving in five cycles, A--E.{\vspace{0.4cm}}}
\end{figure*}

Most spaceborne normalized brightness data presented here were derived by us
from two overlapping~sets{\nopagebreak} of preperihelion apparent magnitudes:\
one from measurements by \mbox{M.-T. Hui} in frames taken with the HI1 imager
on board STEREO-A between Novemver 20.8 and 27.5 UT;\footnote{A nearly complete HI1-A set of apparent magnitudes measured by\,Hui\,is\,posted\,at\,{\tt
http://groups.yahoo.com/neo/groups/comets- ml/conversations/messages/22492}.
Hui kindly provided us with a short extension of this set in a personal
communication.} the other from measurements by Knight \& Battams (2014) in
clear-filter frames taken with the C3 coronagraph on board SOHO between
November 27.1 and 28.6 UT.  These sets were supplemented with several
post-perihelion C3 apparent magnitudes from November 29.38--30.88 UT
measured by Nakano (2013a) and from November 30.23 UT determined by Q.~Ye,
as listed on the ICQ web site (see footnote 1).  We normalized all these
apparent magnitudes to 1~AU from the spacecraft and phase corrected them
using the Marcus (2007) law.\footnote{Our phase corrections refer the
normalized magnitudes to a zero phase angle, not to 90$^\circ$, as does
Marcus (2007).}

The comet's overall light curve, which includes magnitudes from a number of
pre-discovery images, taken between 2011 September 30 and 2012 January 28
at the Pan-STARRS Station on Haleakala, Hawaii (two rather discordant ones,
on November~26 and December~9 averaged), and by the Mt.\,Lemmon Survey on
Catalina near Tucson, Arizona (see the MPC Database in footnote 1), is plotted
as a function of time, reckoned from perihelion, $t_\pi$, in Figure~1.  In
spite of two gaps, an 8-month long pre-discovery one in 2012 and a two-month
long in 2013 due to the comet's conjunction with the Sun, the light curve
appears to show that the comet's normalized brightness evolved in cycles,
each of which started with an {\it ignition (activation) point\/}.  The comet
brightened throughout what we call an {\it expansion stage\/} of the cycle
until a {\it stagnation point\/} was reached.  At this time the comet's
brightness began to stall and subsequently might drop a little in the course
of the cycle's {\it depletion stage\/}.  Eventually each cycle terminated at
the next cycle's ignition point.  The duration of an expansion stage is thus
equal to the time difference between the stagnation and ignition points and
the duration of a depletion stage equals the temporal distance from the
stagnation point to the next ignition point.  In their sum, the lengths
of the expansion and depletion stages make up the duration of the cycle.
Figure~1 shows a total of five cycles, A through E.  The last two are,
however, hard to see, as they fall on the steep portion of the light curve;
they are clearly discernible in Figure~2, a plot of the normalized brightness
against heliocentric distance $r$.

\begin{figure}[ht]
\vspace{-1.265cm}
\hspace{-0.81cm}
\centerline{
\scalebox{0.53}{
\includegraphics{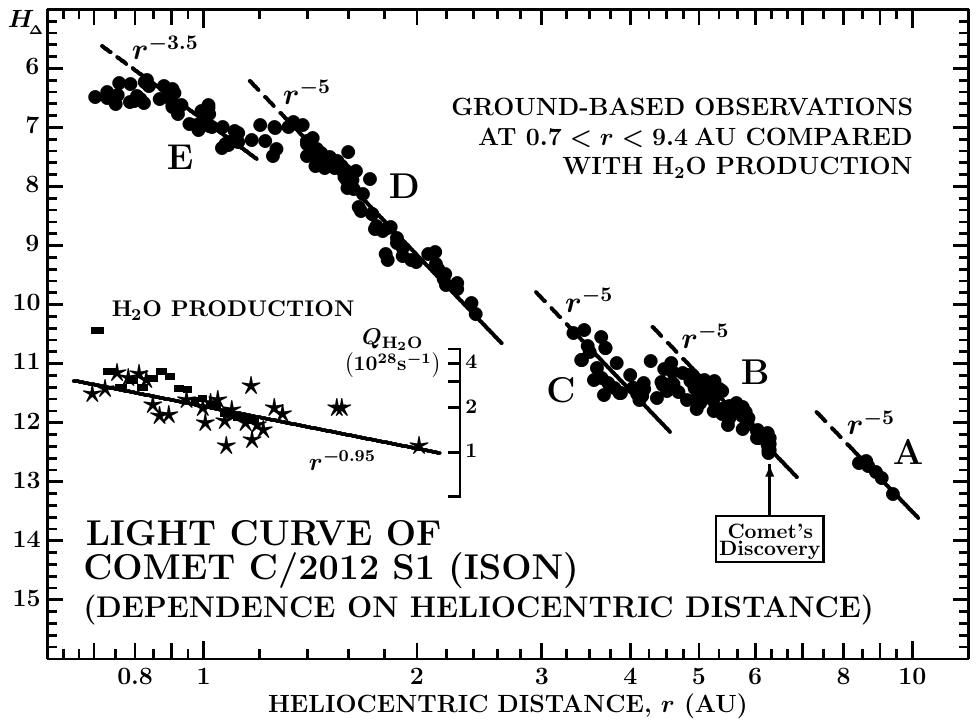}}} 
\vspace{-6.98cm}
\caption{Magnitudes $H_\Delta$ of comet C/2012 S1, normalized to a unit
distance from the observer and to a zero phase angle, plotted against
heliocentric distance.  The plot shows, as solid circles, selected
ground-based brightness observations, made between the distances of 9.4 and
0.7~AU from the Sun.  They are compared with data on the water production
rate, $Q_{\rm H\mbox{$_2$}O}$, represented by stars, with the daily averages,
as presented by Combi et al.\,(2014), shown as thick bars.  Also depicted
are the five activity cycles A--E.{\vspace{0.2cm}}}
\end{figure}

A remarkable property of the expansion stages of the cycles A--D is nearly
the same rate of the comet's brightening, as $r^{-5}$, with a lower rate
only in the cycle E.  The presence of depletion stages reduces the overall
rate of brightening to about $r^{-2.5}$.  Figure~3, which is, with an
overlap, a continuation of Figure~2, covers a range from 1.3~AU to
perihelion and shows that the cycle E terminated at the onset of a brief
precursor to a major outburst.  This event appears to be a sign that the
long, rather orderly era of the comet's evolution had ended and that the
object entered a stormy period, with rambunctious, unpredictable activity
variations.

Since our interest in this study is being focused on the comet's behavior
near the Sun, it is not our objective to examine in detail the nature and
significance of the cycles of activity.  We suggest, however, that their
existence could be related to limited discrete reservoirs of ices ---
sources of activity on and just beneath the nucleus' surface.  Accessed
by the Sun's radiation in due time, each of these sources was activated
(ignition point) and continued to be active over a limited period of time
(expansion stage) until the bulk of the supply became essentially exhausted
(stagnation point).  The brightness then began to subside (depletion stage)
until a new source of activity became available.  The ability to reach the
Sun's radiation and to sublimate profusely enough could be regulated both by
the degree of volatility of icy species and by the depth of the reservoirs
beneath the surface.  In principle, the most volatile ices got released
first, while water ice last.  For example, carbon dioxide sublimates
profusely, at rates greater than 10$^{26}$ molecules per km$^2$ per
second, at a subsolar point at 10~AU from the Sun and it could control
the activity in the cycles A, B, and even C, as suggested by Meech et al.\
(2013b); carbon monoxide outgasses at similar rates still much farther from
the Sun, more than 30~AU (e.g., Sekanina 1992).  Mixtures of ices of uneven
volatility are generally expected to drive the comet's activity cycles.

\begin{figure}[h]
\vspace{-0.92cm} 
\hspace{-0.36cm}
\centerline{
\scalebox{0.49}{
\includegraphics{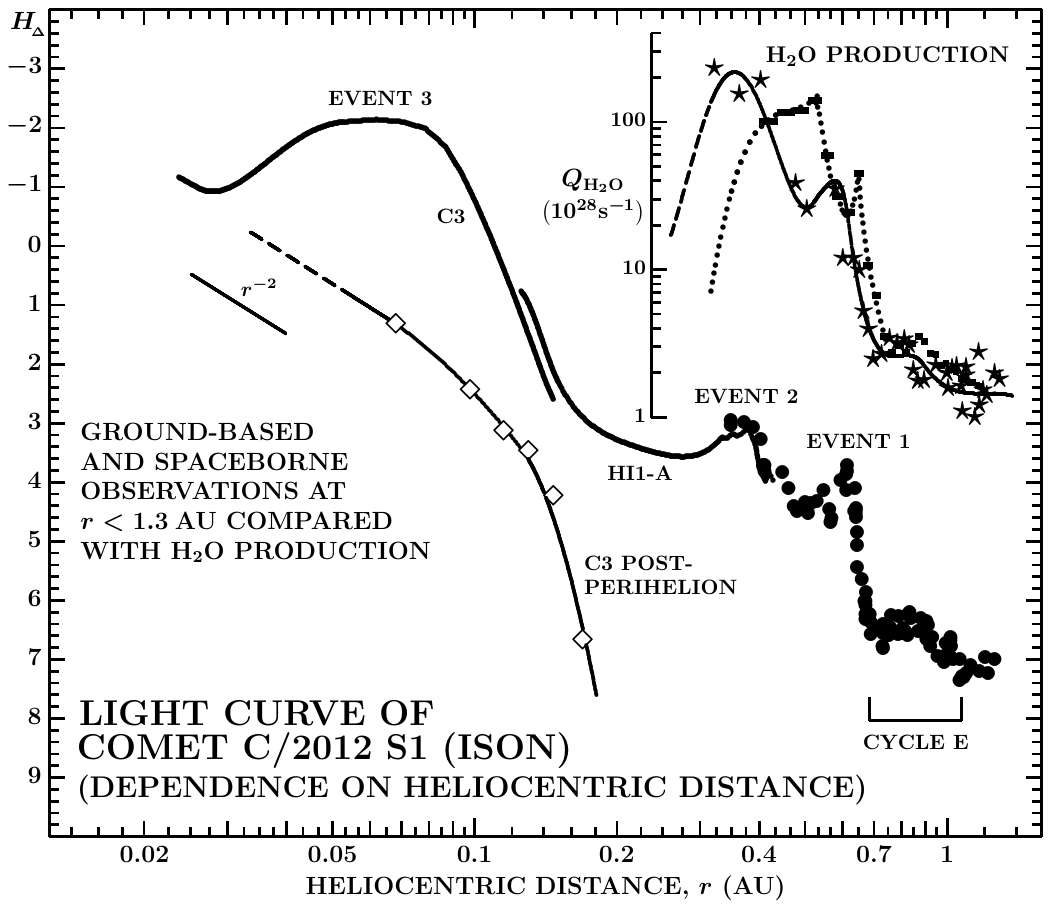}}} 
\vspace{-5.2cm}
\caption{Ground-based and spaceborne normalized magnitudes $H_\Delta$ of
comet C/2012 S1 plotted against heliocentric distance, from 1.3~AU to
perihelion and shortly after perihelion.  The ground-based data are
represented by solid circles, the dense-data sets from preperihelion
imaging with the SOHO's C3 coronagraph (Knight \& Battams 2014) and with
the STEREO-A's HI1 imager (Hui, footnote 2) are shown as thick labeled
curves.  Several C3-based post-perihelion data points by Nakano (2013a)
and by Ye (footnote 1) are depicted with diamonds.  The $r^{-2}$ law of
brightness variations is drawn for reference.  The light curve is compared
with the water production rate, $Q_{\rm H\mbox{$_2$}O}$, plotted with the
same symbols as in Figure~2.  The dotted line is a fit to the daily
averages of $Q_{\rm H\mbox{$_2$}O}$ by Combi et al.\,(2014).{\vspace{0.2cm}}}
\end{figure}

The duration of the activity cycles was getting progressively shorter as the
comet was approaching perihelion, due apparently to the rapidly increasing
rate of the Sun's heating of the comet's nucleus with decreasing heliocentric
distance.  The parameters of the cycles are summarized in Table~1.  For each
cycle, they include the times of the ignition and stagnation points, $t_{\rm
ign}$ and $t_{\rm stg}$, reckoned from perihelion $t_\pi$, the respective
heliocentric distances, $r_{\rm ign}$ and $r_{\rm stg}$, and the calendar
dates, as well as the durations of the expansion and depletion stages.  The
time of the cycle A ignition point is of course unknown, as the comet's
normalized brightness was already increasing in late September of 2011.
The times of the cycle A stagnation point and the cycle B ignition point
are known with large uncertainties because of the long gap between the last
pre-discovery observation and the comet's discovery.  As a result, only
an estimate of a lower limit to the duration of the cycle A expansion stage
and approximate durations of the cycle A depletion stage and the cycle B
expansion stage are tabulated.

\begin{table*}[ht]
\noindent
\vspace{-0.3cm}
\begin{center}
{\footnotesize {\bf Table 1} \\[0.1cm]
{\sc Activity Cycles of Comet C/2012 S1: Ignition and Stagnation Points,
 Expansion and Depletion Stages.}\\[0.15cm]
\begin{tabular}{c@{\hspace{0.65cm}}c@{\hspace{0.5cm}}c@{\hspace{0.4cm}}c@{\hspace{0.75cm}}c@{\hspace{0.5cm}}c@{\hspace{0.5cm}}c@{\hspace{0.65cm}}c@{\hspace{0.3cm}}c}
\hline\hline\\[-0.2cm]
 & \multicolumn{3}{@{\hspace{-0.55cm}}c}{Ignition point}
 & \multicolumn{3}{@{\hspace{-0.5cm}}c}{Stagnation point}
 & \multicolumn{2}{@{\hspace{-0.04cm}}c}{Duration of stage of}\\[-0.02cm]
 & \multicolumn{3}{@{\hspace{-0.55cm}}c}{\rule[0.7ex]{4.65cm}{0.4pt}}
 & \multicolumn{3}{@{\hspace{-0.5cm}}c}{\rule[0.7ex]{4.65cm}{0.4pt}}
 & \multicolumn{2}{@{\hspace{-0.04cm}}c}{\rule[0.7ex]{2.9cm}{0.4pt}}\\[-0.02cm]
Cycle & $t_{\rm ign}\!-\!t_\pi$ & $r_{\rm ign}$ & calendar date
 & $t_{\rm stg}\!-\!t_\pi$ & $r_{\rm stg}$ & calendar date
 & expansion & depletion \\
 & (days) & (AU) & (UT) & (days) & (AU) & (UT) & (days) & (days)\\[0.1cm]
\hline \\[-0.17cm]
A & $<\!-$790 & $>$9.4 & $<$2011 Sept.\,30 & $\sim\!-$610 & $\sim$7.9
  & $\sim$2012 Mar.\,28 & $>$180 & $\!\sim$110 \\
B & $\sim\!-$500 & $\sim$6.9 & $\sim$2012 July 16 & $-$263 & 4.50
  & 2013 Mar.\,10 & $\sim$237 & 51 \\
C & $-$212 & 3.90 & 2013 Apr.\,30 & $-$140 & 2.95 & 2013 July 11
  & 72 & 30 \\
D & $-$110 & 2.51 & 2013 Aug.\,10 & $-$46 & 1.40 & 2013 Oct.\,13
  & 64 & 15 \\
E & $-$31 & 1.07 & 2013 Oct.\,28 & $-$22 & 0.85 & 2013 Nov.\,6
  & 9 & 6 \\[0.05cm]
\hline\\[0.1cm]
%
%
\end{tabular}}
\end{center}
\end{table*}

Figures 3 and 4, the latter representing a closeup of the near-perihelion
light curve plotted against time and both covering practically the same
time span, show that the stormy period of activity commenced in earnest
almost exactly 16 days before perihelion, near November 13.0 UT.  Opitom
et al.\ (2013a) noticed a modest increase in the outgassing of minor
species (such as CN and C$_2$) in early November, but this  was
accompanied by no robust increase in the dust production.  A modest,
about 50~percent, increase in the gas production was reported by Opitom
et al.\ (2013b) during a 24~hour period between November 11.4 and 12.4~UT,
some 17~days before perihelion, with a comparable increase in the integrated
brightness in the infrared, between 3 and 13~microns (Sitko et al.\ 2013),
but again this growth of activity was not particularly significant.  The
cycle E was terminated in the next 24~hours, during which a brightening
was vigorous enough to become apparent in both the light curve and water
production rate.  A sharp upswing began around November~14.0~UT, 14.8~days
before perihelion, the onset of a massive outburst that is called {\it
Event~1\/} in Figure 4.  Boehnhardt et al.\ (2013) reported detection, on
November 14.2 and again two days later, of arclet-like wings extending from
the nucleus in opposite directions, whose appearance is known to have been
associated with fragmentation of the nucleus in other comets in the past
(Boehnhardt 2007); the wings were not detected on November 13.2 UT.  Also
coinciding with the onset of Event 1 was a major change in the comet's
morphology, reported by Ye et al.\ (2013) from comparison of the comet's
images taken by them on November 13.99 and 14.99 UT.

After reaching a peak brightness less than 2~days after the onset of
Event~1, the comet faded a little, but another brightening was detected
as early as November 19.4 UT, or 9.4~days before perihelion (Opitom et
al.\ 2013c).  This appears to have been a precursor to another major
outburst, called {\it Event~2\/}, which, judging from the light curve
based on Hui's data, started two days later, around November~21.2~UT, or
7.6 days before perihelion, exhibited multiple peaks centered on 6~days
before perihelion, and then gradually subsided.  This event proved very
damaging to the integrity of the comet's nucleus, as discussed later.  Yet,
the brightness bottomed out once again 4.3~days before perihelion, followed
by an out-of-control surge at an average rate of as much as 0.2~magnitude
per hour prior to reaching a peak near magnitude $-$2 about 16~hours before
perihelion (Knight \& Battams 2014).  This was {\it Event~3\/}.  The rate
of fading over the first two days after perihelion was almost equally steep.

\begin{figure}[h]
\vspace{-1.6cm}
\hspace{-0.44cm}
\centerline{
\scalebox{0.55}{
\includegraphics{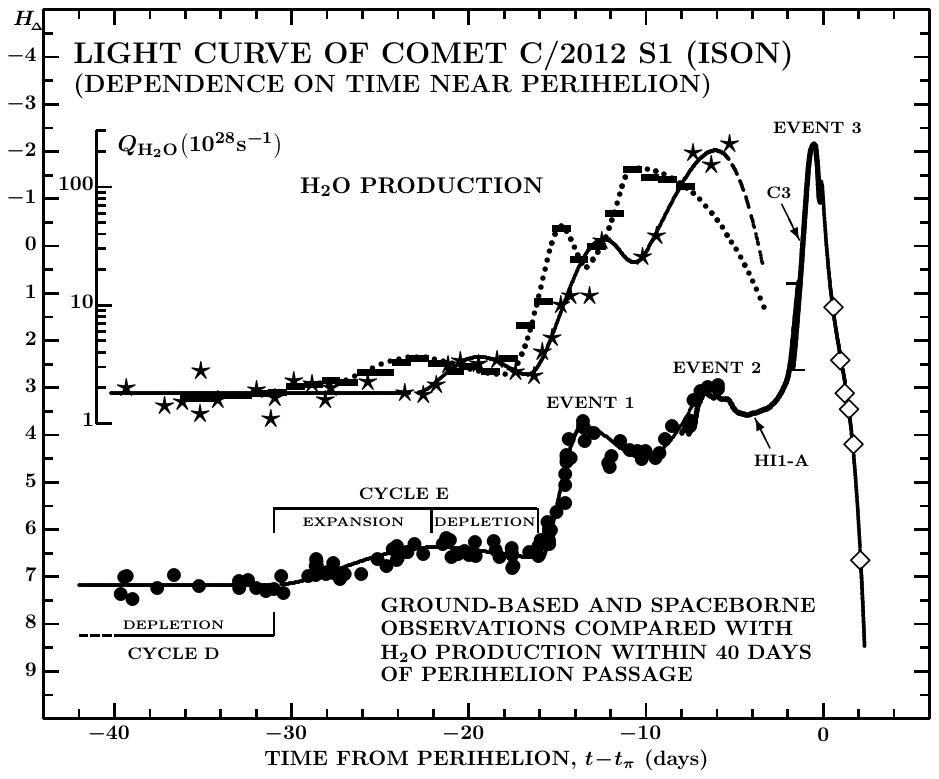}}} 
\vspace{-6.5cm} 
\caption{Light curve of comet C/2012 S1 between 40 days before perihelion
and 2 days after perihelion and its comparison with the comet's water
production rate variations.  The short horizontal ticks show the overlap
of the HI1-A and C3-based light curves.  The other symbols are the same as
those used in Figures 2 and 3.{\vspace{0.5cm}}}
\end{figure}

Starting 8~days before perihelion, there is an excellent agreement between
the light curves derived from ground-based observations and from Hui's HI1-A
data.  The HI1-A curve is, in turn, close to Knight \& Battams' (2014) C3
data: in an overlapping interval of time seen in Figures~3 and 4 this C3
light curve runs in parallel less than 0.5 magnitude below the HI1-A curve.

In general, magnitudes from different instruments on board SOHO and STEREO
are far less consistent.  This topic has been addressed by Knight \& Battams
(2014) in a recent paper, to which the reader is referred for details.
After perihelion, all data show a steep and accelerating fading.
In Figures 1 and 3--4 we illustrate the rate of this drop in brightness on
the data by Nakano (2013a) and by Ye (footnote~1), which are $\sim$2.8
magnitudes brighter than Knight \& Battams' (2014) post-perihelion data.
The discrepancy appears to be an aperture effect:\ Nakano used 27 arcmin,
while Knight \& Battams only 7.5 arcmin, which indeed implies a difference
of \mbox{$5 \log\,(27/7.5) = 2.8$ magnitudes}, suggesting an essentially
constant surface brightness over the measured area.

The normalized magnitude $H_\Delta$ in Figure 4 was next converted to an
intrinsic brightness, independent of the heliocentric distance $r$, using
a formula
\begin{equation}
\Im_0 = r^2 10^{2 - 0.4 H\!\!_{_\Delta}} \! .
\end{equation}
If scattering of sunlight by dust particles dominated, the comet's brightness
was proportional to their total effective cross-sectional area.  For an assumed
geometric albedo of 4~percent, the geometric cross-sectional area (for
back\-scatter) was (in km$^2$)
\begin{equation}
X_{\rm dust} = 3.5 \times 10^5 \Im_0.
\end{equation}
\clearpage
\noindent
The difference between the geometric cross-sectional{\nopagebreak} area and
the cross-sectional area for scattering, which for submicron-sized particles
depends on their optical properties, has been neglected.

The intrinsic brightness $\Im_0$ is displayed in Figure~5 as a function of
time near perihelion.  It had been remarkably constant prior to Event 1,
averaging 0.2~unit, which is equivalent to an effective cross-sectional
area for dust of 70,000~km$^2$.  The two events are comparable in magnitude:\
Event~1 has a higher peak, Event~2 is broader.  Their peaks remain at 14
and 6~days before perihelion, but the dip after Event~2 in the HI1-A data
has moved to 2.4~days before perihelion (see Sec.~4.1).  The peak intrinsic
magnitude, reached $\sim$16~hours before perihelion, is equivalent to a
total geometric cross-sectional area of \mbox{$1.3 \times 10^6$\,km$^2$},
but this is undoubtedly a crude upper limit because of an expected strong
contribution to $\Im_0$ from the doublet of atomic sodium near 5900~{\AA}.

\begin{figure}
\vspace{0.05cm} 
\hspace{-0.31cm}
\centerline{
\scalebox{0.486}{ 
\includegraphics{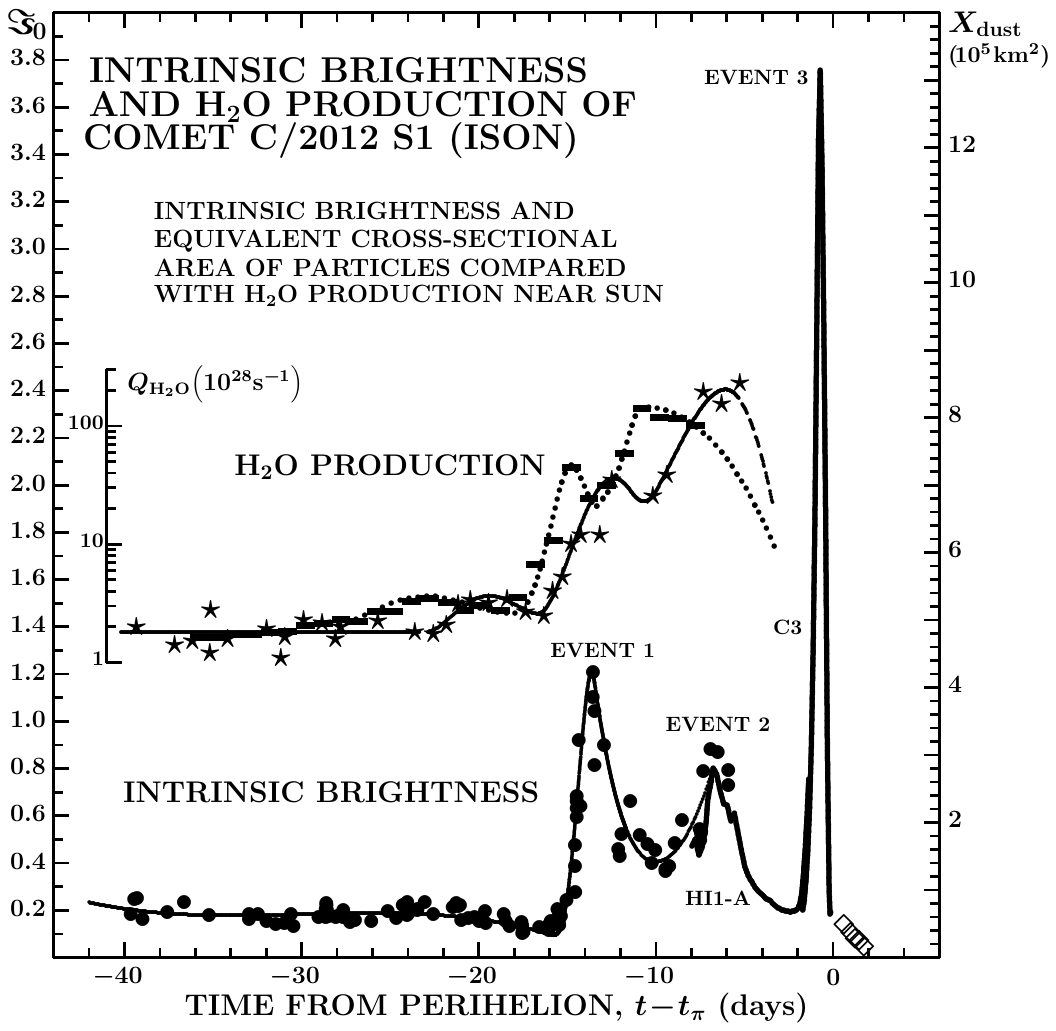}}} 
\vspace{-5.18cm}
\caption{Intrinsic brightness of comet C/2012 S1 and an equivalent
cross-sectional area of dust particles, $X_{\rm dust}$, in the coma
plotted against time near perihelion and compared with the comet's
water production rate variations.  The symbols are the same as those
used in Figures 3 and 4.{\vspace{0.4cm}}}
\end{figure}

\subsection{Water Production Curve, Probable Early Fragmentation, and
Correlations}
Starting with mid-September of 2013, positive detections of the comet's
production of water were reported on many occasions, either from emissions of
its photodissociation products, the hydroxyl in near-UV or radio wavelenghts
and atomic hydrogen in Lyman-$\alpha$, or directly in near-IR.  The
preliminary results have been published by Schleicher (2013a), Bodewits et
al.\ (2013), Opitom et al.\ (2013a, 2013b, 2013c), Weaver et al.\ (2013),
Keane et al.\ (2013), Mumma et al.\ (2013), Combi et al.\ (2013), Dello
Russo et al.\ (2013), Remijan et al.\ (2013), Paganini et al.\ (2013),
Crovisier et al.\ (2013), and Bonev et al.\ (2013).  For the data based
on direct H$_2$O feature measurements, the production rates are referred
to the times of observation, while for the data derived from H or OH
measurements, the rates are plotted at times approximately corrected for
the disscociation lifetime of water (Huebner et al.\,1992) but not for
other effects.  The most systematic set of water production rates has
been published in a detailed paper by Combi et al.\,(2014), who studied
the comet's hydrogen cloud in the full-sky Lyman-$\alpha$ images taken
by the SWAN camera on board SOHO on 22~dates between October~24 and
November~23.  In addition to the production rates calculated from the
individual images, Combi et al.\ also derived daily averages of the
initial water production in the nucleus' proximity, using their
time-resolved model.  This information is particularly useful to
investigations of temporal variations, including initiation of outbursts,
in the comet's outgassing.  Combi et al.'s data are plotted, together with
the others, in Figures~2--5 and compared with the comet's light curve.

Temporal variations in the comet's water production were peculiar.  Since
the first detection, by Schleicher (2013a) 2.5 months before
perihelion,\footnote{A water production rate of \mbox{$3 \times
10^{26}$\,molecules s$^{-1}$} reported by Schleicher (2013b) for 2013 March~5
{\vspace{-0.03cm}}was inferred from a CN production rate of \mbox{$1.3 \times
10^{24}$\,molecules s$^{-1}$}.} the average rate of increase over the next
7--8 weeks followed an $r^{-0.95}$ dependence (Figure 2) and can be
described as sluggish at best.  Because of the noise, we see no obvious
signature of the cycle E's ignition point in the daily averages of the
water production rate, but there is some evidence for the stagnation
point around November~6 (Figures~4--5) and for the depletion stage in the
days up to November~10.  A minor surge began the next day, but by November~14
the production rate was a factor of 15 higher than five days earlier.  This
outburst coincided almost exactly with Event~1 in the light curve, as also
noticed by Combi et al.\,(2014).  After some decline, the water production rate
jumped another factor of 3 on November~18, when it was nearly {\it two orders
of magnitude\/} higher than three weeks earlier.  This sharp peak coincided
with the very beginning of Event~2 in the light curve.  Between November~18
and the end of the SWAN data, the production rate's daily averages show
a gradual, possibly accelerating, drop.  No production of water was detected
by Curdt et al.\ (2014) 0.7~hour before perihelion and by Combi et al.\
(2013, 2014) 5~days after perihelion.

M.\ Drahus reported in two web communications that monitoring the HCN
emission at the IRAM observatory in Spain on November 21--25 indicated
a dramatic {\it production drop\/} by a factor of at least
20.\footnote{These reports are at {\tt http://groups.yahoo.com/neo/groups/
comets-ml/conversations/messages/22461}\,\&\,{\tt 22470};\,and\,on\,IRAM
site: {\tt http://www.iram-institute.org/EN/news/2013/83.html}.}
Because observations show only small variations in the comet's HCN/H$_2$O
production rate ratio (Coulson et al.\ 2013, Meech et al.\ 2013a, Paganini
et al.\ 2013, Biver et al.\ 2013, Ag\'umdez et al.\ 2014), the sharp drop
in the production of HCN was likely to accompany a similarly sharp drop in
the water production during the 5-day period.

It turns out that, just like H$_2$O, a peak production of HCN, a parent
molecule, coincided with Event~1, while its next peak was lagging the peak
of the daily averages of water production, related to Event~2, by some
2--3 days, according to the preliminary findings (see the plot in the
IRAM web news article; footnote~5).  Only a weak HCN signal was detected
with the IRAM dish about 1~day before perihelion.

Short-term violent upturns in the production of water that began two weeks
before perihelion suggest that the comet became prone to fragmentation,
which appears to be the only avenue for activating new, sizable sources of
water ice --- deep in the interior of the nucleus.  In the November~14
update to the first author's ``on-the-go'' investigation of the comet
(Sekanina 2013a), he stated that it was ``unclear whether the \ldots
nature [of Event~1] is benign or cataclysmic.''\footnote{Cometary outbursts,
quite common phenomena, are entirely undiagnostic as to the future evolution
of the afflicted objects:\ after experiencing an outburst, some comets
behave as if nothing happened, some split, and only a tiny minority of them
cataclysmically disintegrate.}  In retrospect, it is likely that the physical
mechanism responsible for Event~1 was the trigger that ``catapulted'' the
comet toward its self-destruction by abruptly opening some of the nucleus'
interior to the Sun's radiation.  Thus, strictly, Event~1 was not
cataclysmic over a period of a few days, but it was damaging in the sense
that the nucleus was unable to recover afterwards.  It is most likely that
Event~1 was accompanying an early breakup of the nucleus (Sec.~4.2),
an inference that is supported by Boehnhardt et al.'s (2013) report of 
arclet-like wings.  As computer tests suggest, one could not expect any
companion(s) to have been detected because by the time of the last
ground-based imaging, on November 22.8 UT, or less than 9~days after
the putative fragmentation event, typical secondary nuclei would have
been separated less than 4 arcsec from the main mass and buried in the
coma.  Later on, when the comet was monitored by instruments on board
SOHO and STEREO, the spacecraft imagers had inadequate spatial resolution.
Judging from the skyrocketing water production rates, the initial breakup
cascaded --- as it usually does --- into a set of fragmentation events
that were fatal.  If Drahus' preliminary report of a rapid decrease in
the gas production by November 25 is (as expected) confirmed, then in
the last three days before perihelion the comet's nucleus must have
continued to orbit the Sun in a formation resembling a cluster of
boulders and pebbles continuously crumbling into ever smaller pieces,
eventually ending up with a {\it dust cloud\/}.

The reader will notice that this evolution led to a paradoxical situation
in the sense that when many in the cometary community rejoiced at the comet
finally brightening substantially, it was ``on the ropes.''  Actually, all
that is needed to make a comet reach apparent magnitude $-$2 at a distance of
0.075~AU from the Sun (the case of C/2012~S1 $\sim$16~hours before perihelion)
is to achieve a cross-sectional area of dust of $\sim$10$^6$\,km$^2$, which is
satisfied by an optically thin cloud of less than 10$^{12}$\,g of independent
moderate-density particles 0.5~micron across!

\subsection{Dimensions and Mass Loss of the Nucleus}

Information on the production of water can be integrated to obtain a fairly
reliable estimate for the mass losses suffered by the comet's nucleus over
a period of at least two months.  Since the nuclear mass --- and therefore
dimensions --- must have been noticeably diminishing with time, a meaningful
exercise in computing the comet's mass loss ought to be anchored to a starting
point at which the nuclear size has been measured to a fairly high degree of
accuracy.  Li et al.\ (2013) observed the comet with a camera on board the
Hubble Space Telescope on 2013 April 10, at a heliocentric distance of 4.15
AU, but were able only to determine that the nucleus was less than 4~km in
diameter.

A much tighter limit on the nuclear size was reported by Delamere et al.\
(2013) from their detection of the comet with the Mars Reconnaissance
Orbiter's High Resolution Imaging Science Experiment (HiRISE) around the time
of the close encounter with Mars on 2013 October~1.7 UT.  On the assumption
that the signal in the brightest pixel was due exclusively to the comet's
nucleus, it is found --- at an adopted geometric albedo of 4 percent and with
a constant correction of 0.04 magnitude per degree of phase angle --- that
the nuclear diameter was \mbox{$1.0 \pm 0.07$}~km,\footnote{An albedo of
4~percent is the authors' assumption; Delamere et al.\ actually assumed an
albedo of 3~percent and the upper limits on the diameter they derived from four
observations were 1.25, 1.12, 1.05, and 1.12 km.} which of course is still
an upper limit on the true diameter.  With a bulk density of 0.4~g~cm$^{-3}$,
the corresponding upper limit on the comet's mass at the time of the encounter
with Mars is \mbox{$2.1 \times 10^{14}$}\,g.

It is now possible to assess the comet's mass loss of water between the close
encounter with Mars on October~1 and the complete disintegration of the
nucleus, which will be shown in Sec.~3 to occur a few hours before perihelion,
on November 28.  For the period October~24 through November~21, the water
losses are given by Combi et al.'s (2014) daily averages of the production
rate, equal to \mbox{$1.81 \times 10^{13}$\,g}.  For the period October~1
through 24, the losses are computed by accepting the dependence of the
production rate on heliocentric distance, $r^{-0.95}$, linked to Combi
et al.'s daily average on October~24.  This contribution equals merely
\mbox{$0.08 \times 10^{13}$\,g}.  Finally, for the period November 21
through 28, the losses of water are estimated by scaling up the HCN production
rate from a preliminary curve in the IRAM web news item (footnote~5), linked
to Combi et al.'s H$_2$O daily-average production rate on November~21.  This
contribution comes out to be \mbox{$0.43 \times 10^{13}$\,g}, with an
estimated uncertainty of some \mbox{$\pm 0.2 \times 10^{13}$\,g}.  The total
mass loss of water (which we equate with the total gas mass loss) over the
entire 58-day period from the Mars encounter on amounts to \mbox{$2.32 \times
10^{13}$\,g}, with an error of about $\pm$15~percent.

Next we estimate the comet's mass loss of dust in~the same period of time.
Using an instrumentation on board the Swift Space Telescope between October~7
and November~7, {\vspace{-0.04cm}}Bodewits et al.\ (2013) derived, from
their measurements of the 3090~{\AA} emission feature of OH, values of the
water production rate in fair agreement with Combi et al.'s (2014) results,
and from their measurements of the continuum, values of \mbox{\it Af}$\rho$,
a dust-production rate proxy.  By converting these values to dust production
rates following A'Hearn et al.\ (1995), we find an average dust-to-water
mass production rate ratio $\Re_0$ to amount to \mbox{$1.5 \pm 0.2$}.
Assuming that it remained nearly constant throughout the 58~day period,
the total mass loss from the comet, which equals the mass of the nucleus
at the Mars encounter, amounts to \mbox{$5.8 \times 10^{13}$\,g} and, at
an assumed bulk density of 0.4~g~cm$^{-3}$, is equivalent to the nucleus'
diameter of 0.65~km. with an estimated uncertainty of about $\pm$0.05~km
at a fixed bulk density and dust-to-water production rate ratio.

It therefore appears that the signal in the brightest pixel of the HiRISE
detector was indeed contaminated significantly by dust.  Considering that
the geometric cross-sectional area of a sphere 1~km across is 0.785~km$^2$
and that Delamere et al.'s (2013) phase correction at the phase angles
between 47$^\circ$ and 51$^\circ$) was between 1.88 and 2.04 magnitudes,
it follows that the total signal detected in the brightest pixel was
equivalent to an observed cross-sectional area of 0.129~km$^2$ on the
assumption of a geometric albedo of 0.04.  Because a nucleus 0.65~km across
accounts for a signal equivalent to an observed cross-sectional area of
0.055~km$^2$, an observed area of 0.074~km$^2$ is to be accounted for by
the contaminating dust.  Noting that the Marcus (2007) phase law for dust
provides in this case a correction of 0.90 magnitude, we conclude that
the geometric cross-sectional area of the dust is equal to 0.17~km$^2$.
Using some simplifying but inconsequential assumptions, we next examine
whether and under what conditions can a self-consistent model account for
this cross-sectional area of dust in the brightest pixel of the HiRISE imager.

\subsection{Accounting for Contamination of the Nucleus' Signal by Dust
Ejecta}

To estimate or constrain a contribution from dust that could contaminate
the nucleus' signal detected by the HiRISE imager, we consider spherical
dust particles with radii between $a$ and \mbox{$a + da$} released from
the nucleus of C/2012~S1 between times $t$ and $t \!+\! dt$,
\mbox{$\dot{N}(t) \, dt f(a) \, da$}, where \mbox{$\dot{N}(t)$} is the
total number of particles ejected per unit time interval at time $t$ and
\mbox{$f(a) \, da$} is a normalized particle-size distribution law.  We
require that particle radii be limited to a range of \mbox{$a_{\rm min}
\leq a \leq a_{\rm max}$} and that therefore
\begin{equation}
\int_{a_{\rm min}}^{a_{\rm max}} \!\!\! f(a) \, da = 1.
\end{equation}
Reckoning $t$ backward from the time of observation, the geometric
cross-sectional area of ejected particles with radii between $a$ and
\mbox{$a + da$} is
\begin{equation}
dX_{\rm dust} = \pi a^2 \! f(a) \, da \!\! \int_{0}^{\infty} \!\!\!
 \dot{N}(t) \, dt.
\end{equation}

Delamere et al.\ (2013) state that at closest approach the HiRISE pixel
scale at the comet was 13~km.  Since the comet's dimensions were computed
from the data at phase angles of 47$^\circ$--51$^\circ$, the pixel size
was then about 17~km and its cross-sectional area was equivalent to that
of a circle 9.6 km in radius.  We call this radius $r_{\rm pix}$ and limit
our investigation to a case in which (i)~the nucleus is at the pixel's
center, (ii)~the dust emission is isotropic, and (iii)~particles move
radially away from the nucleus with a constant expansion velocity that
depends on particle size.  Thus, all particles of the same size and ejected
at the same time populate an expanding spherical surface around the nucleus.
In this approximation of true observational circumstances, the signal from
the nucleus is contaminated by dust particles as long as they stay in the
volume circumscribed by a cylindrical surface whose axis points to the
observers' spacecraft and whose cross-sectional area equals that of the
pixel.  While all dust at distances \mbox{$r_{\rm dust} < r_{\rm pix}$}
from the nucleus contaminates its signal, only a fraction of dust at
\mbox{$r_{\rm dust} > r_{\rm pix}$} does so.  These latter particles move
in directions other than along a perpendicular to the line of sight and
their fraction, decreasing with increasing $r_{\rm dust}$, is determined
by the intersection of the expanding spherical surface with the cylindrical
surface of radius $r_{\rm pix}$.  In general, the cross-sectional area of
particles with radii between $a$ and \mbox{$a + da$} that contaminate the
nucleus' signal is
\begin{equation}
dX_{\rm cntm} = \pi a^2 \! f(a) \, da \!\! \int_{0}^{\infty} \!\!\!
  \dot{N}(t) \, \Phi(a,t) \, dt,
\end{equation}
where the function $\Phi(a,t) \leq 1$.

\begin{table*}[ht]
\vspace{0.1cm}
\begin{center}
{\footnotesize {\bf Table 2} \\[0.1cm]
{\sc Computed Total Geometrical Cross-Sectional Area of Dust
 Ejecta Contaminating Nucleus' Signal.}\\[0.1cm]
\begin{tabular}{c@{\hspace{0.2cm}}c@{\hspace{0.65cm}}c@{\hspace{0.4cm}}c@{\hspace{0.45cm}}c@{\hspace{0.45cm}}c@{\hspace{0.75cm}}c@{\hspace{0.4cm}}c@{\hspace{0.5cm}}c@{\hspace{0.4cm}}c}
\hline\hline\\[-0.2cm]
& & \multicolumn{8}{@{\hspace{-0.15cm}}c}{Total geometric cross-sectional
 area $X_{\rm cntm}^{(\tau)}$ of contaminating dust (km$^2$)}\\[-0.04cm]
& & \multicolumn{8}{@{\hspace{-0.18cm}}c}{\rule[0.6ex]{11.1cm}{0.4pt}}\\[-0.04cm]
Particle bulk & Maximum particle
 & \multicolumn{4}{@{\hspace{-0.4cm}}c}{Size distribution law $a^{-3.5} \, da$
 with $a_{\rm min}$}
 & \multicolumn{4}{@{\hspace{-0.15cm}}c}{Size distribution law $a^{-3.75} \, da$
 with $a_{\rm min}$}\\[-0.04cm]
density, $\rho$ & radius, $a_{\rm max}$
 & \multicolumn{4}{@{\hspace{-0.43cm}}c}{\rule[0.6ex]{5.4cm}{0.4pt}}
 & \multicolumn{4}{@{\hspace{-0.18cm}}c}{\rule[0.6ex]{5.4cm}{0.4pt}}\\[-0.04cm]
(g\,cm$^{-3}$) & (cm) & 0.05\,$\mu$m & 0.1\,$\mu$m & 0.2\,$\mu$m & 0.5\,$\mu$m
   & 0.05\,$\mu$m & 0.1\,$\mu$m & 0.2\,$\mu$m & 0.5\,$\mu$m\\[0.1cm]
\hline \\[-0.2cm]
0.4 & \llap{1}75  & 0.100 & 0.074 & 0.055 & 0.039
                  & 2.396 & 1.455 & 0.890 & 0.472 \\
0.8 & 87\rlap{.5} & 0.074 & 0.055 & 0.042 & 0.030
                  & 1.455 & 0.890 & 0.550 & 0.297 \\
1.2 & 58\rlap{.3} & 0.062 & 0.047 & 0.036 & 0.026
                  & 1.090 & 0.671 & 0.417 & 0.228 \\
1.6 & 43\rlap{.8} & 0.055 & 0.042 & 0.032 & 0.024
                  & 0.890 & 0.550 & 0.344 & 0.190 \\
3.5 & 20\rlap{.0} & 0.041 & 0.031 & 0.025 & 0.019
                  & 0.517 & 0.324 & 0.207 & 0.118 \\[0.1cm]
\hline\\[-0.2cm]
%
%
\end{tabular}}
\end{center}
\end{table*}

More specifically, the computation of the fraction of particles on the
surface of a sphere of radius $r_{\rm dust} > r_{\rm pix}$ that contaminate
the nucleus' signal involves a spherical cap circumscribed by the cylindrical
surface.  The fraction is given by the ratio of the cap's height from its
base of radius $r_{\rm pix}$ to the sphere's radius $r_{\rm dust}$ and equals
\mbox{$1 - \sqrt{1 - (r_{\rm pix}/r_{\rm dust})^2}$}.  It is noted that
the fraction of contaminating dust particles drops rapidly with increasing
radius of the sphere.  When $r_{\rm dust}$ is 1.5 times larger than $r_{\rm
pix}$ the fraction is 0.255; when it is three times larger, the fraction drops
to 0.057; and when $r_{\rm dust}$ is ten times as large as $r_{\rm pix}$, the
fraction is down to 0.005.  Since the expansion velocity of released
particles is allowed to depend on their size, the spherical surface of the
same radius $r_{\rm dust}$ is populated by dust that left the nucleus at
different times; the larger the particles the earlier their release.  For
the particle-size dependence of the expansion velocity $\upsilon(a)$, which
we assume to be independent of the distance from the nucleus, we adopt
(e.g., McDonnell et al.\ 1987)
\begin{equation}
\upsilon(a) = \frac{\upsilon_0}{1 \!+\! \chi \sqrt{a}},
\end{equation}
where $\upsilon_0$ is the peak expansion velocity for $a \rightarrow 0$
and $\chi > 0$ determines the rate of velocity decrease with size and can
be expressed in terms of a radius $a_h$ of particles whose expansion
velocity is ${\textstyle \frac{1}{2}} \upsilon_0$,
\begin{equation}
\chi = a_h^{-\frac{1}{2}}.
\end{equation}
The relation between time, that is, the age of a released particle at
the time of observation, as the sphere's radius $r_{\rm dust}$ is simply
\begin{equation}
r_{\rm dust} = \upsilon(a) \cdot t = \frac{\upsilon_0 t}{1 \!+\! \chi
 \sqrt{a}}.
\end{equation}
Defining
\begin{equation}
t_{\rm pix} = \frac{r_{\rm pix}}{\upsilon_0} (1 \!+\! \chi \sqrt{a}),
\end{equation}
the total cross-sectional area of particles with radii between $a$ and
\mbox{$a + da$} that contaminate the nucleus' signal is from Eq.\,(5)
\begin{eqnarray}
dX_{\rm cntm} & = & \langle \dot{N} \rangle \pi a^2 \! f(a) \, da
 \nonumber \\[0.15cm]
 & & \times \! \left\{ \! t_{\rm pix} \!+\!\! \int_{t_{\rm pix}}^{\infty}
     \!\! \left[ \! 1 \!-\!  \sqrt{1 \!-\!  \left( \! \frac{t_{\rm pix}}{t}
     \!\! \right)^{\!2}} \, \right] \! dt \! \right\} \! ,
\end{eqnarray}
where $\langle \dot{N} \rangle$ is an average value of $\dot{N}(t)$ over the
ejection times of contaminating particles.  This approximation can easily
be justified, because we showed above that the distances $r_{\rm dust}$,
from which large enough fractions of contributing particles come, are only
tens of kilometers, which we will see are even for large particles
equivalent to ejection times that precede the observation by only a day
at the most.  The integral in Eq.\,(10) equals
\begin{eqnarray}
\int_{t_{\rm pix}}^{\infty} \!\! \left[ \! 1 \!-\! \sqrt{1 \!-\! \left( \!
  \frac{t_{\rm pix}}{t} \!\! \right)^{\!2}} \, \right] \! dt & = & t_{\rm pix}
  \!\! \int_{0}^{1} \!\! \frac{x (1 \!-\! x)}{(1 \!-\! x^2)^{\frac{3}{2}}}
  \, dx \nonumber \\
& = & t_{\rm pix} \! \left(\frac{\pi}{2} \!-\! 1 \right) \! ,
\end{eqnarray}
so that
\begin{equation}
dX_{\rm cntm} = \frac{1}{2} \langle \dot{N} \rangle \pi^2
   \frac{r_{\rm pix}}{\upsilon_0} \, a^2 \! f(a) \, da,
\end{equation}
where $\langle \dot{N} \rangle$ is an average number of dust particles
released per unit time around October 1, the time of encounter.

Before integrating over all particle sizes from $a_{\rm min}$ to $a_{\rm max}$,
we express $\langle \dot{N} \rangle$ in terms of an equivalent mass dust
production rate, $\dot{\cal M}_{\rm dust}$, which we already computed from
the water production rate and the dust-to-water mass production rate ratio
in Sec.~2.2.  Since for $\dot{\cal M}_{\rm dust}$ one can write
\begin{equation}
\dot{\cal M}_{\rm dust} = \frac{4}{3} \pi \rho \, \langle \dot{N} \rangle
  \!\! \int_{a_{\rm min}}^{a_{\rm max}} \!\! a^3 \! f(a) \, da,
\end{equation}
a general formula for the total cross-sectional area $X_{\rm cntm}$ of dust
particles that contaminate the signal of the nucleus in the HiRISE images is
\begin{equation}
X_{\rm cntm} = \frac{3\pi}{8} \, \frac{r_{\rm pix}}{\upsilon_0} \,
  \frac{\dot{\cal M}_{\rm dust}}{\rho} \, \frac{{\displaystyle
  \int_{a_{\rm min}}^{a_{\rm max}}} \!\!  (1 \!+\!  \chi \sqrt{a}) \, a^2
  \! f(a) \, da}{{\displaystyle \int_{a_{\rm min}}^{a_{\rm max}}} \!\! a^3
  \!f(a) \, da}.
\end{equation}

In this expression the independent entities are the particle-size distribution
law \mbox{$f(a)\,da$}, the minimum particle size $a_{\rm min}$, and the bulk
density $\rho$ of dust particles.  Once these are prescribed, our exercise
can proceed based on the already determined parametric values,
\mbox{$\dot{\cal M}_{\rm dust} = 5.25 \times 10^5$\,g s$^{-1}$} and
\mbox{$r_{\rm pix} = 9.6$ km}.  The three remaining parameters, $a_{\rm max}$,
$\upsilon_0$, and $\chi$, are related to other physical quantities, as shown
in part by Delsemme \& Miller (1971) in their study of comet C/1959~Y1 and in
part by Sekanina (1981) in his study of comet 109P, which for a particle
expansion velocity employed an expression of the same type as is Eq.\,(6).
The value of $a_{\rm max}$ depends on the water production rate and its
efflux velocity, on the mass of the nucleus, and on the bulk density of dust
particles; the value of $\upsilon_0$ is a function of the thermal velocity
of water molecules and the dust-to-water mass production rate ratio; and the
value of $\chi$ depends on these two and also on the water production rate,
the size of the nucleus, and the bulk density of dust particles.  To limit the
number of variable parameters, we use only the upper limits for the nucleus'
mass and size, and obtain the following expressions for the three
parameters:\ \mbox{$\upsilon_0 = 0.57$ km s$^{-1}$}, \mbox{$\chi = 11.4
\sqrt{\rho}$ cm$^{-\frac{1}{2}}$}, and \mbox{$a_{\rm max} = 70/\rho$ cm}.

Four particle-size distribution laws are employed, of the type
\begin{equation}
f(a)\, da = n_0\! \left( \frac{a_{\rm min}}{a}\!\right)^{\!\tau} \! da,
\end{equation}
where $n_0$ is a normalizing constant [see Eq.\,(3)].  In conformity with
the results of broad studies (e.g., Fulle 1999) of dust-particle size
distributions in large number of comets, we choose for the power index
$\tau$ the values of 3.0, 3.5, 3.75, and 4.0.  After integrating Eq.\,(14),
we obtain a total particle cross-sectional area $X_{\rm cntm}^{(\tau)}$
for \mbox{$\tau = 3.0$}
\begin{equation}
X_{\rm cntm}^{(3.0)} = \frac{3 \pi}{4} \, \frac{r_{\rm pix}}{\upsilon_0} \,
  \frac{\dot{\cal M}_{\rm dust}}{\rho \, a_{\rm min}} \, \frac{\varepsilon}{1
  \!+\! \varepsilon} \! \left[ \chi \sqrt{a_{\rm min}} - \! \frac{\varepsilon}
  {1 \!-\! \varepsilon} \ln \varepsilon \right] \!;
\end{equation}
for \mbox{$\tau = 3.5$}
\begin{equation}
X_{\rm cntm}^{(3.5)} = \frac{3 \pi}{8} \, \frac{r_{\rm pix}}{\upsilon_0} \,
  \frac{\dot{\cal M}_{\rm dust}}{\rho\,a_{\rm min}} \, \varepsilon \! \left[ 1
  \!-\! \frac{\chi \sqrt{a_{\rm min}}}{1 \!-\! \varepsilon} \, \ln \varepsilon
  \right] \!;
\end{equation}
for \mbox{$\tau = 3.75$}
\begin{equation}
X_{\rm cntm}^{(3.75)} = \frac{\pi}{8} \, \frac{r_{\rm pix}}{\upsilon_0} \,
  \frac{\dot{\cal M}_{\rm dust}}{\rho \, a_{\rm min}} \sqrt{\varepsilon}
  \left(1 \!+\! \sqrt{\varepsilon} \!+\! \varepsilon \!+\! 3 \chi
  \sqrt{a_{\rm min}} \, \right)\!;
\end{equation}
and for \mbox{$\tau = 4.0$}
\begin{equation}
X_{\rm cntm}^{(4.0)} = - \frac{3 \pi}{16} \, \frac{r_{\rm pix}}{\upsilon_0} \,
  \frac{\dot{\cal M}_{\rm dust}}{\rho \,  a_{\rm min}} \, \frac{1 \!-\!
  \varepsilon}{\ln \varepsilon} \left(1 \!+\! \varepsilon \!+\! 2 \chi
  \sqrt{a_{\rm min}} \right) \!;
\end{equation}
where
\begin{equation}
\varepsilon = \sqrt{\frac{a_{\rm min}}{a_{\rm max}}} \ll 1.
\end{equation}

\begin{figure*}[ht]
\vspace{-1.85cm} 
\hspace{-0.05cm}
\centerline{
\scalebox{0.85}{
\includegraphics{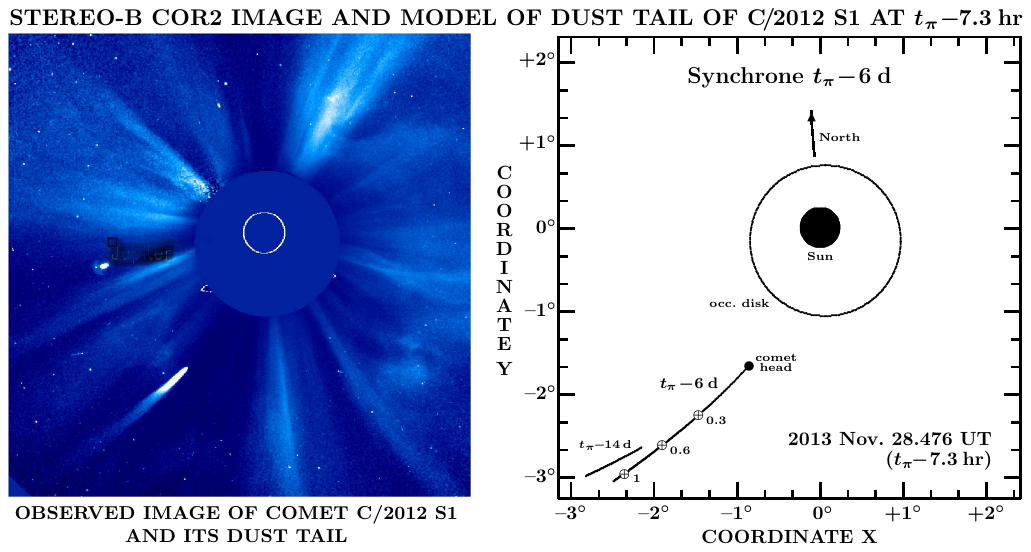}}} 
\vspace{-14.35cm} 
\caption{Appearance of comet C/2012 S1 and its dust tail in an image taken
with the COR2 coronagraph on board the STEREO-B spacecraft on November
28.476 UT, 7.3 hours before perihelion and 8.8\,\,{\Rsun} from the Sun.
The tail, to the{\vspace{-0.06cm}} southeast, is nearly straight and
extends faintly almost to the edge of the frame.  Its observed orientation
and shape conform, in the panel to the right of the image, to a synchrone,
a locus of projected positions of dust particles released from the comet's
nucleus about 6~days before perihelion, \mbox{$t_\pi \!-\! 6$ d}, thus
correlating with Event~2.  The synchrone is calibrated with values of
$\beta$, a particle acceleration by solar radiation pressure, expressed
in units of the Sun's gravitational acceleration.  The brightest part of
the tail corresponds to \mbox{$\beta < 0.2$}, typical for dust grains
a few microns across and larger. The tail becomes extremely faint at
\mbox{$\beta > 0.6$}, submicron-sized absorbing particles (such as metals
or carbon-rich).  Also shown in the panel is a section of a synchrone that
describes the projected positions of dust ejecta from Event~1, 14 days
before perihelion, \mbox{$t_\pi \!-\! 14$ d}.  The orientation of this
synchrone is not in accord with the observed tail.  The panel also offers
information on the scale and orientation of the image. The bright spot to
the northeast of the comet is Jupiter. (Image credit:\ NASA/SECCHI
consortium.){\vspace{0.4cm}}}
\end{figure*}

For no plausible values of $\rho$ and $a_{\rm min}$, did the distribution
laws with $\tau = 3$ and 4 satisfy the condition \mbox{$X_{\rm cntm} =
0.17$ km$^2$}; the calculated area was{\nopagebreak} always near
0.003~km$^2$ with the first law{\nopagebreak} and always greater than
0.6~km$^2$ with the second law.  Table~2 lists the results obtained with
the distribution laws $a^{-3.5}\,da$ and $a^{-3.75}\,da$, showing that
all cross-sectional areas computed with the first law are smaller than
0.17~km$^2$ and, with one exception, all cross-sectional areas for the
second law are greater than 0.17~km$^2$.  Thus, the two laws --- both
in a range typical for size distributions of cometary dust --- provide  
limits for the solutions that are consistent with the condition for the
cross-sectional area of the dust contaminating the nucleus' signal in the
brightest pixel of the HiRISE imager.  Accordingly, we conclude that our
result is self-consistent:\ from now on we adopt 0.65~km as a nominal
diameter of C/2012~S1 at the Mars enounter time on October~1,~2013.

Given the small size of the nucleus, the estimated water production rate of
\mbox{$\sim \! 1.2 \times 10^{28}$}\,molecules s$^{-1}$ at 1.64~AU from the
Sun, at the time of encounter, shows a very impressive level of activity, to
say the least.  As addressed further in Sec.~4.2, the water production from
the entire sunlit hemisphere of a nucleus 0.65~km in diameter should amount
to \mbox{$\sim\!0.3 \times 10^{28}$}\,molecules s$^{-1}$, or only about one
quarter of the observed rate.  It is therefore suggested that most of the
water vapor detected in the comet's atmosphere at distances well over 1~AU
from the Sun was released from ejected icy-dust grains, whose total
surface was more than sufficient to accomplish such a task.  This scenario
does not rule out the sluggish rate of increase in the water production
with decreasing heliocentric distance during October and early November.

%

\section{Morphology of the Comet and Its Tail on SOHO and
STEREO images}
Our primary objective in this section is close inspection of intrinsic
changes in the comet's morphology as detected in the images taken with the
coronagraphs on board SOHO (Brueckner et al.\ 1995) and STEREO-A and B
(Howard et al.\ 2008) near perihelion.  The results confirm that the
correct answer to the question of what happened to the comet was
available {\it shortly\/} after its perihelion passage (see Sekanina 2013b).
The confusion caused by conflicting statements on the comet's survival or
demise, reported in the media for days after perihelion and certainly not
beneficial to the cometary community's reputation, could easily have been
avoided.

Another objective in this section is to examine the manifestations,
in the SOHO and STEREO images, of the history of the comet's
activity and to correlate these results with those in Secs.~2.1--2.2.
We are taking advantage of a powerful stereoscopic capability
provided by the spatial configuration of the three spacecraft.

In the following, we model the morphology of the comet and its
tail in eight particular images taken with the coronagraphs on
board the three spacecraft over a period of time covering some
33 hours around perihelion.  Additional images are then employed
to investigate two more specific issues that have to do with the
morphology of the head and sublimation of dust in the tail.

\begin{figure*}
\vspace{-2.7cm} 
\hspace{0.45cm}
\centerline{
\scalebox{0.895}{
\includegraphics{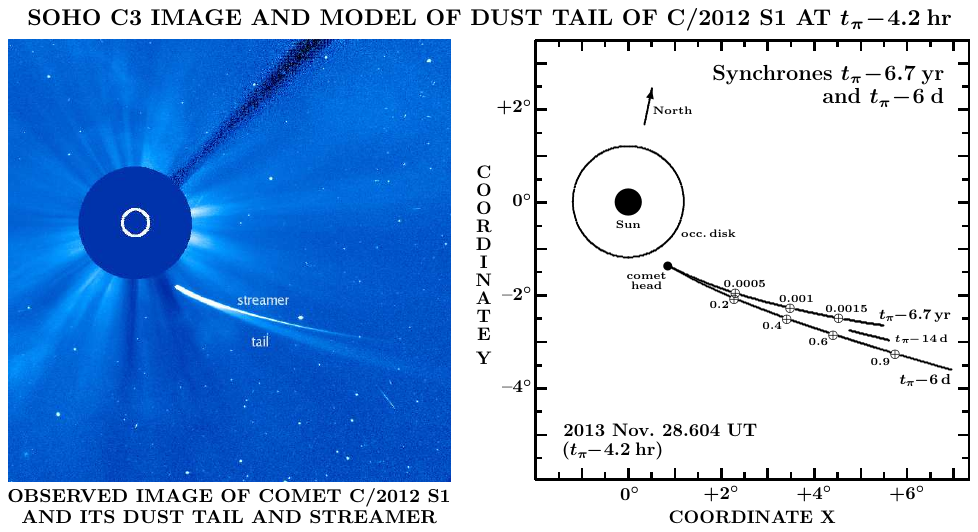}}} 
\vspace{-14.73cm}
\caption{Appearance of comet C/2012 S1 and its dust tail and a streamer in an
image taken with the C3 coronagraph on board the SOHO spacecraft on November
28.604 UT, 4.2 hours before perihelion and 5.9\,\,{\Rsun} from the Sun.  The
tail,{\vspace{-0.04cm}} the broader and less curved part of the bifurcated
feature, extends to the edge of the frame, but becomes faint at distances
greater than about 5$^\circ$ from the head.  As seen from the right panel,
the orientation and shape of this tail conforms to a synchrone for dust
released from the comet's nucleus about 6~days before perihelion, thus again
correlating with Event~2.  As in Figure 6, the synchrone is calibrated with
values of $\beta$, particle accelerations by the Sun's radiation pressure,
expressed in units of the Sun's gravitational acceleration.  The brightest
part of the tail corresponds to \mbox{$\beta < 0.2$}, typical for dust
particles several microns across and larger.  Its faintest part is populated by
submicron-sized absorbing particles with \mbox{$\beta \gg 0.6$}.  Also shown
in the panel is a section of a synchrone for dust ejecta from Event~1, which
appear to contribute to the northern edge of the main tail.  By contrast, the
streamer is made up of sizable grains released from the comet at very large
heliocentric distances; it is modeled by a synchrone for dust ejected from
the comet at a distance of 20~AU from he Sun, 6.7~years before perihelion.
For more description, see the caption to Figure~6. (Image credit:\
ESA/NASA/LASCO consortium.){\vspace{0.35cm}}}
\end{figure*}

\subsection{Image Taken 7.3 Hours Before Perihelion}

To begin with, the left panel of Figure 6 presents the comet's preperihelion
image taken with the COR2 coronagraph on board STEREO-B on November 28.476 UT.
The comet, a little less than 2$^\circ$ from the Sun, displays a narrow,
slightly curved tail that deviates about 15$^\circ$ from the antisolar
direction.  Both this orientation and a smooth appearance indicate the
tail's dust nature.  The right panel of the figure shows that the position
angle of the tail's axis and its curvature are matched by a synchrone
referring to the dust released around 6~days before perihelion, correlating
apparently with Event~2 in Figures~4 and 5.  The tail extends at least
1$^\circ\!$.7 from the head, its brightest part populated --- as shown by
the model in the right panel of Figure~6 --- by dust particles subjected
to a radiation-pressure accelerations \mbox{$\beta < 0.2$} the Sun's
gravitational acceleration.  Such particles are several microns across and
larger.  At a distance of 0$^\circ\!$.5 from the head, the tail's width is
estimated at $\sim$5~arcmin, which, interpreted as a particle-velocity
effect, implies an ejection velocity of $\sim$0.25 km s$^{-1}$, a plausible
value.  On the other hand, the effect of the Event~2's temporal extent on
the tail's width is found to barely exceed 1~arcmin for an estimated
duration of $\sim$3~days, thus contributing little to the observed width.
The phase angle was 59$^\circ$ at the head, but only 42$^\circ$ in the tail
1$^\circ$ away.

\subsection{Image Taken 4.2 Hours Before Perihelion}
The image in Figure 7, showing the tail to be bifurcated on November 28.604
UT, is a representative example of a long series of the comet's observations
made with the SOHO's C3 coronagraph between at least November 27.4 and 28.8
UT that exhibit this feature.  Testing of dust-emission models as well as a
set of C3 images after November 28.8 UT demonstrate conclusively that these
are in fact two independent and --- as it turns out --- very different tails
that, in projection onto the plane of the sky, overlap near the nucleus, but
deviate from one another farther away from it.

The longer, wider tail, making now an angle of almost 40$^\circ$ with the
radius vector, closely fits the synchrone for dust released in the course
of Event 2, centered on $t_\pi \!-\! 6$ days.  The tail's apparent width,
about 15 arcmin at 3$^\circ$ from the nucleus and only slowly increasing
with distance, suggests particle-ejection velocities of up to 0.6 km
s$^{-1}$, at an upper limit of plausible values.  However, it is likely
that the tail also contains a contribution from Event~1, merged with
that from Event~2 into a single feature (Figure 7), which would account for
about one half of the apparent width, and the required ejection velocities
would accordingly be lower, not exceeding 0.3~km~s$^{-1}$, and more in line
with expectation.

\begin{figure*}[ht]
\vspace{-2.8cm} 
\hspace{-0.22cm} 
\centerline{
\scalebox{0.96}{
\includegraphics{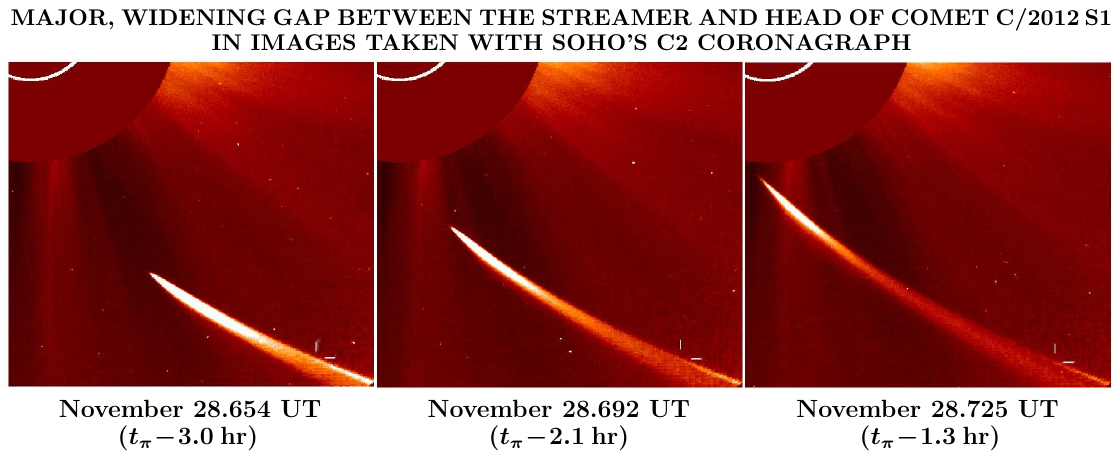}}} 
\vspace{-16.8cm} 
\caption{Examples of a major and widening gap, separating the disconnected
dust streamer from the disintegrating head of comet C/2012 S1 in images
taken with the C2 coronagraph on board SOHO shortly before perihelion. 
The streamer's forward tip, whose approximate location is marked by the short
ticks in the lower right corner of each image, receded rapidly from the
comet's head, but its position was nearly stationary relative to the Sun.
The occulting disk with a contour of the Sun is in the upper left corner.
The direction to the celestial north is essentially the same as in Figure 7.
Each frame is about 95 arcmin wide and 85 arcmin high, which at the Sun's
distance from the spacecraft translates to 4.0 by 3.6 million km. (Image
credit:\ ESA/NASA/LASCO consortium.){\vspace{0.4cm}}}
\end{figure*}

The narrow, shorter tail is a different story.  Referred to as a streamer
in Figure~7, this feature's orientation and curvature practically trace the
projected orbit behind the comet.  The modeling of dust-particle motions
shows that this is a typical property of the loci of sizable grains,
released at near-zero velocities early, at very large heliocentric distances.
The streamer could be called a developing {\it dust trail\/}, if it were
not for this term having been introduced specifically for a persistent
coarse-grained debris distributed along the orbits of periodic comets,
both in front of and behind the parent body (e.g., Sykes et al.\ 1986,
Sykes \& Walker 1992, Reach et al.\ 2000).

\begin{table}[h]
\vspace{-0.15cm}
\noindent
\begin{center}
{\footnotesize {\bf Table 3} \\[0.1cm]
{\sc Computed Position Angles, Radiation Pressure Accelerations, and Sizes
 of Dust Grains at Two Points of the Streamer in the C3 Image of November
 28.604 UT}\\[0.2cm]
\begin{tabular}{@{\hspace{0.3cm}}c@{\hspace{0.1cm}}c@{\hspace{0.15cm}}c@{\hspace{0.25cm}}c@{\hspace{0.12cm}}c@{\hspace{0.22cm}}c@{\hspace{0.25cm}}c@{\hspace{0.12cm}}c}
\hline\hline\\[-0.2cm]
\multicolumn{2}{@{\hspace{0.15cm}}c}{Released$^{\rm a}$}
 & \multicolumn{3}{@{\hspace{-0.05cm}}c}{At 4$^\circ\!$.5 from head$^{\rm b}$}
 & \multicolumn{3}{@{\hspace{0cm}}c}{At 1$^\circ\!$.5 from head$^{\rm b}$}
\\[-0.05cm]
\multicolumn{2}{@{\hspace{0.15cm}}c}{from\,comet}
 & \multicolumn{3}{@{\hspace{-0.05cm}}c}{\rule[0.7ex]{3.1cm}{0.4pt}}
 & \multicolumn{3}{@{\hspace{0cm}}c}{\rule[0.7ex]{3.1cm}{0.4pt}} \\[-0.05cm]
\multicolumn{2}{@{\hspace{0.15cm}}c}{\rule[0.7ex]{1.6cm}{0.4pt}}
 & Posi- & Accel- & Grain & Posi- & Accel- & Grain \\[-0.05cm]
\raisebox{0.05cm}{$t_{\rm rls}\:\;$} & \raisebox{0.05cm}{$r_{\rm rls}$} & tion
 & eration & diam. & tion & eration & diam. \\[-0.05cm]
(yr)$\;\;$ & (AU) & angle & $\beta$ &  (cm) & angle & $\beta$ & (cm) \\[0.05cm]
\hline \\[-0.25cm]
 & & \hspace{0.19cm}$^\circ$ & & & {\hspace{0.19cm}}$^\circ$ & & \\[-0.32cm]
\llap{$-$7}5.\rlap{2} & \llap{1}00 & 265\,.83 & 0.000165 & 1.60 & 259\,.33
 & 0.000043 & 6.49 \\
\llap{$-$2}6.\rlap{6} &  50 & 265.82 & 0.000466 & 0.52 & 259.33 & 0.000120
 & 2.22 \\
\llap{$-$}6.\rlap{7}  &  20 & 265.80 & 0.001836 & 0.11 & 259.32
 & 0.000473 & 0.51 \\
\llap{$-$}2.\rlap{4} & 10 & 265.73 & 0.005173 & 0.03 & 259.28 & 0.001334
 & 0.16 \\
\llap{$-$}0.\rlap{84} & $\;\:$5 & 265.58 & 0.014534 & 0.01 & 259.19 & 0.003746

 & 0.04 \\[0.05cm]
\hline\\[-0.2cm]
\end{tabular}}
\parbox{8.62cm}{\scriptsize $^{\rm a}$\,Time of release, $t_{\rm rls}$, is
 reckoned from the forthcoming perihelion; $r_{\rm rls}$ is heliocentric
 distance at the time of release.}\\[0.08cm]
\parbox{8.62cm}{\scriptsize $^{\rm b}$\,Acceleration $\beta$ by solar
 radiation pressure is expressed in units of the Sun's gravitational
 acceleration; grain diameter is computed from $\beta$ assuming that
 the efficiency for radiation pressure of the grains is unity and their
 bulk density satisfies Eq.\,(21) .} \\[0.25cm]
\end{center}
\end{table}

Table 3 lists the position angles of the streamer and the properties of the
grains that populate it at two distances from the comet's head, the first
being the streamer's observed length in Figure~7.  The table shows that the
time of release cannot be determined from the streamer's position, only
constrained to heliocentric distances larger than $\sim$5~AU, because the
modeled loci of grains released at different times practically overlay
one another.  Columns 3 and 6 demonstrate that differences in the position
angle of grains released between 5 and 100~AU from the Sun are always less
than 0$^\circ\!$.3, whereas the accuracy of orientation measurement is
certainly not better than $\pm$0$^\circ\!$.5.  However, since comets with
perihelia near or beyond 5~AU --- especially the dynamically new ones, from
the Oort Cloud --- are known to possess the same kind of a tail (e.g.,
Sekanina 1975; Meech et al.\ 2009), it is highly unlikely that the streamer
of C/2012~S1 formed at a distance comparable to or less than 5~AU.  Meech et
al.\ observed an Oort-Cloud comet C/2003~A2 {\it already with a tail\/} as
far as 11.5~AU from the Sun preperihelion. Given an extremely low solar
radiation pressure at those heliocentric distances, very extended periods
of time are needed to develop a long enough tail to detect.  The distance
of release of dust grains in the streamer is arbitrarily chosen to be 20~AU
in Figure 7, and the issue is addressed in greater detail in Sec.~5.

Table 3 shows that the {\it earlier\/} the release~of~the~grains that made up
the streamer, the {\it larger\/} they were.~Because of a broad span of grain
sizes, we introduce~a~size~depend\-ent bulk density $\rho$ (expressed in
g cm$^{-3}$),\\[-0.3cm]
\begin{equation}
\rho(\delta) = 3.5 - 3.1 \! \left[ 1 \!+\! \left( \!\frac{10}{\delta} \!
  \right)^{\!0.6} \right]^{-1} \hspace{0.2cm}(0.4 \!<\!
  \rho \!<\! 3.5), \\[-0.1cm]
\end{equation}
where $\delta$ is the grain's diameter in microns.  The streamer thus contained
mostly {\it millimeter-sized and larger\/} dust.

Given the very large heliocentric distances involved, one would expect that
the grains in the streamer were initially rich in ices, including water ice.
However, the streamer's survival to heliocentric distances of less than
0.1~AU rules out the presence of {\it pure or nearly pure\/} icy grains,
which would have sublimated away long ago.

Now why the streamer is not detected in the image in Figure~6?  At
1$^\circ\!$.5 from the nucleus its position angle should be 116$^\circ$,
7$^\circ$ to the north of the synchrone \mbox{$t_\pi \!-\! 14\:$d},
clearly separated from the tail.  The only explanation for its absence
that we can think of is an insufficient exposure time given the relatively
narrow, red-sensitive window of the COR2 coronagraph.  It should be
pointed out that the comet soon began to saturate the detector of the
C3 coronagraph but not of the COR2-B coronagraph.

\begin{figure*}[ht]
\vspace{-1.65cm} 
\hspace{-0.05cm}
\centerline{
\scalebox{0.83}{
\includegraphics{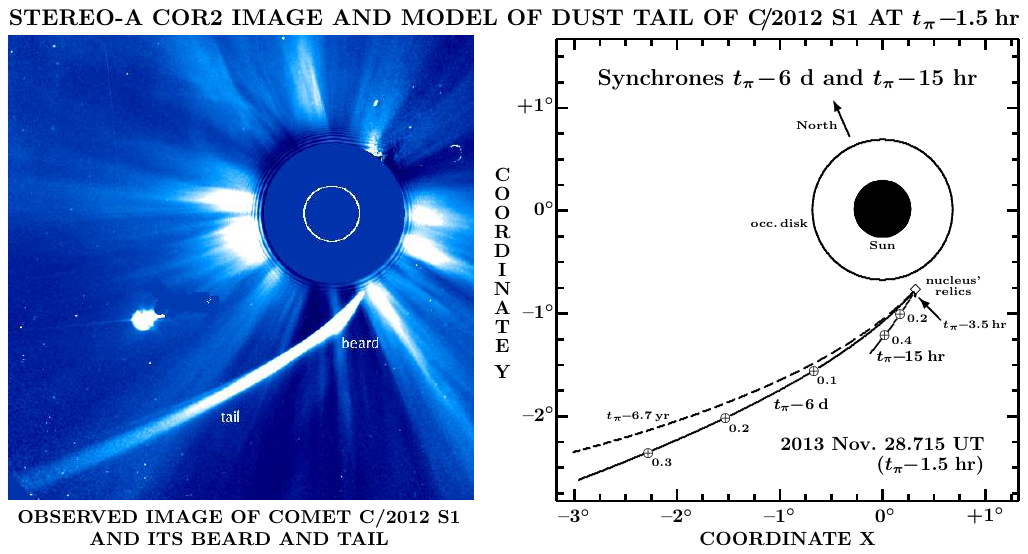}}} 
\vspace{-14.15cm} 
\caption{Appearance of comet C/2012 S1 and its dust tail and a ``beard'' in
an image taken with the COR2 coronagraph on board the STEREO-A spacecraft on
November 28.715 UT, 1.5 hours before perihelion and 3.4\,\,{\Rsun} from the
Sun.{\vspace{-0.05cm}}  A peculiarity of this image is the comet's displaying
a ``beard'' that emanates to the southeast from what was left of the comet's
nucleus and that is made up of microscopic dust released only about
\mbox{13--14} or so hours earlier.  The tail is once again matched best with
a synchrone representing the Event~2 ejecta, which consist of dust particles
several microns in diameter and larger.  The nucleus' relics are depicted by
a diamond.  For more description, see the captions to Figures~6 and 7. The
bright spot above the tail is Venus. (Image credit:\ NASA/SECCHI
consortium.){\vspace{0.42cm}}}
\end{figure*}

A remarkable phenomenon was a major gap, widening with time, that separated
the streamer from the comet's head.  The disconnected streamer shows up in
all images taken with the C2 coronagraph between November 28.65 and at least
29.11 UT, although only a short arc of it fits the instrument's field of view
and it becomes very faint after 28.8~UT.  Three examples are displayed in
Figure~8, covering a period of less than 2~hours.  The disconnection is also
seen in the images taken with the C3 coronagraph between November 28.7 and
at least 29.1 UT.  The forward tip of the streamer remained nearly stationary
relative to the Sun.  The obvious interpretation is in terms of rather sudden,
complete sublimation of grains in the streamer, once they approached the Sun
to a certain critical distance.  In projection this distance is estimated at
about 1$^\circ\!$.9 in Figure~8.  The absence of the streamer in the image
in Figure~6 cannot be explained by this effect.  The fascinating subject of
dust sublimation in the streamer is addressed at some length in Secs.~5.1--5.3.

The phase angle was 82$^\circ$ at the head, but 97$^\circ$ at 1$^\circ\!$.5
from it in the tail and 113$^\circ$ in the streamer, which implies a moderate
effect of forward scattering by the large grains (Marcus 2007).

\subsection{Image Taken 1.5 Hours Before Perihelion}
A peculiar feature immediately apparent in the image of the comet, taken on
November 28.715 UT with the COR2 coronagraph on board STEREO-A and shown in
Figure~9, is a bright ''beard'' extending a little more than 30 arcmin to
the east-southeast from the head.  This feature becomes apparent as early as
November 28.6 UT only in STEREO-A images.  Its sharp southern boundary is
best matched by a synchrone for dust emitted from the nucleus some 15~hours
before perihelion, at 15~{\Rsun} from the Sun.  In Figures 3--5, this
time correlates with Event~3, the last upsurge of brightness before its
eventual sharp decline.  A peak solar radiation pressure \mbox{$\beta_{\rm
peak}$} on particles in the beard is about 0.4 the Sun's gravitational
acceleration, suggesting that the particles are micron-sized and larger.

The comet's most prominent feature in this image is the long, slightly
curving dust tail, as in the images in Figures~6 and 7 (Secs.~3.1--3.2).
It again is matched best by a synchrone for mostly microscopic dust
(\mbox{$\beta < 0.4$}) emitted during Event 2, about 6 days before
perihelion.  As in the STEREO-B image in Figure~6, there is no clear
evidence of the streamer made up of old, massive grains.  The phase
angle decreased from 118$^\circ$ at the disintegrated nucleus and in the
beard to 112$^\circ$ in the tail 2$^\circ$ away.

\subsection{Image Taken 3.3 Hours After Perihelion}
Taken on November 28.918 UT with the C2 coronagraph on board SOHO, this is
the first post-perihelion image of the comet that we model.  It shows, in
Figure 10, an early stage of the most persistent feature observed along the
outbound leg of the orbit --- a fan-shaped cloud of dust.  It emanates from
an approximate site of the {\it nuclear condensation that is no longer
apparent\/}.  The eastern boundary of the fan is at this time --- unlike later
--- much fainter than the western boundary, and at first sight the fan may
appear to be narrower than it actually is.  The direction of the eastern
boundary provides a {\it key piece of information\/} on the comet --- the time
when the {\it emission of dust irrevocably ceased\/}.  From the boundary's
measurements on a number of images, this occurred {\it 3.5 hours before
perihelion\/} with an estimated uncertainty of $\pm$0.3 hour.  At that time,
the {\it disintegration\/} of the nucleus was apparently {\it completed\/}
and the {\it existence of C/2012~S1 was over\/} (Sec.~4.2).  Dust particles
that made up the eastern boundary of the fan were a few microns across and
larger (\mbox{$\beta < 0.3$}).  

\begin{figure*}[ht]
\vspace{-2.7cm} 
\hspace{0.43cm} 
\centerline{
\scalebox{0.9}{
\includegraphics{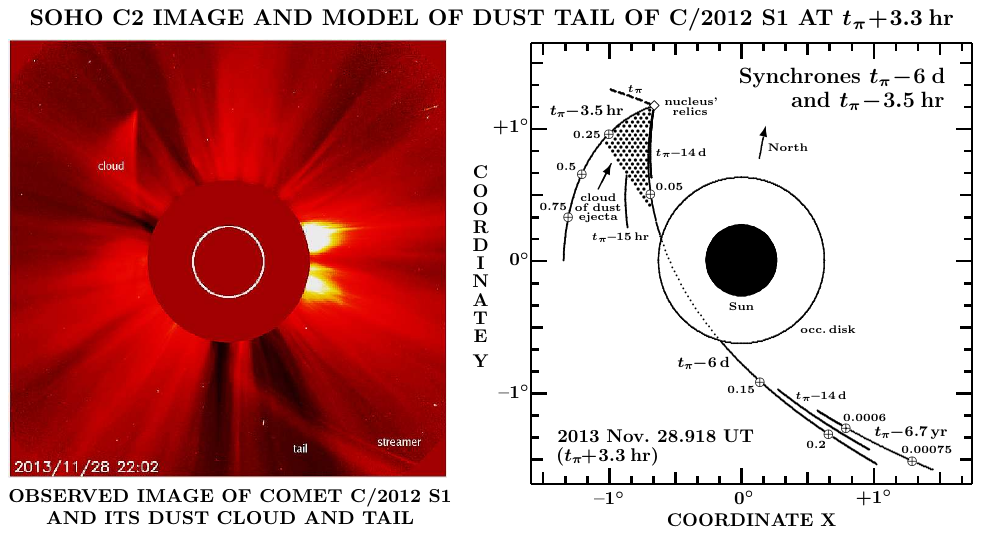}}} 
\vspace{-14.75cm}
\caption{Appearance of comet C/2012 S1 and its dust cloud and tail in an image
taken with the C2 coronagraph on board the SOHO spacecraft on November 28.918
UT, 3.3 hours after perihelion and 5.0\,\,{\Rsun} from the Sun.  This image
shows{\vspace{-0.04cm}} a dust cloud emanating from the site of the nuclear
condensation that is no longer apparent.  The dust cloud is a fan-shaped
feature that contains all the dust ejected before, and surviving the passage
through, perihelion.  The eastern boundary of the cloud marks the emission of
dust 3.5~hours before perihelion, at which point the activity ceased.  The
synchrone fitting the beard in Figure~9 (\mbox{$t_\pi \!-\! 15$\,hr}) is
shown to be in the western part of the fan.  A synchrone, to the east, for
dust that would have been released at perihelion is seen to correspond to
no feature in the image.  On the other hand, a tail is shown made up of dust
particles ejected mostly during Event~2, which did not pass their perihelion
points as yet; and a barely detectable disconnected streamer, terminating ---
also before perihelion --- at a smaller angular distance from the Sun than in
the images in Figure~8.  For other description, see the captions to Figures~6,
7, and 9.  (Image credit:\ ESA/NASA/LASCO consortium.){\vspace{0.35cm}}}
\end{figure*}

The dust that made up the beard in Figure~9 is now closer to the cloud's sharp
western boundary, which consists of dust ejecta from a wide range of times,
including both outbursts --- Event~1 and Event~2.  The temporal resolution of
this boundary is poor, permitting no more definite conclusions.  Fortunately,
a continuation of this boundary in Figure~10 is a tail to the southwest of the
Sun.  Its position allows us to establish that it was made up of dust mainly
from Event~2, with a probable, unresolved contribution from Event~1.  Slightly
to the north of the westernmost part of the tail one can barely detect very
faint remnants of the streamer, described in Sec.~3.2.  The western boundary
of the cloud contained, regardless of the ejecta's age, only fairly large
particles, not smaller than approximately 10~microns in diameter (\mbox{$\beta
< 0.05$}), while the tail to the southwest of the Sun was composed mostly of
particles a few microns across (\mbox{$\beta \simeq 0.2$}).  The surviving
segment of the disconnected streamer appears to terminate at a smaller angular
distance from the Sun than in the images in Figure~8 (Secs.~5.1--5.3).

Much of the emission fan in Figure 10 contains dust ejected at times of
major preperihelion activity.  It is a common occurrence, especially among
comets with small perihelion distances, that a ``sequence'' of significant
preperihelion dust-production events shows up as a fan-shaped feature
shortly after perihelion.  By reading and interpreting its orientation, the
comet's preperihelion dust-emission history can often be unequivocally
established.  With more information provided by the tail on the other side
from the Sun, one finds out from Figure~10 that the dust that ``counts''
was all released over a period of $\sim$6~days terminating 3.5~hours before
perihelion.  Only an {\it earlier\/} preperihelion history of dust emission
can at best be determined from images taken shortly before perihelion, as
illustrated in Figure~9:\ the 6-day old ejecta make up the tail there as well,
but the termination of activity 3.5~hours before perihelion (or 2~hours before
that image was taken) presents itself as an unrecognizable blip near the head.

Also plotted in Figure 10 are (i)~a synchrone confirming that no dust was
ejected at perihelion, thus reiterating that by that time the comet was dead;
and (ii)~a synchrone fitting the streamer, seen with difficulties from the
image's edge to no closer than $\sim$1$^\circ\!$.5 from the Sun.

The phase angle at the site of the disintegrated nucleus in Figure~10 was
96$^\circ$, while along both boundaries of the fan and in the tail it was in a
general range from 50$^\circ$ to 90$^\circ$, ruling out any forward-scattering
effect.

\subsection{Image Taken 5.7 Hours After Perihelion}

{\vspace{0.2cm}}
In the image in Figure 11, taken with the COR2 coronagraph on board STEREO-A
on November 29.017 UT, the fan-shaped cloud of preperihelion dust ejecta is
located a little less than 2$^\circ$ to the north of the Sun, with the tail
trailing to the southeast of it.  The boundaries of the cloud are again
matched well by the synchrones for dust released, respectively, 6~days (the
southern boundary) and 3.5~hours (the northern boundary) before perihelion.
The cloud is limited to only \mbox{$\beta \: {\lapeq} \: 0.2$} or so along
the latter boundary that marks the end of activity --- thus suggesting that
the smallest particles were a few microns across --- but along its southern
boundary it extends across much of the image and is transformed into a tail
made up of micron- and submicron-sized particles (\mbox{$\beta \: {\lapeq}
\: 0.7$}).  The tail appears fairly narrow; as it is gradually fading with
time, its edges may no longer be bright enough to show up in the image, as
suggested by a much wider separation between the short segments of the
synchrones for dust released 8 and 4~days before perihelion.  The synchrone
for dust released at perihelion lies once again entirely outside the
ejecta's fan.

\begin{figure*}[ht]
\vspace{-1.7cm} 
\hspace{-0.1cm}
\centerline{
\scalebox{0.837}{
\includegraphics{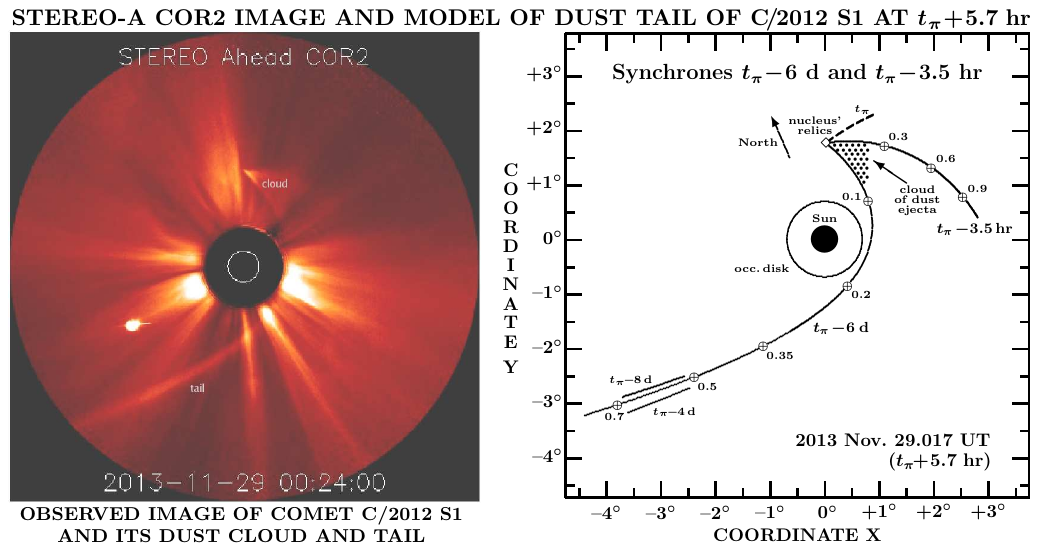}}} 
\vspace{-14.35cm} 
\caption{Appearance of comet C/2012 S1 and its dust cloud and tail in an
image taken with the COR2 coronagraph on board the STEREO-A spacecraft
on November 29.017 UT, 5.7 hours after perihelion and 7.3\,\,{\Rsun} from
the Sun.  The{\vspace{-0.04cm}} cloud of dust is again confined to a sector
bounded by the synchrones for dust released, respectively, 6~days (southern
boundary) and 3.5 hours (northern) before perihelion.  To the southeast of
the Sun, the southern boundary becomes a tail made up of dust ejected during
Event~2.  Segments of the synchrones for dust released 4 and 8~days before
perihelion are plotted for comparison.  While only particles several microns
across and larger make up the northern boundary, the tail also contains  
submicron-sized grains.  A synchrone for dust released at perihelion
corresponds to no feature in the image.  For more description, see the
captions to Figures~6--7 and 9. (Image credit:\ NASA/SECCHI consortium.)}
\vspace{-1.4cm} 
\hspace{-0.09cm}
\centerline{
\scalebox{0.842}{
\includegraphics{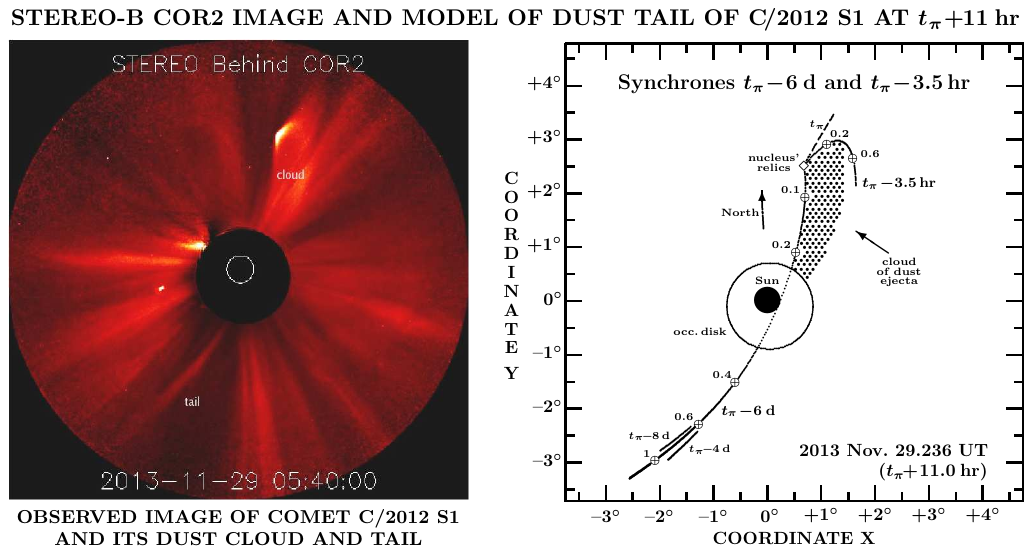}}} 
\vspace{-14.35cm} 
\caption{Appearance of comet C/2012 S1 and its dust cloud and tail in an image
taken with the COR2 coronagraph on board the STEREO-B spacecraft on November
29.236 UT, 11.0 hours after perihelion and 11.9\,\,{\Rsun} from the Sun.  The
cloud{\vspace{-0.06cm}} of dust, now seen from a different perspective, makes
up a very wide fan, whose brightness is enhanced by a significant
forward-scattering effect, in contrast to the subdued tail.  The fan consists
again of dust ejecta from 6~days to 3.5~hours before perihelion, with no
emission from times closer to perihelion.  For more description, see the
captions to Figures~6--7 and 9.  (Image credit:\ NASA/SECCHI consortium.)
{\vspace{0.3cm}}}
\end{figure*}

The phase angle at the site of the disintegrated nucleus was 61$^\circ$.  It
increased to 67$^\circ$ at a distance of 0$^\circ\!$.7 along the fan's nothern
boundary and to 79$^\circ$, 118$^\circ$, and 110$^\circ$ at distances
1$^\circ\!$.3, 2$^\circ\!$.6, and 3$^\circ\!$.5, respectively, along the
southern boundary.  Only a limited enhancement of brightness is seen in the
tail's middle parts due to forward scattering.  The phase angle eventually
decreased to 94$^\circ$ at 6$^\circ$ from the disintegrated nucleus, near
the edge of the field of view.  Very close to the occulting disk there was
much interference from solar features.

\begin{figure*}[ht]
\vspace{-2.6cm} 
\hspace{0.48cm}
\centerline{
\scalebox{0.9}{
\includegraphics{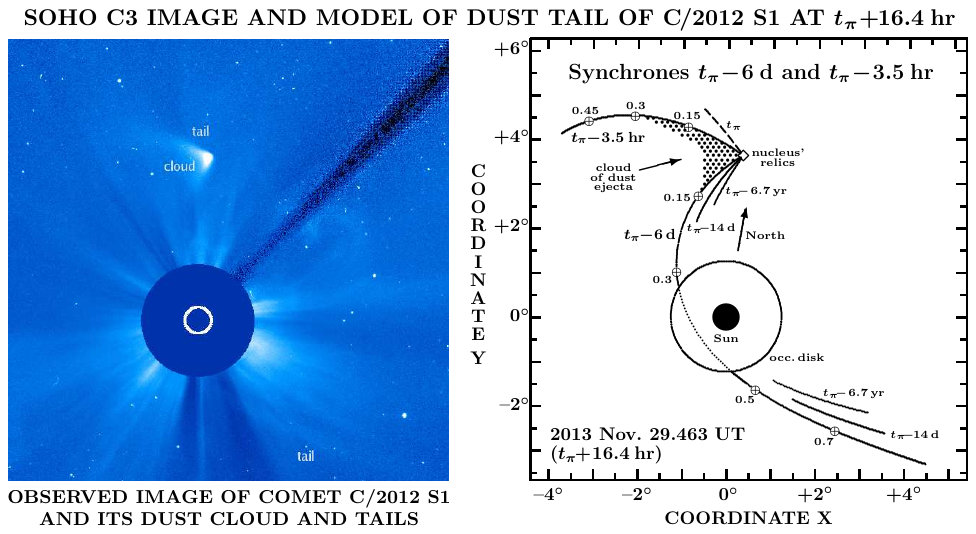}}} 
\vspace{-14.75cm}
\caption{Appearance of comet C/2012 S1 and its dust cloud and tails in an image
taken with the C3 coronagraph on board the SOHO spacecraft on November 29.463
UT, 16.4 hours after perihelion and 16.1\,\,{\Rsun} from the Sun.  The northern
branch{\vspace{-0.04cm}} of what now looks like a winged object, has become
longer and brighter, looking like an independent tail.  This is the branch that
signals the end of the comet's activity 3.5~hr before perihelion.  The fuzzy
southern branch is no longer diagnostic of the age of the ejecta it contains,
but the tail on the other side of the Sun is made up of the dust ejected in
Event~2.  The pseudo-condensation near the location of the disintegrated
nucleus is due to effects of forward scattering.  For more description, see
the caption to Figures~6.  (Image credit:\ ESA/NASA/LASCO consortium.)
{\vspace{0.4cm}}}
\end{figure*}

\subsection{Image Taken 11.0 Hours After Perihelion}
The image in Figure 12, obtained with the COR2 coronagraph on board STEREO-B,
looks unusual in part because the spacecraft was only 13$^\circ$ above the
comet's orbit plane at the time of observation, on November 29.236 UT.  The
dust cloud, which is in part superposed on some solar features, appears to be
considerably brighter and to have much larger dimensions than in Figure~11,
while the tail is now fainter.  The two features are consistent with the
scenario established from the previous images --- they are products of the
comet's activity between 6~days and 3.5~hours before perihelion, as shown
in the right-hand side panel of Figure~12, but the apparent brightening is
an effect of forward scattering of sunlight by dust particles in the cloud.
While the phase angle reached 62$^\circ$ at the location of the disintegrated
nucleus and mostly between 80$^\circ$ and 90$^\circ$ along the fan's northern
boundary, it was steadily increasing along the southern boundary to
$\sim$120$^\circ$ close to the occulting disk.  Inside the fan, the phase
angles were even higher and often exceeded 140$^\circ$.  For example, for
submicron-sized particles having \mbox{$\beta \simeq 0.6$}, located more than
1$^\circ$ to the south of the disintegrated nucleus, right in the middle of
the bright blob in Figure~12, the phase angle was 144$^\circ$, which implies
a forward-scattering driven brightness enhancement by 2~magnitudes.  The
phase angle variations along the tail were, on the other hand, in a range
from 40$^\circ$ to 90$^\circ$.  This example plainly shows that there is
no need for a tempting but untenable hypothesis explaining this major
brightening in terms of a surge of new, post-perihelion activity.

\subsection{Image Taken 16.4 Hours After Perihelion}

Forward scattering of sunlight by dust similarly explains the comet's apparent
brightening in many images taken after perihelion with the C3 coronagraph on
board SOHO.  One of these, displayed in Figure~13, was obtained on November
29.463 UT.  Comparison with Figure~10, exposed with the C2 coronagraph on
board the same spacecraft suggests that there is now a pseudo-condensation
at the site of the disintegrated nucleus, where there was none some 13~hours
earlier.  Even though different instruments were used, the change in the
comet's appearance is rather startling.  The phase angle at the location of
the disintegrated nucleus was 96$^\circ$ in the C2 image in Figure~10, but it
increased to 124$^\circ$ in the C3 image in Figure 13, a difference equivalent
to more than 1~magnitude in a relative phase effect due to forward-scattering.
In addition, the phase angle peaked in Figure~13 at the location of the
pseudo-condensation, decreasing with increasing distance from it in what
now looks like a winged object and reaching merely 116$^\circ$ just 1$^\circ$
from the pseudo-condensation along both wings.

There are other differences in the comet's appearance in Figure~13 compared
to Figure~10.  The northern branch, referring to the end of activity
3.5~hours before perihelion, has properties of an independent tail.  It is
now longer and more pronounced than the southern branch, which has become
fuzzy, providing no longer a constraint on the age of the debris in it.
A dust tail projecting on the other side of the Sun from the winged feature
is, however, still visible, especially in the section made up of
submicron-sized particles with \mbox{$\beta \simeq 0.7$}.  The position
of this tail shows that, once again, it is a product of Event~2 centered
on 6~days before perihelion.  The phase angles in the brightest parts of
this tail in Figure~13 were in a range from about 90$^\circ$ to 100$^\circ$.
\begin{figure*}[ht]
\vspace{-2.6cm} 
\hspace{0.46cm}
\centerline{
\scalebox{0.9}{
\includegraphics{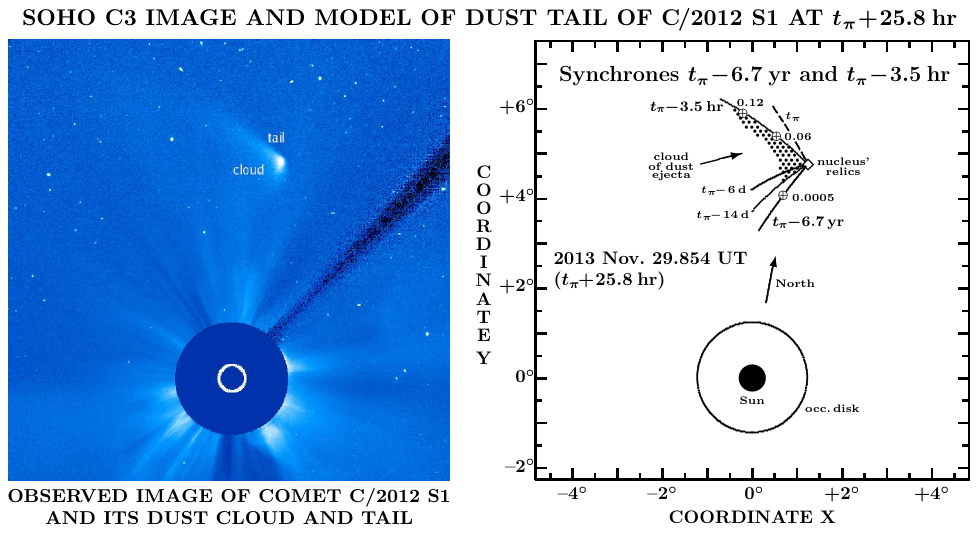}}} 
\vspace{-14.75cm}
\caption{Appearance of comet C/2012 S1 and its dust cloud and tail in an image
taken with the C3 coronagraph on board the SOHO spacecraft on November 29.854
UT, 25.8 hours after perihelion and 22.4\,\,{\Rsun} from the Sun.  The
pseudo-condensation{\vspace{-0.04cm}} is noticeably fainter than in the image
in Figure~13.  The northern branch now completely dominates the southern one
and becomes the only tail (making $\sim$70$^\circ$ with the antisolar
direction), because the old, preperihelion tail to the south of the Sun is
no longer convincingly recognized.  For other description, see the captions to
Figures 9 and 13.  (Image credit:\ ESA/NASA/LASCO consortium.){\vspace{0.4cm}}}
\end{figure*}

\subsection{Image Taken 25.8 Hours After Perihelion}

The final image that is modeled was taken on November 29.854 UT, also with
the C3 coronagraph.  Presented in Figure~14, it resembles in many respects
the image from Figure~13.  However, the pseudo-condensation faded, while
the tail to the northeast lengthened.  The brightness of the cloud to the
south of this tail also subsided.  The other tail, to the south and southwest
of the Sun, an area not shown in Figure~14, has disappeared.  The phase
angle at the site of the disintegrated nucleus still slightly increased,
to 127$^\circ$, helping to keep the pseudo-condensation alive.  Along the
northeastern tail, the phase angle, dropping a little, was equal to
119$^\circ$ at 1$^\circ\!$.5 from the pseudo-condensation.

\subsection{Summary of Results from Modeled Images}

The eight modeled images of comet C/2012 S1 cover a period of about 33 hours.
The three preperihelion images differ from one another in that they show
diverse features, although all dusty as indicated by their large deviations
from the antisolar direction and a generally smooth appearance.  The first
image (Figure~6), taken with the COR2-B coronagraph 7.3~hours before
perihelion when the comet was 8.8~{\Rsun} from the Sun, displays only a
narrow, slightly curved tail, a product of Event~2.  Evidence for this
outburst is repeatedly confirmed, by one feature or another, in six of
the remaining seven images in Figures~7 and 9--14.  Only in the last
image a signature of this event is missing, apparently too obliterated due
to dispersion in space.

The second image (Figure 7), taken with the SOHO's C3 corona\-graph 4.2~hours
before perihelion when the comet was 5.9~{\Rsun} from the Sun, shows a
bifurcated tail, a phenomenon caused by a partial overlap of two very different
features.  One is the dust tail from Event~2, probably with a contribution from
Event~1, the other is a streamer that closely traces the projected orbit behind
the comet.  It is a product of activity at very large heliocentric distances
and is disconnected from the head.  The process of grain release may have
proceeded over a long, rather indeterminate period of time.  However, the
age of the grains is more than $\sim$1~year.  The comet was active at 9.4~AU
from the Sun, the earliest images taken 26~months before perihelion.  The
streamer is not seen in the STEREO's COR2 images (both B and A), probably
because of insufficient exposure times.

The third modeled image (Figure 9), taken with the COR2-A coronagraph 1.5~hours
before perihelion when the head of the comet was only 3.4~{\Rsun} from the
Sun, displays --- in addition to the usual tail --- a prominent ``beard,'' made
up of dust released during Event~3, only a little more than 12~hours before
observation.

\begin{table*}[ht]
\vspace{0.1cm}
\begin{center}
{\footnotesize {\bf Table 4} \\[0.1cm]
{\sc Orbital Elements of Dust Particles Released from Comet C/2012 S1}\\[0.1cm]
\begin{tabular}{c@{\hspace{0.1cm}}c@{\hspace{0.5cm}}c@{\hspace{0.35cm}}c@{\hspace{0.45cm}}c@{\hspace{0.38cm}}c@{\hspace{0.6cm}}c@{\hspace{0.4cm}}c@{\hspace{0.4cm}}c@{\hspace{0.4cm}}c@{\hspace{0.6cm}}c@{\hspace{0.35cm}}c@{\hspace{0.15cm}}c@{\hspace{0.1cm}}c}
\hline\hline\\[-0.2cm]
Radiation & & \multicolumn{12}{@{\hspace{-0.02cm}}c}{Orbital elements of
released dust particles$^{\rm c}$ as function of the time of release before
perihelion} \\[-0.04cm]
pressure &
& \multicolumn{12}{@{\hspace{-0.02cm}}c}{\rule[0.7ex]{14.28cm}{0.4pt}} \\[-0.04cm]
acceler- & Particle & \multicolumn{4}{@{\hspace{-0.4cm}}c}{Perihelion distance
 ({\Rsun})}
 & \multicolumn{4}{@{\hspace{-0.35cm}}c}{Orbit eccentricity$^{\rm d}$}
 & \multicolumn{4}{@{\hspace{0.04cm}}c}{Perihelion time (hr)$^{\rm
 e}$} \\[-0.04cm]
ation$^{\rm a}$ & diameter$^{\rm b}$
 & \multicolumn{4}{@{\hspace{-0.4cm}}c}{\rule[0.7ex]{3.75cm}{0.4pt}} 
 & \multicolumn{4}{@{\hspace{-0.35cm}}c}{\rule[0.7ex]{4.55cm}{0.4pt}}
 & \multicolumn{4}{@{\hspace{0.04cm}}c}{\rule[0.7ex]{4.85cm}{0.4pt}} \\[-0.04cm]
$\beta$ & (mm/$\mu${\it m}) & 3.5\,hr & 6\,d & 14\,d & 6.7\,yr & 3.5\,hr
 & 6\,d & 14\,d & 6.7\,yr & $\;\;$3.5\,hr & 6\,d & 14\,d & \,6.7\,yr \\[0.05cm]
\hline \\[-0.25cm]
0.0001 & \llap{2}6.9 & 2.68 & 2.68 & 2.68 & 2.68 & 1.0001 & 1.0000 & 1.0000
       & 1.0000 & $\;\;\:$0.00 & $\;\;\:$0.00 & +0.01 & $\;\:$+1.18 \\
0.0003 &  8.4 & 2.68 & 2.68 & 2.68 & 2.68 & 1.0003 & 1.0000 & 1.0000 & 1.0000
       & $\;\;\:$0.00 & +0.01 & +0.02 & $\;\:$+3.55 \\
0.001$\;\:$ & 2.2 & 2.68 & 2.68 & 2.68 & 2.68 & 1.0010 & 1.0001 & 1.0000
       & 1.0000 & $\;\;\:$0.00 & +0.03 & +0.07 & +11.82 \\
0.003$\;\:$ & 0.5\rlap{9} & 2.68 & 2.68 & 2.68 & 2.68 & 1.0031 & 1.0002
       & 1.0001 & 1.0000 & $\;\;\:$0.00 & +0.10 & +0.22 & +35.51 \\
0.01$\;\;\:\:$ & 0.1\rlap{2} & 2.69 & 2.70 & 2.70 & \ldots & 1.0105 & 1.0007
 & 1.0004 & \ldots & $\;\;\:$0.00 & +0.34 & +0.75 & \,\,\,\dots \\
0.03$\;\;\:\:$ & \llap{\it 2}{\it 4.9} & 2.72 & 2.76 & 2.76 & \ldots & 1.0323
       & 1.0023 & 1.0013 & \ldots & $-$0.01 & +1.03 & +2.26 & \,\,\,\ldots \\
0.1$\;\;\;\:\:\:$ & {\it 5.1} & 2.81 & 2.96 & 2.97 & \ldots & 1.1199 & 1.0087
       & 1.0049 & \ldots & $-$0.05 & +3.55 & +7.81 & \,\,\,\ldots \\
0.2$\;\;\;\:\:\:$ & {\it 2.2} & 2.93 & 3.31 & 3.32 & \ldots & 1.2820 & 1.0219
       & 1.0123 & \ldots & $-$0.12 & +7.51 & +16.51$\;\:$ & \,\,\,\ldots \\
0.3$\;\;\;\:\:\:$ & {\it 1.4} & 3.05 & 3.74 & 3.78 & \ldots & 1.5036 & 1.0425
       & 1.0240 & \ldots & $-$0.20 & +11.96$\;\:$ & +26.29$\;\:$
       & \,\,\,\ldots \\
0.5$\;\;\;\:\:\:$ & {\it 0.7}\rlap{\it 8} & \ldots & 5.02 & 5.15 & \ldots
       & \ldots & 1.1330 & 1.0766 & \ldots & \,\,\ldots & +22.81$\;\:$
       & +50.40$\;\:$ & \,\,\,\ldots \\
0.7$\;\;\;\:\:\:$ & {\it 0.5}\rlap{\it 4} & \ldots & 7.28 & 7.85 & \ldots
       & \ldots & 1.4503 & 1.2721 & \ldots & \,\,\ldots & +36.87$\;\:$
       & +83.82$\;\:$ & \,\,\,\ldots \\
1$\;\;\;\;\;\:\:\:\:$ & {\it 0.3}\rlap{\it 7} & \ldots & \llap{1}4.21
       & \llap{1}8.98 & \ldots & \ldots & $\infty$ & $\infty$ & \ldots
       & \,\,\ldots & +57.70$\;\:$ & +148.74$\;\;\:\:$ & \,\,\,\ldots\\[0.05cm]
\hline\\[-0.2cm]
\end{tabular}}
\parbox{17.4cm}{\scriptsize $^{\rm a}$\,Measured in units of the Sun's
 gravitational acceleration.} \\[-0.04cm]
\parbox{17.4cm}{\scriptsize $^{\rm b}$\,Assuming the bulk density varies with
 particle size according to Eq.\,(21).  Millimeters are in roman font, microns
 in italics.} \\
\parbox{17.4cm}{\scriptsize $^{\rm c}$\,Release (ejection) velocity assumed
to be zero.} \\[-0.05cm]
\parbox{17.4cm}{\scriptsize $^{\rm d}$\,For $\beta = 1$ (motion in a straight
 line), the eccentricity is by definition infinitely large regardless of the
 time of release.} \\ [0cm]
\parbox{17.4cm}{\scriptsize $^{\rm e}$\,Reckoned from the time of perihelion
passage of the comet (Sec.~6); minus sign means before, and vice versa.} \\
\end{center}
\end{table*}

The fourth modeled and the first post-perihelion image (Figure 10), taken
with the C2 coronagraph 3.3~hours after closest approach when the comet was
5.0~{\Rsun} from the Sun, shows two major morphological changes.  One is a
{\it total loss of the head with the nuclear condensation\/}, the other is
a fan-shaped, sideways-pointing tail.  The head's diappearance was recorded
earlier in real time by the C2 coronagraph on board SOHO (Sec.~4.1).  The
leading boundary of the emission fan, oriented toward the southeast in this
image, measured the {\it time of termination of the comet's dust production\/}.
 Determined independently from a number of images, this singular condition
is found to have occurred {\it \mbox{${\it 3.5}\pm{\it 0.3}$} hours before
perihelion\/}; at that time {\it all activity ceased, never again to be
resuscitated, the nucleus' disintegration was completed, and the existence
of the comet as such was over\/}.  This result refines an earlier preliminary
determination (\mbox{Sekanina} 2013b) and is consistent with Curdt et al.'s
(2014) conclusion, based on the scattered-light observations of the comet's
dust tail in Lyman-$\alpha$ with the SUMER spectrometer on board SOHO merely
0.7~hour before perihelion.  Their images suggest an outburst of dust about
8.5 hours before perihelion, followed by a sharp decline in the dust
production over the next hours.

The source of much confusion in early media reports was the re-appearance
of the comet's relics from behind the occulting disk of the C3 coronagraph
(Figures~13--14).  A bright pseudo-condensation was apparent in these images,
but {\it no nuclear condensation\/}; the brightening was due to forward
scattering of sunlight by microscopic dust in a relatively dense cloud,
contrary to the comet's re-appearance from behind the occulting disk in
the C2 coronagraph (Figure~10), which preceded that in the C3 coronagraph.
The reason for the discrepancy is this timing:\ in C2 the post-perihelion
images covered a period from about 0.7 to 5 hours after perihelion, while
in C3 the post-perihelion show did not start until about 3.5~hours after
perihelion, with the pseudo-condensation steadily brightening up over
another 3.5~hours as the heliocentric distance continued to increase.
The phase angles at the site of the disintegrated nucleus varied from
55$^\circ$ to 106$^\circ$ for the C2 post-perihelion images, but from
97$^\circ$ to nearly 130$^\circ$ for the C3 post-perihelion images.  The
pseudo-condensation was not very impressive until the phase angle reached
about 113$^\circ$.  The maximum effect was observed between $\sim$10
and $\sim$22 hours after perihelion, when the phase angle ranged from
119$^\circ$ to 126$^\circ$.  Marcus' phase law predicts a brightness
deficit, relative to backscatter, of 0.9 magnitude at a phase angle
55$^\circ$ and 0.3 magnitude at 97$^\circ$, no enhancement at 106$^\circ$,
and enhancements of 0.25, 0.51, and 0.85 magnitude at, respectively,
113$^\circ$, 119$^\circ$, and 126$^\circ$.

With the exception of the last modeled image in Figure 14, not all dust
ejecta observed after perihelion were confined to the space enclosed
by the emission fan.  As a rule, the fan was accompanied, on the other
side of the Sun, by a narrow tail.  The tail was in fact an extension
of the fan's trailing boundary, but because it usually appeared
decoupled from the fan, it could be regarded as a separate feature.  The
dust along the trailing boundary and in the tail, although both originating
in Event~2 (in some images merging with the ejecta from Event~1), differed
from one another in the particle size and in that those in the tail did
not as yet pass their perihelion points at the times of observation.  All
grains in the streamer, barely detected in the post-perihelion image in
Figure~10, likewise were still approaching perihelion (Sec.~5).

A range of perihelion times of dust particles, released at four different
times and subjected to various solar radiation pressure accelerations, is
depicted in Table 4, which also lists the perihelion distance and
eccentricity of the particles' orbits.  The release times refer, respectively,
to the termination of dust production, to Events~2 and 1, and to ejection
at 20~AU from the Sun.  No orbital elements are listed for particles
which our modeling showed remained undetected in all eight images:\
those with \mbox{$\beta \ge 0.5$} for the end of dust production and
with \mbox{$\beta \ge 0.01$} for the ejecta from a heliocentric distance
of 20~AU.

A few examples illustrate the meaning of the~\mbox{difference} between
the perihelion times of a particle and the disintegrated nucleus.  In
order that the ejecta from Event~2 arrive at perihelion at the time
the image in Figure~10 was taken, 3.3~hours after perihelion, they
should have been subjected to a radiation-pressure acceleration of
\mbox{$\beta_{\rm per} = 0.0933$}; their perihelion distance should
have been 2.94~{\Rsun} and the eccentricity 1.0080.  All Event~2 ejecta
with \mbox{$\beta > \beta_{\rm per}$} arrived at their perihelia later,
their perihelion distances were greater than 2.94~{\Rsun}, and their
eccentricities greater than 1.0080; and vice versa.  Since the
Event~2 ejecta that made up the trailing boundary of the fan in
Figure~10 had \mbox{$\beta \ll 0.0933$}, they already passed their
perihelia, while the ejecta from the same event contained in the tail
had \mbox{$\beta \gg 0.0933$} and did not pass their perihelia as yet.
Similarly one finds that $\beta_{\rm per}$ discriminating the two
categories of Event~2 ejecta is 0.1557 for a perihelion time coinciding
with the time when the image in Figure~11 was taken, 0.2794 for
Figure~12, and 0.3884 for Figure~13.

The fact that in the post-perihelion images, such as in Figures 10--12,
dust particles that had already passed perihelion populated the same
synchrone as dust particles that in the same images were still moving
toward perihelion offers evidence that the former particles had survived
the passage with no obvious harm, as otherwise the critical segments
of the synchrone would be void of dust.  The survival of such particles
was facilitated by their perihelion distances, which were noticeably
greater than 2.67~{\Rsun}.  On the other hand, since the perihelion
distances of more sizable grains, in the range from many tens of
microns across up, were always close to 2.67~{\Rsun}, the impact of
the sublimation process --- which is an extremely steep function of
heliocentric distance close to the Sun --- on these grains was much
greater, so much so that their larger size might not have been enough
to protect them.  It thus becomes, counterintuitively, possible that
smaller particles are more likely to survive than more sizable ones.

The leading boundary of the post-perihelion emission fan in Figures~10--14,
a product of the terminating dust production 3.5~hours before perihelion,
demonstrates the presence of particles larger than about 1~micron
across (\mbox{$\beta < 0.3$}) but not the smaller ones.  All such
particles, released at a distance of 5.2~{\Rsun} from the Sun,
passed perihelion before the comet's disintegrated nucleus.  The
absence of submicron-sized particles (with \mbox{$\beta > 0.3$}) in
the ejecta released 3.5~hours before perihelion may be either due to
their absence in these ejecta from the very beginning, or due to their
sublimation very shortly before or at perihelion.  Because of what has
just been said about the perihelion distances of such tiny grains, the
first option appears to be more likely.

In the ejecta released during Event~2, submicron-sized particles were
unquestionably present, as documented by the radiation-pressure
accelerations we found in the lagging tail.  However, it is not possible
to say conclusively whether these grains survived their passage through
perihelion or not because they did not reach that point until many days
or weeks after the head did, by which time the comet's relics were too faint
to observe.

Submicron-sized particles were certainly not present in the pseudo-condensation
apparent in a number of C3 post-perihelion images, such as those in Figures~13
or 14.  Their absence is dictated by the dynamics:\ if large grains did
fragment profusely into submicron-sized particles many hours after perihelion, 
there would be a tail pointing away from the Sun.  The absence of such tiny
particles is also consistent with the strong forward-scattering effect already
commented on.  As summarized by Marcus (2007, and references therein),
particles about 0.1~micron in diameter do not exhibit any forward scattering
in the visible light and those around 0.5~micron across do so only to a very
limited extent.  It is the grains of microns to tens of microns in diameter
and of high porosity that are the most efficient forward scatterers at
phase angles near 130$^\circ$.

The complete disintegration of the comet's nucleus and the termination of
all activity before perihelion is supported by other evidence as well.  In
all the images we modeled, the synchrone predicted for dust emitted at
perihelion is consistently located way outside the fan-shaped cloud of
ejecta.  No trace of the comet was detected by the Solar Dynamics
Observatory (SDO) in immediate proximity of perihelion.\footnote{See {\tt
http://sdoisgo.blogspot.com/2013/11/where-was-comet -ison.html}.}  The
conclusion is likewise corroborated not only by additional images, not modeled
here, taken with the SOHO's C3 coronagraph until November~30 and with the
STEREO-A's HI1 imager until December~7, but also by numerous unsuccessful
searches via ground-based deep-imaging efforts between December~7 and 16 by
Sako et al.\ (2013) and, as reported by Nakano (2013b), also by K.~Kadota,
by K.~Yoshimoto, and by Y.~Ikari, some of them down to a limiting magnitude
19.0.  Last but not least, the Hubble Space Telescope failed to find any
remnants or debris of the comet\footnote{See {\tt
http://hubblesite.org/hubble\_discoveries/comet\_ison.}}
on December~18.  As for a possible meteor shower associated with the comet,
Sekhar \& Asher (2014) concluded that not even an ejection velocity of
1~km s$^{-1}$ would be sufficient to deflect a meteoroid from the comet's
path to impact the atmosphere of the Earth around the time of transit of the
comet's orbit plane on 2014 January~16.  Under such circumstances one
must be extremely skeptical about any reports of possible detection
(Golubev et al.\ 2014).

\begin{figure*}
\vspace{-2.55cm} 
\hspace{-0.27cm}
\centerline{
\scalebox{0.915}{ 
\includegraphics{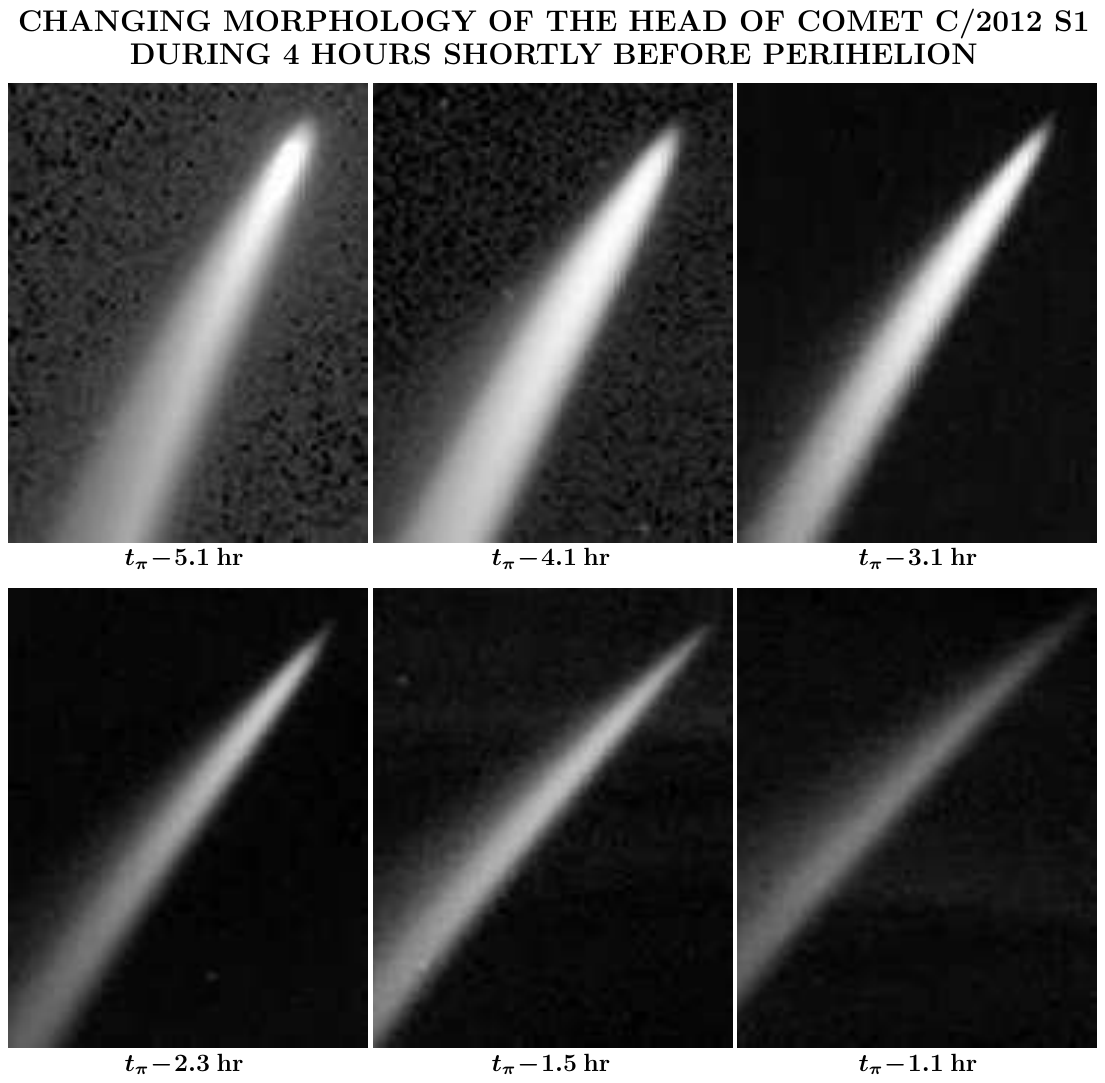}}} 
\vspace{-6.35cm}
\caption{Rapidly changing morphology of comet C/2012~S1 shortly before
perihelion, at time $t_\pi$, resulting in the disappearance of the head
in the images taken with the C2 coronagraph on board SOHO.  The images
were taken, respectively, on November 28.567, 28.608, 28.649, 28.683,
28.717, and 28.733 UT, when{\vspace{-0.05cm}} the comet was 6.9, 5.9,
5.0, 4.2, 3.5, and 3.2~{\Rsun} from the Sun.  The phase angles ranged
from 87$^\circ$ to 51$^\circ$ from the upper left to the lower right.
Along the diagonal each image measures 15.7 arcmin, or, on the average,
670\,000~km, increasing by 5600~km between the first and the last images.
The Sun is to the right.  Note a faint, lengthening extension sticking
out of the ever fuzzier head on the sunward side; a truncated leading
boundary of part of the tail, starting at the tip and becoming
progressively more conspicuous with time; and the tail's concurrently
decreasing width. (Image credit:\ ESA/NASA/LASCO consortium.)
{\vspace{0.5cm}}}
\end{figure*}

\section{Disappearance of the Head and a Timeline of Nucleus' Fragmentation}
A period of time that includes the time of termination of dust
production, 3.5~hours before perihelion, is covered by a series of images
taken with the C2 coronagraph on board SOHO.  They show a rapidly changing
morphology of the comet's head and tail over a span of 4~hours.

\subsection{Disappearance of the Head}
Six selected images from this set are enlarged in Figure~15, in which the
first was exposed on November 28.567 UT, about 1.5~hours after the comet
entered the field of view of the coronagraph.  The general trend is from
a bright and well-defined, practically symmetrical round head that
gradually widens into a slightly curved tail, to a faint, sharply tipped
head that becomes a narrow tail, brightening, up to a point, with
increasing distance from the head.  Although not apparent from the figure,
the tail's axis makes a large angle with the antisolar direction, which
keeps increasing with time, amounting to 25$^\circ$ in the first image
and 51$^\circ$ in the last.

From cursory inspection, the most rapid fading appears to have taken place
in the time span between the second and the fourth images in Figure 15,
which includes the critical time 3.5~hours before perihelion (Sec.~3.9).
A faint, arrow-shaped extension, steadily lengthening with time, is seen
in the images to be protruding from the head on the forward side.  It is
barely recognizable in the first image, but becomes more prominent beginning
with the third image.  Concurrently, the leading boundary of the tail
(on the left) becomes truncated at the head as if cut off on its outside
along a nearly straight line.  Again, this feature is marginally detected
in the first image, but the resulting asymmetry of the tail relative to
its axis is quite apparent from the second image on.

\begin{table}[b]
\noindent
\vspace{0.35cm}
\begin{center}
{\footnotesize {\bf Table 5} \\[0.1cm]
{\sc Tail Orientation and Morphology in Images Taken with\\SOHO's C2
Coronagraph Shortly Before Perihelion.}\\[0.2cm]
\begin{tabular}{c@{\hspace{0.1cm}}c@{\hspace{0.3cm}}c@{\hspace{0.2cm}}c@{\hspace{0.15cm}}c@{\hspace{0.15cm}}c@{\hspace{0.2cm}}c}
\hline\hline\\[-0.2cm]
\multicolumn{2}{@{\hspace{0.1cm}}c}{Time of imaging}
 & \multicolumn{4}{@{\hspace{0cm}}c}{Position angle} & Time of \\
\multicolumn{2}{@{\hspace{0.1cm}}c}{\rule[0.7ex]{2.7cm}{0.4pt}}
 & \multicolumn{4}{@{\hspace{0cm}}c}{\rule[0.7ex]{3.6cm}{0.4pt}}
 & release of \\
2013$\:$Nov. & relative$\:$to & tail's & Event\,2 & Differ- & tail's
 & truncating \\
(UT) & perihelion & axis & synchr. & ence & cutoff
 & ejecta \\[0.08cm]
\hline \\[-0.21cm]
 & & & \hspace{0.31cm}$^\circ$ & \hspace{0.24cm}$^\circ$ & & \\[-0.32cm]
28.567 & $t_\pi \!-\! 5.1\,$hr & 254\rlap{$^\circ$} & 253\,.5 & +0\,.5
       & \,\ldots & \,\ldots\ldots\ldots \\
28.608 & $t_\pi \!-\! 4.1\,$hr & 250 & 250.8 & $-$0.8 & 239\rlap{$^\circ$}
       & $t_\pi \!-\! 15.3\,$hr \\
28.649 & $t_\pi \!-\! 3.1\,$hr & 247 & 247.3 & $-$0.3 & 237
       & $t_\pi \!-\! 16.3\,$hr \\
28.683 & $t_\pi \!-\! 2.3\,$hr & 244 & 243.3 & +0.7 & 234
       & $t_\pi \!-\! 16.6\,$hr \\
28.717 & $t_\pi \!-\! 1.5\,$hr & 238 & 237.7 & +0.3 & 228
       & $t_\pi \!-\! 12.9\,$hr \\
28.733 & $t_\pi \!-\! 1.1\,$hr & 234 & 234.2 & $-$0.2 & 226
       & $t_\pi \!-\! 15.8\,$hr \\[0.08cm]
\hline 
\end{tabular}}
\end{center}
\end{table}

Also evident from Figure 15 is the tail's narrowing down with time.  We
measured the orientation and curvature of the tail's axis in all six images
in Figure~15.  The position angles at $\sim$0$^\circ\!$.1 from the head are
in column~3 of Table~5.  We found that both the orientation and the bending
are in excellent agreement with those of the synchrone for the Event~2 ejecta.
This is not surprising because this match consistently resulted from all
images taken before perihelion and less than 1~day after perihelion that were
modeled in Sec.~3.  The identity of the tail with the debris from Event~2
allowed us to estimate the expansion velocity of dust particles.  From the
width near the lower left corner of each image, about 10~arcmin from the tip
of the head, we find, after accounting for a small contribution from the
$\sim$3-day duration of Event~2, the following velocities:\ (i)~96~m~s$^{-1}$
for particles whose \mbox{$\beta = 0.019$}, that is, about 47~microns in
diameter according to Eq.\,(21) from the image{\vspace{-0.03cm}} taken
5.1~hours before perihelion; (ii)~64~m~s$^{-1}$ for particles 66~microns
in diameter from the image taken 3.1~hours before perihelion; and
(iii)~40~m~s$^{-1}$ for particles 91~microns in diameter from the image
taken 1.5~hours before perihelion.  However, these last particles are also
situated about 6.4~arcmin from the head in the image in the upper left
corner of Figure~15, and that tail's width suggests {\vspace{-0.03cm}}that
their expansion velocity was at least 75~m~s$^{-1}$, a value nearly twice
as high.  This discrepancy must mean that the tail's observed width is
a product of at least two separate components that partly overlap one another.
This argument also explains the truncated nature of the tail in the rest
of the images, which we address next, and the apparent failure of the tail,
far from the head, to widen with increasing distance from it in the last
two images in Figure~15.

The measured position angles of the line, along which the tail is truncated
from the tip to the lower left in the images in Figure~15, are listed in
column~6 of Table~5.  The measurement could not be done in the first
image, in which the truncation is hardly at all visible, but was
relatively easy in the rest of them.  This exercise could also have been
performed using a dozen or so additional C2 images that are available,
but the results for the five cases are unequivocal enough that there was
no need to extend the sample.  The last column of Table~5 presents the
release times for the synchrones that correspond to the measured position
angles.  An average of the five entries is \mbox{$15.4 \pm 1.5\:$hours}
before perihelion, equivalent to a heliocentric distance of 0.07~AU or
$\sim$15~{\Rsun}.  This is the time of termination of dust production
during an outburst, which we call {\it Event~3\/} and which accounts
for the second tail in Figure~15 that overlaps the tail made up of the
Event~2 ejecta.  The beginning of this third event cannot be ascertained
from the features in the figure because of the overlap.

Two correlations should be mentioned.  One, the feature that causes the
truncated tail in Figure~15 is undoubtedly identical with what we called
a ``beard'' in Sec.~3.3, a feature so conspicuously apparent in Figure~9.
Note that the STEREO-A image in that figure and the fifth of the six C2
images in Figure~15 were taken only a couple of minutes apart, so the same
feature is expected to show in both under at least moderately favorable
projection conditions.  Thus, indeed, it is the beard that causes in
Figure~15 the tail to be truncated in five images and to fail to widen,
in the last two images, far from the head with increasing distance from it.
Two, the time of termination of this third emission event practically
coincides with the peak on the intrinsic-brightness curve in close
proximity of perihelion in Figure~5, based on Knight \& Battams' (2014)
C3 clear-filter magnitudes.  From this correlation one can expect that
Event~3 began at the time of the last upswing on the intrinsic-brightness
curve in Figure~5, that is, 2.4~days before perihelion, when the comet
was 40~{\Rsun} from the Sun and its nucleus already extensively
fragmented (Sec.~4.2).

The sharp forward pointing extension from the head's tip cannot be
interpreted as a product of near edge-on projection, because the SOHO
spacecraft was about 42$^\circ$ out of the comet's orbital plane.  The
feature is generally too short to measure its position angle with adequate
accuracy.  However,  the last two images suggest that it was aligned with
the axis of the Event~2's tail.  Being ahead of the rest of the comet in
the orbit, the forward extension was necessarily the location of the {\it
most massive fragments\/} left from the original nucleus during its
extensive fragmentation in the course of Events~1 and 2 (Sec.~4.2).  This
implies that it is this {\it extension's forward tip\/} --- rather than the
head's sunward end --- that pinpointed the site of the original nucleus in
Figure~15, implying that the lengthening of the extension with time was due
to the motion of the head relative to the extension's tip, a conclusion that
is consistent with our orbital results (Sec.~6).  The extension should have
thus contained a lined-up procession of boulder- and pebble-sized objects
and other coarse debris with near-zero differential velocities, imaged at
a low spatial resolution.  The feature is reminiscent of the elongated
cloud of subnuclei of comet C/1999~S4 detected, at much higher resolution,
by the Hubble Space Telescope (Weaver et al.\ 2001).  The head itself
contained a debris of less sizable dust particles whose dimensions may
not have exceeded millimeters.  The head lagged behind the extension
primarily because of the differential effects of solar radiation pressure,
just as the tail lagged behind the head.  As dust was driven away from
the Sun, the tail gradually contained ever larger particles; their lower
ejection velocities may explain the mentioned narrowing down of the tail.  A
near ``dissolution'' of the entire object, best apparent in the last image of
Figure~15, is evidence of the disintegration of the debris itself, after the
termination of dust production.  We suggest that dust particle sublimation
may have played, in this near-perihelion stage of the comet's disintegration,
a more important role than fragmentation.  The sharp boundary of the truncated
tail, signaling the termination of the major dust production some 15~hours
before perihelion, could represent a precursor to the final breakup of more
sizable debris, which continued to produce dust, at progressively lower rates,
for another half day, until $\sim$3.5~hours before perihelion, and which
accounted for the leading boundary of the fan-shaped feature that showed
up in all post-perihelion images taken with the SOHO and STEREO coronagraphs.

\subsection{Timeline of the Comet's Disintegration}
The fragmentation sequence that led to the complete disintegration of the
comet looks like a puzzle whose individual pieces finally appear to fall
into place.  We suggest below that the comet's highly variable water
production rate during November played a pivotal role in determining the
timeline and properties of this process.

To interpret the observed water production rate as a function of time,
we introduce an averaged water sublimation rate, $\langle Z_{\rm subl}
\rangle$, from a unit surface area on the sunlit hemisphere of an intact
(unaffected by fragmentation) active icy nucleus as a function of the
distance from the Sun by integrating over all zenith distances of the
Sun, normalizing it in terms of the subsolar sublimation rate, scaling
it by the size of the nucleus, and comparing it with observation.  The
sublimation rate is derived from a model in which the surface insolation
is assumed to be spent on water sublimation and thermal reradiation, but
not on heat conduction into the nucleus.  With a Bond albedo near zero
and a unit emissivity of the surface, a solution for the sublimation rate
per unit surface area at the Sun's zenith distance $\vartheta$ and a
heliocentric distance $r$, $Z_{\rm subl}(\vartheta,r)$, is found in an
empirical form (Sekanina 1988)
\begin{equation}
Z_{\rm subl}(\vartheta,r) = Z_0(r) \, \Omega(\vartheta,r),
\end{equation}
where $Z_0(r)$ is the sublimation rate at a subsolar point at distance $r$,
and \mbox{$0 \leq \Omega(\vartheta,r) \leq 1$} is a relative sublimation
rate at a zenith angle $\vartheta$,
\begin{eqnarray}
\Omega(\vartheta,r) & = & \cos \vartheta \!-\! f_0(r) \sin^2 \vartheta
 \hspace{0.45cm} {\rm for}\;\, 0 \leq \vartheta \leq \vartheta_0(r),\nonumber\\
 & = & 0 \hspace{3.01cm} {\rm for} \;\, \vartheta > \vartheta_0(r).
\end{eqnarray}
The expressions for $Z_0$, $f_0$ and $\vartheta_0$, which allow one to
compute the model's expected sublimation rates with a relative accuracy
to better than $\pm$1~percent at heliocentric distances \mbox{$r < 1$}~AU,
are presented in Sekanina (1988).

The averaged rate $\langle Z_{\rm subl} \rangle$ per unit surface area at
time $t$ and a distance $r$ is given by
\begin{eqnarray}
\langle Z_{\rm subl} \rangle & = & Z_0 \! \int_{0}^{\frac{1}{2}\pi} \!\!
  \sin \vartheta \; \Omega(\vartheta,r) \, d\vartheta \nonumber \\[0.1cm]
& = & Z_0 \! \left[ \sin^2 \! \vartheta_0 \!-\! {\textstyle \frac{4}{3}} f_0
  (1 \!-\! \cos \vartheta_0)^2 \!\left( 1 \!+\! {\textstyle \frac{1}{2}}
  \cos \vartheta_0 \right) \right]\!. \;\;\;\;\;\;\;
\end{eqnarray}
If $D_0(t)$ is the nucleus' diameter and $Q_{\rm obs}(t)$ the observed
production rate of water at time $t$, the modeled production rate from
the nucleus' entire sunlit hemisphere is \mbox{$Q_{\rm subl}(t) = {\textstyle
\frac{1}{2}} \pi D_0^2 \langle Z_{\rm subl} \rangle (t)$} and the effective
observed sublimation area is \mbox{$X_{\rm obs}(t) = Q_{\rm obs}(t)/\langle
Z_{\rm subl} \rangle (t)$}.  A convenient measure of the outgassing rate is
a dimensionless quantity $\Theta_{\rm subl}(t)$, equal to the observed
sublimation area expressed in units of the hemispherical area of the
nucleus or to the ratio of the observed and modeled production rates,
\begin{equation}
\Theta_{\rm subl}(t) = \frac{X_{\rm obs}(t)}{{\textstyle \frac{1}{2}} \pi
 D_0^2(t)} = \frac{Q_{\rm obs}(t)}{Q_{\rm subl}(t)}.
\end{equation}

In the following we compare this model with the measured water production
and total mass-loss rates of C/2012~S1 between the time of close encounter
with Mars on October~1 and the time of complete disintegration on November~28,
as referred to in Sec.~3.9.  The model depends on three parameters:\ the
nucleus' initial diameter (at the time of Mars encounter), its bulk
density, and the dust-to-water mass production rate ratio.  To examine a
range of plausible solutions, we present in an extensive Table~6 not only the
nominal case (Sec.~2.4), referred to as {\it Model G\/}, but also scenarios
for seven other combinations of the three parameters.  The table is organized
in a manner as follows.  Relevant parameter-independent quantities are listed
for a selected set of dates between October~1 and November~28 in the first
part of the table.  Each date represents actually a 24-hour interval that
starts 12 hours before and ends 12 hours after the tabulated time.  This
arrangement complies with Combi et al.'s (2014) daily water production
averages, whose sequence, extrapolated beyond the original 28-day interval,
is tabulated as an {\it observed production rate\/} $Q_{\rm obs}$.  Because
the production rates between October~1 and November~10 varied very slowly
with time, this period is represented by only four entries.  Besides the
rates, the first part of Table~6 also presents, as a function of time, a
{\it modeled averaged sublimation rate per unit surface area\/}, $\langle
Z_{\rm subl} \rangle (t_{\rm obs})$, derived from Eq.\,(24), and an {\it
effective sublimation area\/}, \mbox{$X_{\rm subl}(t_{\rm obs})$}, derived
from the expression in the text between Eqs.\,(24) and (25).

The second part of Table~6 is divided into sections, one section per scenario.
Each scenario satisfies the basic condition, requiring that the combination
of the three parameters --- the initial diameter, $D_0(t_{\rm enc})$, the bulk
density, $\rho$, and the dust-to-water mass production rate ratio, $\Re_0$
--- satisfy an equality between the mass of the nucleus, ${\cal M}_0 (t_{\rm
enc})$, at the time of encounter with Mars, $t_{\rm enc}$, and the mass-loss
rate integrated from $t_{\rm enc}$ to the time of the comet's complete
disintegration, {\tend}, 3.5~hours before perihelion (Sec.~3.9):
\begin{eqnarray}
{\cal M}_0 (t_{\rm enc}) = {\textstyle \frac{1}{6}} \pi \rho D_0^3(t_{\rm
 enc}) & = & \int_{t_{\rm enc}}^{\mbox{\ttend}} \!\! \mu_0 Q_{\rm obs}(t)
 \, (1 + \Re_0) \, dt
 \nonumber \\[0.1cm]
 & = & 2.32 \times 10^{13}\,(1 + \Re_0) ,
\end{eqnarray}
where ${\cal M}_0(t_{\rm enc})$ is in grams and $\mu_0 = 2.99 \times
10^{-23}$\,gram is the mass of a water molecule.

For each scenario Table 6 provides (i)~a modeled water production rate,
$Q_{\rm subl}$, at the middle of each time interval, $t_{\rm obs}$, from the
entire sunlit hemisphere of an {\it intact\/} spherical nucleus of the given
(gradually dwindling) dimensions; (ii)~its remaining mass, ${\cal M}_0(t_{\rm
fin})$, at the end of each 24-hour{\vspace{-0.03cm}} interval, \mbox{$t_{\rm
fin} = t_{\rm obs} \!+\! 0.5$ day}, expressed by \mbox{${\cal M}_0(t_{\rm fin})
= \mu_0 (1\!+\!\Re_0) \int_{t_{\rm fin}}^{\mbox{\ttend}}\!Q_{\rm obs}(t)\,dt$};
{\vspace{-0.03cm}}and (iii)~a water production rate ratio, $\Theta_{\rm
subl}(t_{\rm obs})$, defined by Eq.\,(25).  The water sublimation rate
$Q_{\rm subl}$ represents an ideal, hypothetical case, whereas a ratio
$\Theta_{\rm subl}$ measures a ``degree'' of deviation from this case.  If
\mbox{$\Theta_{\rm subl} \!<\!  1$}, only a fraction of the sunlit hemisphere's
surface is active.  On the other hand, a ratio \mbox{$\Theta_{\rm subl}
\!>\! 1$}, varying smoothly with time, typically indicates a major contribution
to the total water production from sublimating icy-dust grains in the comet's
atmosphere.  Finally, a ratio \mbox{$\Theta_{\rm subl} \!\gg\! 1$}, varying
erratically with time, suggests an explosion, a likely result of fragmentation
that opens up the interior of the nucleus.

The first two scenarios in Table 6, {\it Models~A\/} and {\it B\/}, are based
on the assumption of the comet's maximum nuclear diameter at the time of Mars
encounter (Sec.~2.3); in addition, a nominal dust-to-water production rate
ratio of \mbox{$\Re_0 = 1.5$} is adopted in the first case, a nominal bulk
density of \mbox{$\rho = 0.4$ g cm$^{-3}$} in the second case.  In the next two
scenarios, {\it C\/} and {\it D\/}, the ratio $\Re_0$ is varied by a factor
of two either way, with the bulk density kept at its nominal value.  In the
remaining four scenarios, {\it E\/} through {\it H\/}, the bulk density is
varied from 0.2 to 0.5 g cm$^{-3}$, with $\Re_0$ kept at its nominal value.

\begin{table*}[p]
\vspace{0.1cm}
\begin{center}
{\footnotesize {\bf Table 6} \\[0.1cm]
{\sc Water Sublimation Models for Comet C/2012 S1 (2013 October 1--November
 16).}\\[0.2cm]
\begin{tabular}{l@{\hspace{0.5cm}}c@{\hspace{0.45cm}}c@{\hspace{0.45cm}}c@{\hspace{0.45cm}}c@{\hspace{0.45cm}}c@{\hspace{0.45cm}}c@{\hspace{0.45cm}}c@{\hspace{0.45cm}}c@{\hspace{0.25cm}}c}
\hline\hline\\[-0.1cm]
& \multicolumn{9}{@{\hspace{-0.05cm}}c}{Midtime, $t_{\rm obs}$, of 24-hour
 interval, from $t_{\rm obs}\!-\!12$\,hr to $t_{\rm obs}\!+\!12$\,hr (2013
 UT)} \\
& \multicolumn{9}{@{\hspace{-0.05cm}}c}{\rule[0.7ex]{12.02cm}{0.4pt}} \\
Quantity & \llap{O}ct.\,1\rlap{.0} & \llap{O}ct.\,24\rlap{.0}
 & \llap{N}ov.\,4\rlap{.0} & \llap{N}ov.\,10\rlap{.0}
 & \llap{N}ov.\,12\rlap{.0} & \llap{N}ov.\,13\rlap{.0}
 & \llap{N}ov.\,14\rlap{.0} & \llap{N}ov.\,15\rlap{.0}
 & \llap{No}v.\,1\rlap{6.0} \\[0.1cm]
\hline \\[-0.2cm]
Time from perihelion, $t_{\rm obs} \!- \! t_\pi$ (days) & $-$58.78 & $-$36.78
 & $-$24.78 & $-$18.78 & $-$16.78 & $-$15.78 & $-$14.78 & $-$13.78
 & $-$12.78 \\
Heliocentric distance, $r_{\rm obs}$ (AU) & 1.651 & 1.204 & 0.923 & 0.765
 & 0.709 & 0.680 & 0.650 & 0.620 & 0.589 \\
Observed production rate (10$^{28}$\,s$^{-1}$)$^{\rm a}$ & 1.17 & 1.58 & 2.69
 & 2.76 & 6.67 & 10.66 & 44.86 & 24.20 & 31.34 \\
Modeled averaged sublimation rate & & & & & & & & & \\[-0.03cm]
{\hspace{0.2cm}}per unit surface area (10$^{28}$km$^{-2}$s$^{-1}$) & 0.43
 & 0.94 & 1.72 & 2.59 & 3.06 & 3.35 & 3.69 & 4.09 & 4.56 \\
Effective sublimation area (km$^2$) & 2.70 & 1.68 & 1.57 & 1.06 & 2.18 & 3.18
 & 12.15 & 5.92 & 6.87 \\[0.1cm]
\hline \\[-0.15cm]
 & & & \llap{M{\scriptsize ODEL} A:{\hspace{0.45cm}}Nucleus' diameter at Mars
 encounter = 1.00 km{\hspace{0.5cm}}}\rlap{Bulk density = 0.11 g cm$^{-3}$
 {\hspace{0.4cm}}Dust-to-water ratio by mass = 1.5} & & & & & & \\[0.15cm]
\hline \\[-0.2cm]
Nucleus' production rate (10$^{28}$\,s$^{-1}$)$^{\rm b}$ & 0.68 & 1.44
 & 2.59 & 3.86 & 4.54 & 4.94 & 5.41 & 5.85 & 6.34 \\
Remaining mass of nucleus (10$^{14}$\,g)$^{\rm c}$       & 0.581 & 0.562
 & 0.546 & 0.534 & 0.527 & 0.521 & 0.492 & 0.476 & 0.456 \\
Sublimation rate ratio, $\Theta_{\rm subl}(t_{\rm obs}$) & 1.7  & 1.1
 & 1.0  & 0.7  & 1.5  & 2.2  & 8.3  & 4.1  & 4.9  \\[0.1cm]
\hline \\[-0.15cm]
 & & & \llap{M{\scriptsize ODEL} B:{\hspace{0.45cm}}Nucleus' diameter at Mars
 encounter = 1.00 km{\hspace{0.5cm}}}\rlap{Bulk density = 0.40 g cm$^{-3}$
 {\hspace{0.4cm}}Dust-to-water ratio by mass = 8.0} & & & & & & \\[0.15cm]
\hline \\[-0.2cm]
Nucleus' production rate (10$^{28}$\,s$^{-1}$)$^{\rm b}$ & 0.68 & 1.44
 & 2.59 & 3.86 & 4.53 & 4.94 & 5.41 & 5.85 & 6.34 \\
Remaining mass of nucleus (10$^{14}$\,g)$^{\rm c}$       & 2.093 & 2.024
 & 1.967 & 1.924 & 1.900 & 1.876 & 1.771 & 1.715 & 1.642 \\
Sublimation rate ratio, $\Theta_{\rm subl}(t_{\rm obs}$) & 1.7 & 1.1
 & 1.0  & 0.7  & 1.5  & 2.2  & 8.3  & 4.1  & 4.9  \\[0.1cm]
\hline \\[-0.15cm]
 & & & \llap{M{\scriptsize ODEL} C:{\hspace{0.42cm}}Nucleus' diameter at Mars
 encounter = 0.58 km{\hspace{0.45cm}}}B\rlap{ulk density = 0.40 g cm$^{-3}$
 {\hspace{0.4cm}}Dust-to-water ratio by mass = 0.75} & & & & & & \\[0.15cm]
\hline \\[-0.2cm]
Nucleus' production rate (10$^{28}$\,s$^{-1}$)$^{\rm b}$ & 0.23 & 0.48
 & 0.87 & 1.30 & 1.52 & 1.66 & 1.81 & 1.96 & 2.13 \\
Remaining mass of nucleus (10$^{14}$\,g)$^{\rm c}$       & 0.406 & 0.393
 & 0.382 & 0.374 & 0.369 & 0.364 & 0.344 & 0.333 & 0.319 \\
Sublimation rate ratio, $\Theta_{\rm subl}(t_{\rm obs}$) & 5.1 & 3.3
 & 3.1  & 2.1  & 4.4  & 6.4  & 24.7 & 12.3 & 14.7 \\[0.1cm]
\hline \\[-0.15cm]
 & & & \llap{M{\scriptsize ODEL} D:{\hspace{0.45cm}}Nucleus' diameter at Mars
 encounter = 0.76 km{\hspace{0.5cm}}}\rlap{Bulk density = 0.40 g cm$^{-3}$
 {\hspace{0.4cm}}Dust-to-water ratio by mass = 3.0} & & & & & & \\[0.15cm]
\hline \\[-0.2cm]
Nucleus' production rate (10$^{28}$\,s$^{-1}$)$^{\rm b}$ & 0.39 & 0.84
 & 1.51 & 2,24 & 2.63 & 2.87 & 3.14 & 3.40 & 3.68 \\
Remaining mass of nucleus (10$^{14}$\,g)$^{\rm c}$       & 0.926 & 0.895
 & 0.870 & 0.851 & 0.841 & 0.830 & 0.784 & 0.759 & 0.726 \\
Sublimation rate ratio, $\Theta_{\rm subl}(t_{\rm obs}$) & 3.0 & 1.9
 & 1.8  & 1.2  & 2.5  & 3.7  & 14.3 & 7.1  & 8.5 \\[0.1cm]
\hline \\[-0.15cm]
 & & & \llap{M{\scriptsize ODEL} E:{\hspace{0.47cm}}Nucleus' diameter at Mars
 encounter = 0.82 km{\hspace{0.5cm}}}\rlap{Bulk density = 0.20 g cm$^{-3}$
 {\hspace{0.4cm}}Dust-to-water ratio by mass = 1.5} & & & & & & \\[0.15cm]
\hline \\[-0.2cm]
Nucleus' production rate (10$^{28}$\,s$^{-1}$)$^{\rm b}$ & 0.46 & 0.97
 & 1.75 & 2.60 & 3.06 & 3.33 & 3.65 & 3.94 & 4.27 \\
Remaining mass of nucleus (10$^{14}$\,g)$^{\rm c}$       & 0.579 & 0.560
 & 0.544 & 0.532 & 0.526 & 0.519 & 0.490 & 0.475 & 0.454 \\
Sublimation rate ratio, $\Theta_{\rm subl}(t_{\rm obs}$) & 2.6  & 1.6
 & 1.5  & 1.1  & 2.2  & 3.2  & 12.3 & 6.1 & 7.3 \\[0.1cm]
\hline \\[-0.15cm]
 & & & \llap{M{\scriptsize ODEL} F:{\hspace{0.42cm}}Nucleus' diameter at Mars
 encounter = 0.72 km{\hspace{0.45cm}}}B\rlap{ulk density = 0.30 g cm$^{-3}$
 {\hspace{0.4cm}}Dust-to-water ratio by mass = 1.5} & & & & & & \\[0.15cm]
\hline \\[-0.2cm]
Nucleus' production rate (10$^{28}$\,s$^{-1}$)$^{\rm b}$ & 0.35 & 0.74
 & 1.33 & 1.99 & 2.33 & 2.54 & 2.78 & 3.01 & 3.26 \\
Remaining mass of nucleus (10$^{14}$\,g)$^{\rm c}$       & 0.579 & 0.559
 & 0.544 & 0.532 & 0.525 & 0.518 & 0.490 & 0.474 & 0.454 \\
Sublimation rate ratio, $\Theta_{\rm subl}(t_{\rm obs}$) & 3.3 & 2.1
 & 2.0  & 1.4  & 2.9  & 4.2  & 16.1 & 8.0  & 9.6 \\[0.1cm]
\hline \\[-0.15cm]
 & & & \llap{M{\scriptsize ODEL} G$^{\rm d}$:{\hspace{0.38cm}}Nucleus' diameter
 at Mars encounter = 0.65 km{\hspace{0.4cm}}}\rlap{Bulk density = 0.40 g
 cm$^{-3}$ {\hspace{0.35cm}}Dust-to-water ratio by mass = 1.5} & & & & &
 & \\[0.15cm]
\hline \\[-0.2cm]
Nucleus' production rate (10$^{28}$\,s$^{-1}$)$^{\rm b}$ & 0.29 & 0.61
 & 1.10 & 1.64 & 1.93 & 2.10 & 2.30 & 2.49 & 2.70 \\
Remaining mass of nucleus (10$^{14}$\,g)$^{\rm c}$       & 0.580 & 0.561
 & 0.545 & 0.533 & 0.527 & 0.520 & 0.491 & 0.476 & 0.455 \\
Sublimation rate ratio, $\Theta_{\rm subl}(t_{\rm obs}$) & 4.0 & 2.6
 & 2.4 & 1.7 & 3.5 & 5.1 & 19.5 & 9.7 & 11.6 \\[0.1cm]
\hline \\[-0.15cm]
 & & & \llap{M{\scriptsize ODEL} H:{\hspace{0.42cm}}Nucleus' diameter at Mars
 encounter = 0.61 km{\hspace{0.45cm}}}B\rlap{ulk density = 0.50 g cm$^{-3}$
 {\hspace{0.4cm}}Dust-to-water ratio by mass = 1.5} & & & & & & \\[0.15cm]
\hline \\[-0.2cm]
Nucleus' production rate (10$^{28}$\,s$^{-1}$)$^{\rm b}$ & 0.25 & 0.53
 & 0.95 & 1.41 & 1.66 & 1.81 & 1.98 & 2.14 & 2.32 \\
Remaining mass of nucleus (10$^{14}$\,g)$^{\rm c}$       & 0.579 & 0.560
 & 0.544 & 0.533 & 0.526 & 0.519 & 0.490 & 0.475 & 0.455 \\
Sublimation rate ratio, $\Theta_{\rm subl}(t_{\rm obs}$) & 4.7 & 3.0 
 & 2.8 & 2.0 & 4.0 & 5.9 & 22.7 & 11.3 & 13.5 \\[0.1cm]
\hline\\[-0.1cm] 
\multicolumn{10}{l}{\parbox{17.4cm}{\scriptsize $^{\rm a}$Daily averages
 of the water production rate from{\vspace{-0.01cm}} Combi et al.\,(2014)
 for the dates October~24 through November~16; extrapolated to October~1
 by applying a law $r^{-0.95}$ (Figure 2) linked to Combi et al.'s daily
 average for October~24.}} \\[0.24cm]
\multicolumn{10}{l}{\parbox{17.4cm}{\scriptsize $^{\rm b}$Derived from the
 described water sublimation model for an intact spherical nucleus of
 given dimensions at time $t_{\rm obs}$, outgassing from the entire sunlit
 hemisphere.}} \\[0.28cm]
\multicolumn{10}{l}{\parbox{17.4cm}{\scriptsize $^{\rm c}$Derived from the
 described water sublimation model and the adopted dust-to-water mass
 production rate ratio; the mass refers to the end of the 24-hour period
 of time.  From November 13--14 on, the nucleus was made up of a
 progressively increasing number of fragments.}} \\[0.18cm]
\multicolumn{10}{l}{\parbox{15cm}{\scriptsize $^{\rm d}$Nominal model
 (Sec.~2.4).}}\\[0.5cm]
\end{tabular}}
\end{center}
\end{table*}
\begin{table*}[p]
\vspace{0.1cm}
\begin{center}
{\footnotesize {\bf Table 6 (continued)} \\[0.1cm]
{\sc Water Sublimation Models for Comet C/2012 S1 (2013
 November 17--28).}\\[0.2cm]
\begin{tabular}{l@{\hspace{0.5cm}}c@{\hspace{0.45cm}}c@{\hspace{0.45cm}}c@{\hspace{0.45cm}}c@{\hspace{0.45cm}}c@{\hspace{0.45cm}}c@{\hspace{0.45cm}}c@{\hspace{0.45cm}}c@{\hspace{0.25cm}}c}
\hline\hline\\[-0.1cm]
& \multicolumn{9}{@{\hspace{-0.1cm}}c}{Midtime, $t_{\rm obs}$, of 24-hour
 interval, from $t_{\rm obs}\!-\!12$\,hr to $t_{\rm obs}\!+\!12$\,hr (2013
 UT)} \\
& \multicolumn{9}{@{\hspace{-0.1cm}}c}{\rule[0.7ex]{12.02cm}{0.4pt}} \\
Quantity & \llap{N}ov.\,17\rlap{.0} & \llap{N}ov.\,18\rlap{.0}
 & \llap{N}ov.\,19\rlap{.0} & \llap{N}ov.\,20\rlap{.0}
 & \llap{N}ov.\,21\rlap{.0} & \llap{N}ov.\,22\rlap{.0}
 & \llap{N}ov.\,24\rlap{.0} & \llap{N}ov.\,26\rlap{.0}
 & \llap{No}v.\,2\rlap{8.0} \\[0.1cm]
\hline \\[-0.2cm]
Time from perihelion, $t_{\rm obs} \!- \! t_\pi$ (days) & $-$11.78 & $-$10.78
 & $\:-$9.78$\;$ & $\:-$8.78$\;$ & $\:-$7.78$\;$ & $\:-$6.78$\;$
 & $\:-$4.78$\;$ & $\:-$2.78$\;$ & $\:-$0.78$\;$ \\
Heliocentric distance, $r_{\rm obs}$ (AU) & 0.557 & 0.525 & 0.491 & 0.456
 & 0.420 & 0.382 & 0.300 & 0.206 & 0.083 \\
Observed production rate (10$^{28}$\,s$^{-1}$)$^{\rm a}$ & 59.17 & 139.7
 & 119.2 & 116.7 & 101.1 & (80) & (15) & (15) & (5) \\
Modeled averaged sublimation rate & & & & & & & & & \\[-0.03cm]
{\hspace{0.2cm}}per unit surface area (10$^{28}$km$^{-2}$s$^{-1}$) & 5.14
 & 5.84 & 6.73 & 7.88 & 9.38 & 11.5 & 19.1 & 42.3 & 287 \\
Effective sublimation area (km$^2$) & 11.5 & 23.9 & 17.7 & 14.8 & 10.8
 & \llap{(}6.97\rlap{)} & \llap{(}0.78\rlap{)} & \llap{(}0.35\rlap{)}
 & \llap{(}0.02\rlap{)} \\[0.1cm]
\hline \\[-0.15cm]
 & & & \llap{M{\scriptsize ODEL} A:{\hspace{0.45cm}}Nucleus' diameter at Mars
 encounter = 1.00 km{\hspace{0.5cm}}}\rlap{Bulk density = 0.11 g cm$^{-3}$
 {\hspace{0.4cm}}Dust-to-water ratio by mass = 1.5} & & & & & & \\[0.15cm]
\hline \\[-0.2cm]
Nucleus' production rate (10$^{28}$\,s$^{-1}$)$^{\rm b}$ & 6.97 & 7.58 & 7.87
 & 7.78 & 7.57 & 7.11 & 5.90 & 9.00 & 34.4 \\
Remaining mass of nucleus (10$^{14}$\,g)$^{\rm c}$   & 0.418 & 0.328 & 0.252
 & 0.177 & 0.112 & 0.060 & 0.031 & 0.017 & $<$0.001 \\
Sublimation rate ratio, $\Theta_{\rm subl}(t_{\rm obs}$)$^{\rm b}$ & 8.5 & 18.4
 & 15.2 & 15.0 & 13.3 & 11.2 & 2.5 & 1.7 & 0.15 \\[0.1cm]
\hline \\[-0.15cm]
 & & & \llap{M{\scriptsize ODEL} B:{\hspace{0.45cm}}Nucleus' diameter at Mars
 encounter = 1.00 km{\hspace{0.5cm}}}\rlap{Bulk density = 0.40 g cm$^{-3}$
 {\hspace{0.4cm}}Dust-to-water ratio by mass = 8.1} & & & & & & \\[0.15cm]
\hline \\[-0.2cm]
Nucleus' production rate (10$^{28}$\,s$^{-1}$)$^{\rm b}$ & 6.97 & 7.58
 & 7.86 & 7.77 & 7.55 & 7.08 & 5.80 & 8.73 & 31.9 \\
Remaining mass of nucleus (10$^{14}$\,g)$^{\rm c}$       & 1.505 & 1.181
 & 0.904 & 0.633 & 0.398 & 0.212 & 0.108 & 0.057 & $<$0.001 \\
Sublimation rate ratio, $\Theta_{\rm subl}(t_{\rm obs}$)$^{\rm b}$ & 8.5 & 18.4
 & 15.2 & 15.0 & 13.4 & 11.3 & 2.6 & 1.7 & 0.16 \\[0.1cm]
\hline \\[-0.15cm]
 & & & \llap{M{\scriptsize ODEL} C:{\hspace{0.42cm}}Nucleus' diameter at Mars
 encounter = 0.58 km{\hspace{0.45cm}}}B\rlap{ulk density = 0.40 g cm$^{-3}$
 {\hspace{0.4cm}}Dust-to-water ratio by mass = 0.75} & & & & & & \\[0.15cm]
\hline \\[-0.2cm]
Nucleus' production rate (10$^{28}$\,s$^{-1}$)$^{\rm b}$ & 2.34 & 2.54
 & 2.64 & 2.61 & 2.54 & 2.38 & 1.97 & 2.99 & 11.3 \\
Remaining mass of nucleus (10$^{14}$\,g)$^{\rm c}$       & 0.292 & 0.229
 & 0.176 & 0.123 & 0.078 & 0.042 & 0.022 & 0.012 & $<$0.001 \\
Sublimation rate ratio, $\Theta_{\rm subl}(t_{\rm obs}$)$^{\rm b}$ & 25.3
 & 55.0 & 45.2 & 44.7 & 39.9 & 33.6 & 7.6 & 5.0 & 0.44 \\[0.1cm]
\hline \\[-0.15cm]
 & & & \llap{M{\scriptsize ODEL} D:{\hspace{0.45cm}}Nucleus' diameter at Mars
 encounter = 0.76 km{\hspace{0.5cm}}}\rlap{Bulk density = 0.40 g cm$^{-3}$
 {\hspace{0.4cm}}Dust-to-water ratio by mass = 3.0} & & & & & & \\[0.15cm]
\hline \\[-0.2cm]
Nucleus' production rate (10$^{28}$\,s$^{-1}$)$^{\rm b}$ & 4.05 & 4.40
 & 4.56 & 4.51 & 4.38 & 4.10 & 3.34 & 4.99 & 17.7 \\
Remaining mass of nucleus (10$^{14}$\,g)$^{\rm c}$       & 0.666 & 0.522
 & 0.399 & 0.279 & 0.175 & 0.093 & 0.047 & 0.024 & $<$0.001 \\
Sublimation rate ratio, $\Theta_{\rm subl}(t_{\rm obs}$)$^{\rm b}$ & 14.6
 & 31.8 & 26.1 & 25.9 & 23.1 & 19.5 & 4.5 & 3.0 & 0.28 \\[0.1cm]
\hline \\[-0.15cm]
 & & & \llap{M{\scriptsize ODEL} E:{\hspace{0.47cm}}Nucleus' diameter at Mars
 encounter = 0.82 km{\hspace{0.5cm}}}\rlap{Bulk density = 0.20 g cm$^{-3}$
 {\hspace{0.4cm}}Dust-to-water ratio by mass = 1.5} & & & & & & \\[0.15cm]
\hline \\[-0.2cm]
Nucleus' production rate (10$^{28}$\,s$^{-1}$)$^{\rm b}$ & 4.70 & 5.11
 & 5.30 & 5.24 & 5.09 & 4.77 & 3.89 & 5.84 & 21.0 \\
Remaining mass of nucleus (10$^{14}$\,g)$^{\rm c}$       & 0.416 & 0.327
 & 0.250 & 0.175 & 0.110 & 0.058 & 0.030 & 0.015 & $<$0.001 \\
Sublimation rate ratio, $\Theta_{\rm subl}(t_{\rm obs}$)$^{\rm b}$ & 12.6
 & 27.4 & 22.5 & 22.3 & 19.9 & 16.8 & 3.9 & 2.6 & 0.24 \\[0.1cm]
\hline \\[-0.15cm]
 & & & \llap{M{\scriptsize ODEL} F:{\hspace{0.42cm}}Nucleus' diameter at Mars
 encounter = 0.72 km{\hspace{0.45cm}}}B\rlap{ulk density = 0.30 g cm$^{-3}$
 {\hspace{0.4cm}}Dust-to-water ratio by mass = 1.5} & & & & & & \\[0.15cm]
\hline \\[-0.2cm]
Nucleus' production rate (10$^{28}$\,s$^{-1}$)$^{\rm b}$ & 3.58 & 3.89
 & 4.04 & 3.99 & 3.88 & 3.63 & 2.95 & 4.40 & 15.5 \\
Remaining mass of nucleus (10$^{14}$\,g)$^{\rm c}$       & 0.416 & 0.326
 & 0.249 & 0.174 & 0.109 & 0.058 & 0.029 & 0.015 & $<$0.001 \\
Sublimation rate ratio, $\Theta_{\rm subl}(t_{\rm obs}$)$^{\rm b}$ & 16.5
 & 35.9 & 29.5 & 29.2 & 26.1 & 22.0 & 5.1 & 3.4 & 0.32 \\[0.1cm]
\hline \\[-0.15cm]
 & & & \llap{M{\scriptsize ODEL} G$^{\rm d}$:{\hspace{0.38cm}}Nucleus' diameter
 at Mars encounter = 0.65 km{\hspace{0.4cm}}}\rlap{Bulk density = 0.40 g
 cm$^{-3}$ {\hspace{0.35cm}}Dust-to-water ratio by mass = 1.5} & & & & &
 & \\[0.15cm]
\hline \\[-0.2cm]
Nucleus' production rate (10$^{28}$\,s$^{-1}$)$^{\rm b}$ & 2.96 & 3.22
 & 3.34 & 3.31 & 3.22 & 3.02 & 2.49 & 3.77 & 14.1 \\
Remaining mass of nucleus (10$^{14}$\,g)$^{\rm c}$       & 0.417 & 0.328
 & 0.251 & 0.176 & 0.111 & 0.059 & 0.031 & 0.016 & $<$0.001 \\
Sublimation rate ratio, $\Theta_{\rm subl}(t_{\rm obs}$)$^{\rm b}$ & 20.0
 & 43.4 & 36.7 & 35.3 & 31.4 & 26.5 & 6.0 & 4.0 & 0.35 \\[0.1cm]
\hline \\[-0.15cm]
 & & & \llap{M{\scriptsize ODEL} H:{\hspace{0.42cm}}Nucleus' diameter at Mars
 encounter = 0.61 km{\hspace{0.45cm}}}B\rlap{ulk density = 0.50 g cm$^{-3}$
 {\hspace{0.4cm}}Dust-to-water ratio by mass = 1.5} & & & & & & \\[0.15cm]
\hline \\[-0.2cm]
Nucleus' production rate (10$^{28}$\,s$^{-1}$)$^{\rm b}$ & 2.55 & 2.77
 & 2.88 & 2.84 & 2.76 & 2.59 & 2.12 & 3.19 & 11.6 \\
Remaining mass of nucleus (10$^{14}$\,g)$^{\rm c}$       & 0.417 & 0.327
 & 0.250 & 0.175 & 0.110 & 0.059 & 0.030 & 0.016 & $<$0.001 \\
Sublimation rate ratio, $\Theta_{\rm subl}(t_{\rm obs}$)$^{\rm b}$ & 23.2
 & 50.4 & 41.4 & 41.0 & 36.6 & 30.9 & 7.1 & 4.7 & 0.43 \\[0.1cm]
\hline\\[-0.1cm] 
\multicolumn{10}{l}{\parbox{17.4cm}{\scriptsize $^{\rm a}$Daily averages of
 the water production rate from {\vspace{-0.01cm}}Combi et al.\,(2014) for
 the dates November 17 through 21; extrapolated to November 28 by mimicking
 the variations in the HCN production rate (see the IRAM website in
 footnote~5) and linked to Combi et al.'s daily average for November~21;
 these numbers are parenthesized.}} \\[0.33cm]
\multicolumn{10}{l}{\parbox{17.4cm}{\scriptsize $^{\rm b}$Derived from
 the described water sublimation model for an intact spherical nucleus
 of given dimensions at time $t_{\rm obs}$, outgassing from the
 entire sunlit hemisphere.}} \\[0.28cm]
\multicolumn{10}{l}{\parbox{17.4cm}{\scriptsize $^{\rm c}$Derived from
 the described water sublimation model and the adopted dust-to-water mass
 production rate ratio by mass; the mass refers to the end of the 24-hour
 period of time.  During the entire period of November~17--28, the nucleus
 was made up of a progressively increasing number of fragments.}} \\[0.31cm]
\multicolumn{10}{l}{\parbox{15cm}{\scriptsize $^{\rm d}$Nominal model
 (Sec.~2.4).}}
\end{tabular}}
\end{center}
\end{table*}

Inspection of Table 6 suggests that the adoption of the maximum diameter
of 1~km for the nucleus at the time of Mars encounter leads to an improbably
low density for Model~A and to extremely high dust-to-water production rate
ratio for Model~B; in addition, from Sec.~2.4 the particle size distribution
function would have to be extraordinarily flat, with \mbox{$\tau \,\lapeq
\,3$}.  We doubt that these two are realistic scenarios and we ignore them
in the following.  Since all remaining scenarios, Models~C--H, require that
the nucleus at Mars encounter be less than 1~km across, a significant
contamination by dust of the brightest pixel in the HiRISE images, already
suggested in Sec.~2.4 on the basis of what is now called Model~G, appears
to be corroborated.

On the other hand, the diameter $D_0(t_{\rm enc})$ in a range of 0.6 to
0.8~km, implied by Models C--H, requires that the observed water production
signature at the time of Mars encounter was due mostly to icy-dust grains
in the atmosphere, since the ratio \mbox{$\Theta_{\rm subl} \approx
2$.6--5.1}.  This ratio is seen to be systematically decreasing between
October~1 and November~10, which is due largely to the sluggish rate of
increase in the water production rate in the period October~1--24.  As
late as 18 days before perihelion, we still fail to see any sign of the
comet's impending cataclysmic demise.

About 17 days before perihelion, on November~\mbox{11--12}, the downturn
in $\Theta_{\rm subl}$ was suddenly reversed and on November~14 the ratio
reached a peak between 12 and 25 in Models~C through H, declining back
to between 6 to 15 during the next two days.  A new surge of water release
began November~17, when $\Theta_{\rm subl}$ jumped by a factor of 1.7,
and the next day by another factor of 2.2.  For the nominal model, G, the
increase in $\Theta_{\rm subl}$ between November 10 and 18 amounted to 41.7,
up a factor of more than 25.  Afterwards, the ratio began to decrease,
quite possibly at an accelerated rate.  The behavior of $\Theta_{\rm subl}$
is somewhat reminiscent, in the period from November~11 on, of the activity
cycles (Sec.~2.1), with an ignition point on November~11, an expansion stage
terminated at a stagnation point on November~18, and a depletion stage
ending on November~28.

However, two enormous peaks and the temporal extent suggested that much
more was at stake in this case than an injection of a cloud of icy grains
into the atmosphere.  The persistence of exceptionally high levels of
water production and the erratic variations in its rate were diagnostic
of a major augmentation of the sublimation surface, which is hard to achieve
without suddenly opening the nucleus' interior by its severe fragmentation.  

Let us assume that sublimating fragments of the~nucleus satisfy a
size-distribution law, such that the number of fragments whose effective
diameters are between $D$ and \mbox{$D \!+\! dD$} is \mbox{$f(D) dD$} and
the minimum and maximum diameters are, respectively, $D_{\rm min}$ and
$D_{\rm max}$.\footnote{The definition of the minimum fragment diameter,
$D_{\rm min}$, excludes of course dust, whose mass is expressed through
the ratio $\Re_0$.  However, in an advanced stage of the fragmentation
process, the difference between the smallest fragments and the largest
dust particles may become blurred.}  Accordingly, the total number, $N_{\rm
frg}$, of fragments is
\begin{equation}
\int_{D_{\rm min}}^{D_{\rm max}} \!\! f(D) \, dD = N_{\rm frg}.
\end{equation}
From Eq.\,(25), the fragments' total sublimation area is
\begin{equation}
\int_{D_{\rm min}}^{D_{\rm max}} \!\! {\textstyle \frac{1}{2}} \pi D^2 \!
  f(D) \, dD =  {\textstyle \frac{1}{2}} \pi D_0^2 \Theta_{\rm subl}.
\end{equation}
Assuming that all fragments were sublimating, a sum of their masses,
${\cal M}_{\rm frg}(D)$, should equal the mass of the nucleus,
${\cal M}_0$, at the onset of its fragmentation, $t_{\rm frg}$,
\begin{equation}
\int_{D_{\rm min}}^{D_{\rm max}} \!\! {\cal M}_{\rm frg}(D) \, f(D) \, dD =
 {\cal M}_0,
\end{equation}
where, as before, ${\cal M}_0 = {\textstyle \frac{1}{6}} \pi \rho D_0^3$ for
an intact nucleus at $t_{\rm frg}$, with $\rho$ being a constant bulk density.
Similar relation also applies to any fragment.

To illustrate the implications of these conditions in general and of the
magnitude of $\Theta_{\rm subl}$ in particular, we adopt for the size
distribution a law
\begin{equation}
f(D) \, dD = C_{\rm D} D^{-3.5} \, dD,
\end{equation}
where $C_{\rm D}$ is a constant.  This law is identical with the size
distribution law for a collisional model of asteroids and their debris,
derived by Dohnanyi (1969).  Although the fragmentation mechanisms are
different, explosive phenomena are involved in both cases.{\pagebreak}

The integrals (27) to (29) now become
\begin{eqnarray}
N_{\rm frg} & = & 0.4 \, C_{\rm D} D_{\rm min}^{-2.5} \!\left[ 1 \!-\! \left(
 \! \frac{D_{\rm min}}{D_{\rm max}} \! \right)^{\!\! \frac{5}{2}} \right] \!,
 \nonumber \\[0.15cm]
D_0^2 \, \Theta_{\rm subl} & = & 2 \, C_{\rm D} D_{\rm min}^{-\frac{1}{2}} \!
 \left[ 1 \!-\! \left( \! \frac{D_{\rm min}}{D_{\rm max}} \!
 \right)^{\!\! \frac{1}{2}} \right] \!, \nonumber \\[0.15cm]
D_0^3 & = & 2 \, C_{\rm D} D_{\rm max}^{\frac{1}{2}} \!\! \left[ 1 \!-\! \left(
 \! \frac{D_{\rm min}}{D_{\rm max}} \! \right)^{\!\! \frac{1}{2}} \right] \!.
\end{eqnarray}
Dividing the third equation by the second, we obtain
\begin{equation}
\frac{D_{\rm min}}{D_0} \, \frac{D_{\rm max}}{D_0} =
  \frac{1}{\Theta_{\rm subl}^2}.
\end{equation}
Dividing the first equation by the second, we find, after inserting from
Eq.\,(32) for $D_{\rm min}/D_0$,
\begin{equation}
N_{\rm frg} = 0.2\,\Theta_{\rm subl}^5 \! \left( \! \frac{D_{\rm max}}{D_0}
 \! \right)^{\!\! 2} \! \left[ 1 \!+\! \sum_{k=1}^{4} \! \left( \!
 \Theta_{\rm subl} \, \frac{D_{\rm max}}{D_0} \! \right)^{\!\! -k} \right]
 \! .
\end{equation}

We note that these conditions are independent of the nucleus' size, because
the dimensions of fragments enter the equations only as fractions of the
parent's dimensions.  However, we have two equations for three unknowns,
$N_{\rm frg}$, $D_{\rm min}$, and $D_{\rm max}$, so one of them is still
a free parameter.  We choose $D_{\rm max}$, the diameter of the largest
fragment.

We now examine two extreme possibilities with major implications.  One is
a case in which the largest fragment has a diameter considerably (orders
of magnitude) greater than the smallest fragment, \mbox{$D_{\rm min}\!\ll\!
D_{\rm max}$}, but comparable to the parent's diameter, \mbox{$D_{\rm max} =
\Gamma_{\rm frg} D_0$}, where $\Gamma_{\rm frg}$ is a number moderately
smaller than unity.  The diameter of the smallest fragment is then found
to equal \mbox{$D_{\rm min} = D_0/(\Theta_{\rm subl}^2 \Gamma_{\rm frg})$}
and, since \mbox{$\Theta_{\rm subl} \Gamma_{\rm frg} \! \gg \! 1$},
\begin{equation}
N_{\rm frg} \simeq 0.2 \, \Theta_{\rm subl}^5 \, \Gamma_{\rm frg}^2 
 \left[ 1 \!+\! (\Theta_{\rm subl} \, \Gamma_{\rm frg})^{-1} \right] .
\end{equation}

At the other extreme, let the largest fragment be much smaller than the
nucleus, \mbox{$D_{\rm max}\!\ll\!D_0$}, but just moderate\-ly greater
than the smallest {\vspace{-0.04cm}}fragment, \mbox{$D_{\rm min} \!=\!
\Gamma_{\rm frg} D_{\rm max}$}, where again \mbox{$\Gamma_{\rm frg}
\!<\!  1$}.~Then~\mbox{$D_{\rm min} \!=\!  D_0 \sqrt{\Gamma_{\rm frg}}/
\Theta_{\rm subl}$} and
\begin{equation}
N_{\rm frg} \simeq 0.2 \, \frac{\Theta_{\rm subl}^3}{\Gamma_{\rm frg} \!
 \left( 1 \!-\! \sqrt{\Gamma_{\rm frg}} \, \right)}.
\end{equation}
Thus, the number of fragments varies with the 5th power of the ratio
$\Theta_{\rm subl}$ in the first case, but only with its cube in the
second case.

Before we assess the extent of fragmentation in the period of time beginning
in mid-November, we remark on one fundamental difference between Event~1 and
Event~2.  At the end of Event~1, the comet still retained about 85~percent
of its water-ice reservoir available at the time of Mars encounter, whereas
at the end of Event~2 the reservoir was nearly gone.  This difference in
the outcome suggests that after Event~1 the nucleus' debris included very
large fragments (carrying the ice supplies), which, however, collapsed into
much smaller, ``dry'' pieces during Event~2.  Thus, Events~1 and 2 resemble,
respectively, the first and the second extreme cases described above. 

We now use Model G to assess the effects of fragmentation in Event~1 and 2.
The other acceptable models, C--F and H, would provide fairly similar results.
The nucleus' mass remaining after Event~1 is in Table~6 equivalent to an
intact nucleus 0.62~km across.  The ratio \mbox{$\Theta_{\rm subl} = 19.5$},
but an enhancement over the pre-outburst level of this ratio is estimated at
only $\sim$17.  A choice of, for example, 300~meters for the diameter of the
largest fragment requires \mbox{$\Gamma_{\rm frg} \approx \frac{1}{2}$},
which implies that the smallest fragments are $\sim$4~meters across and that
\mbox{$N_{\rm frg} \approx 80\,000$} from Eq.\,(34).  This is a meaningful,
even though crude, order-of-magnitude estimate, all that we are interested in.

The nucleus' mass, $0.06 \times 10^{14}$\,g, remaining after Event~2 (Model~G)
is equivalent to a nucleus 0.30~km across, and we estimate the corrected ratio
at \mbox{$\Theta_{\rm subl} \simeq 41$}.  In order to get the same dimensions
of the smallest fragments as before, one needs to adopt \mbox{$\Gamma_{\rm frg}
\approx 0.35$}, in which case the largest fragment is about 12~meters across
and \mbox{$N_{\rm frg} \approx 100\,000$} from Eq.\,(35), a truly cataclysmic
event.  For comparison, the mass of the sublimated water ice and ejected dust,
integrated over the duration of Event~2, amounted to $0.36 \times 10^{14}$\,g,
six times the mass in the cloud of remaining fragments.  They were in the next
few days after Event~2 subjected to ever increasing temperatures and therefore
to further fragmentation, and also to sublimation, especially of sodium
compounds.

The applied model could be refined in several ways, for example, by
distinguishing between sublimating and ``dry'' fragments, by varying the
dust-to-water sublimation rate ratio with time, etc.  Such a refinement
would, however, require a number of additional assumptions and/or
unavailable parameters.  Our aim has been not to describe the process
of the comet's disintegration in detail, but, rather, to illustrate what
information can be extracted by applying a basic sublimation model.

To summarize, we propose that the timeline of the disintegration of comet
C/2012~S1 went like this: The cycle E was definitely over by November~12,
16~days before perihelion, when a relatively minor increase in the
production of water occurred, a precursor to Event~1.  The peak water
production rate, 16 times the rate four days earlier, coincided with a
peak on the light curve.  The comet tapped a limited source of water,
evidently from its interior, that had been unavailable before.  To make
that happen, the nucleus had to be subjected to major fragmentation.  If
the production rate went back to normal in a matter of a day or two, Event~1
might have had no major effect on the comet's health.  Instead, several days
later the water-production rate went up another factor of 3.  The mass of the
already fragmented nucleus was in this explosion, closely related to Event~2
on the light curve, shattered into an estimated hundred thousand or so
boulders, none much larger than several meters across.  A major
difference between Events~1 and 2 was in the degree of retention of
water ice supplies by sizable fragments:\ it was fairly high in Event~1 but
close to nil in Event~2, indicating a failure of these fragments to survive
this latter event.  Even much smaller boulders were still falling apart as
late as hours before perihelion, as shown by a fading narrow extension
protruding from the head in a series of SOHO's C2 preperihelion images.

Some 3 days before perihelion came a dramatic drop in the production of
gas, as already discussed in Sec.~2.2.  Although this development has not
as yet been independently confirmed, the available results appear to
imply that the comet's reservoirs of ices, including water ice, were
at this point nearly or completely exhausted.  The comet {\it de facto\/}
ceased to exist already at this time, but the debris continued to orbit
the Sun.  The loss of ices deprived the comet of gas-driven activity,
but by now, less than 0.2~AU from the Sun, the sublimation rate of
sodium from dust was still increasing.  In the meantime, the process of
cascading fragmentation continued.  The sodium sublimation rate was
increasing not only because of the growing temperature, but also due to
continuing fragmentation that was multiplying the total cross-sectional
area of the debris.  The flare-up during Event~3 was probably brought
about in this fashion, as the peak near 15~{\Rsun} compares favorably
with similar peaks for the Kreutz minicomets.  Curdt et al.\ (2014) placed
a dust-emission event at 10~{\Rsun}, based on their observations made with
the SUMER spectrometer on board SOHO.  The life or death by sublimation
was for each dust particle decided by its perihelion distance.
Images taken merely an hour or so before perihelion suggest that the most
resistant component of the disrupted nucleus, boulder-sized fragments, were
also succumbing to the hostile environment, as the thin extension which they
populated and which pointed from the head in the direction of motion in
earlier images, was fading rapidly.  The process of cascading fragmentation
continued down to about 5~{\Rsun}, as shown by evidence from early
post-perihelion images on the ultimate termination of all activity
3.5~hours before perihelion (Sec.~3.9).

\subsection{Estimated Size of the Largest Surviving Fragments of the Nucleus}
The question of what are the comet's largest surviving fragments has two
sides:\ active vs inert ones.  Knight \& Battams (2014) remarked that ``any
remaining active nucleus was $<$10~m in radius.''  Our conclusion that all
activity ceased 3.5~hours before perihelion implies that {\it no active\/}
fragment of the nucleus survived.  Our argument is consistent with no
detection of any trace of a tail made up of perihelion or post-perihelion
ejecta in the HI1 imager and in the coronagraphs C2, C3, COR2-A, and COR2-B.
It also fits Curdt et al.'s (2014) failure to find any Lyman-$\alpha$ emission
less than 1~hour before perihelion.

The problem of the largest inert fragments is more difficult and an answer
more uncertain.  Knight \& Battams (2014) argue that the ``limiting
magnitudes of the SOHO and STEREO telescopes do not set meaningful upper
limits on any surviving inactive fragments.''  While this is true, our
examination of the comet's post-perihelion light curve (Figures~3--5) and
morphology of its nucleus' relics (Figures~10--14) offers an approximate
solution, based on an indirect method.

A necessary prerequisite for this approach is a selection of a
size-distribution function of surviving inert material.  As in Sec.~4.2, we
employ the law from Eq.\,(30), which now covers both the nucleus' fragments
and dust ejecta.  Integrating it again from a minimum particle diameter,
$D_{\rm min}$, to the maximum diameter, $D_{\rm max}$, that we search for,
we find the relationship between the total mass, ${\cal M}_{\rm surv}$, of
surviving particulates and their total cross-sectional area, $X_{\rm surv}$,
in the form
\begin{equation}
{\cal M}_{\rm surv} = {\textstyle \frac{2}{3}} \rho_{\rm eff} X_{\rm surv}
 \sqrt{D_{\rm min} D_{\rm max}},
\end{equation}
where $\rho_{\rm eff}$ is an effective bulk density of the fragments.  If
the bulk density is related to the fragment size via Eq.\,(21), a plausible
approximation for $\rho_{\rm eff}$ is
\begin{equation}
\rho_{\rm eff} = \sqrt{\rho_{\rm min} \, \rho_{\rm max}},
\end{equation}
where $\rho_{\rm min}$ is the bulk density of the smallest fragments,
whose diameter is $D_{\rm min}$, and $\rho_{\rm max}$ the bulk density
of the largest fragments.  The diameter of these largest fragments is
equal
\begin{equation}
D_{\rm max} = \left( \frac{3 {\cal M}_{\rm surv}}{2 \rho_{\rm eff}
 X_{\rm surv}} \! \right)^{\!2} \!\! \cdot \! D_{\rm min}^{-1}.
\end{equation}

The cross-sectional area of surviving fragments of the nucleus can be
estimated from an intrinsic brightness $\Im_{\rm surv}$ of the comet's
rocky relics in early post-perihelion images measured in a large aperture.
The most appropriate data are the apparent magnitudes published by Nakano
(2013a) and plotted in Figures~3--5.

Figure 3 shows that up to at least Nakano's first data point, referring
to November 29.38 UT, or 0.60~day after perihelion, the light curve
follows essentially an inverse-square law of heliocentric distance, so
that $\Im_{\rm surv}$ and the total cross-sectional area of the nucleus'
relics are practically constant.  From an apparent magnitude of +0.5 that
the object had at that time according to Nakano (2013a) in a 27-arcmin
aperture, it follows that after accounting for the effects of SOHO-centric
distance and the phase angle, \mbox{$H_\Delta = +1.3$}.  From Eq.\,(1)
with \mbox{$r = 0.0681$ AU} we obtain \mbox{$\Im = 0.140$} and from
Eq.\,(2) \mbox{$X_{\rm surv} = 5 \times 10^4$ km$^2$}, assuming the cloud
was optically thin.  Furthermore, the calculations show that the smallest
particles that fitted a 27-arcmin aperture at the time had
\mbox{$\beta_{\rm min} = 0.052$} along the cloud's leading boundary and
0.055 along its trailing boundary (cf.\ Figure~13 for the object's
appearance 2~hours later).  On the average, these radiation-pressure
accelerations indicate that \mbox{$D_{\rm min} \simeq 11$ microns} and
\mbox{$\rho_{\rm min} \simeq 1.9$ g cm$^{-3}$}.  The largest fragments
should according to Eq.\,(21) have $\rho_{\rm max}$ close to 0.4~g~cm$^{-3}$,
so that \mbox{$\rho_{\rm eff} \simeq 0.9$ g cm$^{-3}$}.

The mass ${\cal M}_{\rm surv}$ of the surviving debris in the 27-arcmin
aperture can be estimated only very approximately.  Our guess for a crude
upper limit is \mbox{$5 \times 10^{13}$\,g}, from Model G in Table~6, equal
to the nucleus' mass just before Event~1.  However, since much of that mass
sublimated near perihelion (Sec.~4.1), a tighter estimate should be closer
to \mbox{$5 \times 10^{12}$\,g} or less.  Inserted with the other numbers
into Eq.\,(38), the two mass limits lead to, respectively, \mbox{$D_{\rm max}
\!\ll 0.25\:$m} and \mbox{$D_{\rm max} \!  < 0.25\:$cm}.  We thus find  that
the most sizable surviving inert fragments of the nucleus were unlikely to be
larger than pebbles and may have been just subcentimeter-sized grains.  The
comet's disintegration was apparently quite complete, with no boulders left
intact.

\section{Sublimation of Dust in the Streamer}
In Sec.~3.2 we briefly investigated the nature of the sharp, narrow dust
streamer, noting that a typical grain size was in the millimeter range and
larger and that the release of grains from the nucleus dated back to the
times when the comet was very far from the Sun. Because of the crowding
of synchrones on top of each other, the times of release could not be
determined with any degree of accuracy from the measured position angles.

%
The streamer's disconnection from the head and its likely physical trigger
--- dust-grain sublimation near the Sun --- were already mentioned in
Sec.~3.2.  In the following we describe our measurements of the point of
disconnection in a number of images and our investigation of the phenomenon.

\subsection{Measurement of the Streamer's Disconnection in Images Taken
with the C2 Coronagraph}
Even though the gap between the comet's head and streamer appears at first
sight to end abruptly, inspection under magnification shows that there in
fact is a steep but smooth transition.  Measurements of the streamer's
point of disconnection are therefore affected by errors that depend on
the measurer's perception and judgment.

\begin{table}[t]
\noindent
\vspace{-0.23cm}
\begin{center}
{\footnotesize {\bf Table 7} \\[0.08cm]
{\sc Measurements of Streamer's Disconnection in Images\\Taken with the C2
Coronagraph.} \\[0.14cm] 
\begin{tabular}{c@{\hspace{0.6cm}}c@{\hspace{0.5cm}}c@{\hspace{0.2cm}}c@{\hspace{0.2cm}}c}
\hline\hline\\[-0.2cm]
\multicolumn{2}{@{\hspace{-0.1cm}}c}{Time of imaging}
 & \multicolumn{3}{@{\hspace{-0.05cm}}c}{Streamer's tip} \\[-0.08cm]
\multicolumn{2}{@{\hspace{-0.1cm}}c}{\rule[0.7ex]{3.36cm}{0.4pt}}
 & \multicolumn{3}{@{\hspace{-0.05cm}}c}{\rule[0.7ex]{4.2cm}{0.4pt}} \\[-0.08cm]
2013 Nov. & relative to & separation & position & separation \\
(UT) & perihelion & from head & angle & from Sun \\[0.05cm]
\hline \\[-0.24cm]
 & & \hspace{-0.15cm}$^\circ$ & & \hspace{-0.17cm}$^\circ$ \\[-0.33cm]
%
28.654 & $t_\pi \!-\! 3.00$\,hr & 0\,.71 & 253\rlap{$^\circ$} & 1\,.88 \\
28.660 & $t_\pi \!-\! 2.84$\,hr & 0.76 & 252 & 1.89 \\
28.669 & $t_\pi \!-\! 2.65$\,hr & 0.88 & 252 & 1.92 \\
28.675 & $t_\pi \!-\! 2.50$\,hr & 0.94 & 251 & 1.92 \\
28.683 & $t_\pi \!-\! 2.30$\,hr & 1.01 & 251 & 1.91 \\
28.692 & $t_\pi \!-\! 2.10$\,hr & 1.06 & 250 & 1.88 \\
28.700 & $t_\pi \!-\! 1.90$\,hr & 1.22 & 250 & 1.95 \\
28.708 & $t_\pi \!-\! 1.70$\,hr & 1.28 & 249 & 1.93 \\
28.717 & $t_\pi \!-\! 1.50$\,hr & 1.38 & 249 & 1.94 \\
28.725 & $t_\pi \!-\! 1.30$\,hr & 1.47 & 248 & 1.94 \\
\hline\\[-0.3cm]
28.825 & $t_\pi \!+\! 1.10$\,hr & 2.10 & 232 & 1.48 \\
28.842 & $t_\pi \!+\! 1.52$\,hr & 2.26 & 229 & 1.48 \\
28.858 & $t_\pi \!+\! 1.90$\,hr & 2.40 & 227 & 1.48 \\
28.884 & $t_\pi \!+\! 2.52$\,hr & 2.63 & 225 & 1.52 \\
28.908 & $t_\pi \!+\! 3.10$\,hr & 2.77 & 222 & 1.49 \\[0.05cm]
\hline\\[-0.1cm]
\end{tabular}}
%
%
\end{center}
\end{table}

Because of the large pixel size (56 arcsec) of the C3 coronagraph, our
measurements were limited to images taken with the C2 coronagraph.  The
results, presented in Table~7, show that most data points come from the
preperihelion images in which the streamer was much brighter.  Only
every second image showing the disconnection point was measured in
the post-perihelion images (in which however the streamer itself did
not yet pass perihelion).  The strong tendency for the disconnection
point to stay at the same angular distance from the Sun is clearly
demonstrated; errors of measurement are estimated at less than
$\pm$0$^\circ\!$.1.  Surprisingly, however, this critical distance differs
in the pre- and post-perihelion images, amounting to, on the average,
\mbox{$1^\circ\!.92 \pm 0^\circ\!.03$} and \mbox{$1^\circ\!.49
\pm 0^\circ\!.02$}, respectively,  This difference cannot be due to errors
of measurement.   Next, we develop and apply a sublimation
model in an effort to estimate the thermophysical properties of the dust in
the streamer.

\subsection{Sublimation Rate and Its Integrated Effect}
The streamer's termination on its sunward side allows one to estimate the
sublimation heat of the material that made up the streamer and thus to
confirm or refute the preliminary conclusion that water-ice grains could not
be involved (Sekanina 2013c; also Sec.~3.2).

For a dust grain located at the point of disconnection of the streamer,
the radius $a_{\rm rls}$ at the time $t_{\rm rls}$ of its release from
the comet should equal a linear loss rate by sublimation, $da_{\rm
subl}/dt$, integrated over the period of time from $t_{\rm rls}$ to the
time of the grain's complete sublimation, $t_{\rm subl}$:
\begin{equation}
\Delta a_{\rm subl} = a_{\rm rls} = \int_{t_{\rm rls}}^{t_{\rm subl}}
  \frac{da_{\rm subl}}{dt} \, dt.
\end{equation}
The loss rate of a grain's radius by sublimation is a function of the	
equilibrium temperature $T$ and can be written in terms of the mass
sublimation rate per unit area, $d{\cal Z}_{\rm subl}(T)/dt$, and the
grain's bulk density, $\rho$,
\begin{equation}
\frac{da_{\rm subl}}{dt} = \frac{1}{\rho} \frac{d}{dt} {\cal Z}_{\rm
  subl}(T) ,
\end{equation}
where
\begin{equation}
\frac{d}{dt} {\cal Z}_{\rm subl}(T) = \wp \sqrt{\frac{\mu}{2\pi \epsilon
  \:\! {\Re}T}},
\end{equation}
with $\Re$ being the gas constant (in cal mol$^{-1}$K$^{-1}$),
a conversion factor $\epsilon = 4.1854 \times 10^7$\,erg cal$^{-1}$,
$\mu$ the molar weight (in g mol$^{-1}$), and $\wp$ the vapor pressure
of the sublimating material (in dyn cm$^{-2}$), for which we write
\begin{equation}
\wp = \Lambda \exp \!\left(\!-\frac{L}{{\Re}T} \!\right).
\end{equation}
In this expression $L$ is the latent heat of sublimation of the material
(in cal mol$^{-1}$).  The coefficient $\Lambda$ (in dyn cm$^{-2}$) is
calculated from an approximate formula
\begin{equation}
\Lambda = \exp \left( 22.105 \!+\! 0.956 \!\times\! 10^{-4} L \right).
\end{equation}
Inserting from (40), (41), and (42) into (39), and replacing, as a variable,
time $t$ with heliocentric distance $r$, we find with a parabolic
approximation and neglect of radiation pressure effects (because only
large dust is involved), for the grain's radius (in cm) that sublimated
away
\begin{equation}
\Delta a_{\rm subl} = C_0 \!\! \int_{r_{\rm subl}}^{r_{\rm rls}} \!
  r^{\frac{1}{2}} \!\left( 1 \!-\! \frac{q}{r} \right)^{-\frac{1}{2}}
  \! T^{-\frac{1}{2}} \exp \!\left( \! - \frac{L}{{\Re}T} \!\right) dr,
\end{equation}
where $q$ is the perihelion distance (in AU) of the grain's orbit, the
temperature is a function of heliocentric distance, \mbox{$r_{\rm subl} =
r(t_{\rm subl})$}, \mbox{$r_{\rm rls} = r(t_{\rm rls})$},
\begin{equation}
C_0 = \frac{\Lambda \sigma}{2k_{\rm G} \rho}\sqrt{\frac{\mu}{\pi \epsilon
 \:\!{\Re}}},
\end{equation}
$k_{\rm G} = 0.0172021$ AU$^{\frac{3}{2}}$ day$^{-1}$, and $\sigma$ is
a conversion factor, $\sigma = 0.864 \times 10^5$\,s day$^{-1}$.

In general, the equilibrium temperature $T(r)$ of a dust grain of radius $a$
in the radiation field of the Sun is derived from the balance between the
solar energy absorbed and the energy reradiated at the given heliocentric
distance $r$ (e.g., Sekanina et al.\ 2001):
\begin{equation}
\frac{\pi a^2}{r^2} \!\!\int_{0}^{\infty}\!\!\!\!\!{\cal Q}_{\rm abs}(a,\lambda)
 S_0(\lambda) d\lambda = 4 \pi a^2 \!\! \int_{0}^{\infty} \!\!\!\!\!
 {\cal Q}_{\rm abs}(a,\lambda) \pi {\cal B_\lambda}(T) d\lambda,
\end{equation}
where the integration is carried out over all wavelengths $\lambda$,
${\cal Q}_{\rm abs}(a,\lambda)$ in the grain's absorption efficiency 
at $\lambda$, $S_0(\lambda)$ is the solar flux at 1~AU, and $\pi{\cal
B}_\lambda(T)$ is the Planck function.  Because we deal with grains
much larger than the wavelength, ${\cal Q}_{\rm abs}$ is practically
constant (e.g., van de Hulst 1957), so that {\vspace{-0.03cm}}Eq.\,(46)
becomes independent of the particle's cross-sectional area for absorption,
\mbox{$\pi a^2 {\cal Q}_{\rm abs}$}.  The integral on the right side is
proportional to $T^4$ and
\begin{equation}
T(r) = T_0 \sqrt{r_0/r},
\end{equation}
where $r_0 = 1$ AU and $T_0 = 280$ K is a blackbody approximation.  However,
because of the particle's sublimation, only part of the absorbed solar
radiation is spent on the reradiation of the energy and the temperature
could increase with decreasing heliocentric distance somewhat less steeply
than indicated by Eq.\,(47).  With this caveat in mind (see below), we
{\vspace{-0.04cm}}nevertheless insert it into Eq.\,(44) and substitute
\mbox{$z = \sqrt{r/q}$}.  Recognizing that substantial particle sublimation
occurs only near $t_{\rm subl}$ and that therefore \mbox{$r_{\rm rls}
\rightarrow \infty$} represents an excellent approximation, the expression
for $\Delta a_{\rm subl}$ becomes
\begin{equation}
\Delta a_{\rm subl} = C \! \int_{\sqrt{r_{\rm subl}/q}}^{\infty}
  z^{\frac{5}{2}} \!  \left(\! 1 \!-\! \frac{1}{z^2} \!
  \right)^{\!-\frac{1}{2}} \! \exp \:\!(-B z) \, dz,
\end{equation}
where
\begin{equation}
C = 2 \:\! C_0 T_0^{-\frac{1}{2}} r_0^{-\frac{1}{4}} q^{\frac{7}{4}}
\end{equation}
and
\begin{equation}
B = \frac{L}{{\Re} T_0} \sqrt{\frac{q}{r_0}}.
\end{equation}
Equation (48) can easily be integrated numerically, after substituting
a new variable, \mbox{$\zeta = 1/z$},
\begin{equation}
\Delta a_{\rm subl} = C \! \int_{0}^{\sqrt{q/r_{\rm subl}}} \!
  \zeta^{-\frac{9}{2}} \!\left( 1 \!-\! \zeta^2 \right)^{-\frac{1}{2}}
  \exp \! \left( \! -\frac{B}{\zeta} \right) d \zeta.
\end{equation}

\begin{table*}[ht]
\noindent
\vspace{-0.25cm}
\begin{center}
{\footnotesize {\bf Table 8} \\[0.1cm]
{\sc Dust Grain Sublimation Parameters for Streamer in Images Taken witth
the C2 Coronagraph.} \\[0.2cm] 
\begin{tabular}{c@{\hspace{0.05cm}}c@{\hspace{0.2cm}}c@{\hspace{0.25cm}}c@{\hspace{-0.06cm}}c@{\hspace{0.15cm}}c@{\hspace{0.22cm}}c@{\hspace{-0.03cm}}c@{\hspace{0.15cm}}c@{\hspace{0.22cm}}c@{\hspace{-0.03cm}}c@{\hspace{0.15cm}}c@{\hspace{0.22cm}}c@{\hspace{-0.03cm}}c}
\hline\hline\\[-0.2cm]
 & Subli- & \multicolumn{12}{@{\hspace{0.01cm}}c}{Sublimation parameters of
   dust grains for a preperihelion time of release, $t_{\rm rls}$, from the
   comet} \\[-0.04cm]
Time & mation
& \multicolumn{12}{@{\hspace{0.01cm}}c}{\rule[0.7ex]{15.25cm}{0.4pt}}\\[-0.04cm]
of & distance & \multicolumn{3}{@{\hspace{0cm}}c}{at 10 AU from Sun}
   & \multicolumn{3}{@{\hspace{0cm}}c}{at 20 AU from Sun}
   & \multicolumn{3}{@{\hspace{0cm}}c}{at 50 AU from Sun}
   & \multicolumn{3}{@{\hspace{0cm}}c}{at 100 AU from Sun} \\[-0.04cm]
\raisebox{0.3ex}{imaging} & from
          & \multicolumn{3}{@{\hspace{0cm}}c}{\rule[0.7ex]{3.68cm}{0.4pt}}
          & \multicolumn{3}{@{\hspace{0cm}}c}{\rule[0.7ex]{3.68cm}{0.4pt}}
          & \multicolumn{3}{@{\hspace{0cm}}c}{\rule[0.7ex]{3.68cm}{0.4pt}}
& \multicolumn{3}{@{\hspace{0cm}}c}{\rule[0.7ex]{3.68cm}{0.4pt}} \\[-0.04cm]
2013 & Sun, & Grain & Grain & Sublima-
            & Grain & Grain & Sublima-
            & Grain & Grain & Sublima-
            & Grain & Grain & Sublima- \\[-0.04cm]
Nov. & \raisebox{0.4ex}{$r_{\rm subl}$} & acceler- & diam.  & tion heat
     & acceler- & diam. & tion heat & acceler- & diam. & tion heat
     & acceler- & diam. & tion heat \\[-0.04cm]
(UT) & ({\Rsun}) & ation $\beta$ & (cm) & (cal\,mol$^{-1}\!$)
                 & ation $\beta$ & (cm) & (cal\,mol$^{-1}\!$)
                 & ation $\beta$ & (cm)\, & (cal\,mol$^{-1}\!$)
                    & ation $\beta$ & (cm)\, & (cal\,mol$^{-1}\!$) \\[0.05cm]
\hline \\[-0.21cm]
%
28.654 & 6.91 & 0.000510 & 0.47 & 90\,100 
              & 0.000181 & 1.45 & 85\,800
              & 0.0000459 & 6.00 & 80\,100
              & 0.0000163 & 17.3 & 75\,800 \\
28.660 & 6.89 & 0.000541 & 0.44 & 90\,500
              & 0.000192 & 1.36 & 86\,200
              & 0.0000487 & 5.65 & 80\,500
              & 0.0000172 & 16.3 & 76\,100 \\
28.669 & 7.03 & 0.000623 & 0.38 & 90\,000
              & 0.000221 & 1.17 & 85\,700
              & 0.0000561 & 4.89 & 80\,000
              & 0.0000199 & 14.1 & 75\,800 \\
28.675 & 7.04 & 0.000660 & 0.36 & 90\,100
              & 0.000234 & 1.10 & 85\,900
              & 0.0000594 & 4.61 & 80\,200
              & 0.0000210 & 13.3 & 75\,900 \\
28.683 & 7.01 & 0.000700 & 0.33 & 90\,600
              & 0.000249 & 1.03 & 86\,300
              & 0.0000630 & 4.34 & 80\,700
              & 0.0000223 & 12.5 & 76\,400 \\
28.692 & 6.90 & 0.000722 & 0.32 & 91\,700
              & 0.000256 & 1.00 & 87\,300
              & 0.0000650 & 4.20 & 81\,600
              & 0.0000230 & 12.1 & 77\,300 \\
28.700 & 7.15 & 0.000831 & 0.27 & 90\,200
              & 0.000295 & 0.86 & 86\,000
              & 0.0000748 & 3.63 & 80\,400
              & 0.0000265 & 10.5 & 76\,200 \\
28.708 & 7.06 & 0.000857 & 0.26 & 91\,100
              & 0.000304 & 0.83 & 86\,800
              & 0.0000771 & 3.52 & 81\,100
              & 0.0000273 & 10.2 & 76\,900 \\
28.717 & 7.10 & 0.000913 & 0.25 & 91\,000
              & 0.000324 & 0.78 & 86\,700
              & 0.0000822 & 3.30 & 81\,100
              & 0.0000291 & 9.56 & 76\,900 \\
28.725 & 7.09 & 0.000960 & 0.23 & 91\,300
              & 0.000341 & 0.74 & 87\,000
              & 0.0000865 & 3.13 & 81\,400
              & 0.0000306 & 9.08 & 77\,100 \\
\hline\\[-0.3cm]
28.825 & 5.51 & 0.001139 & 0.19 & \llap{1}08\,400
              & 0.000405 & 0.61 & \llap{1}03\,300
              & 0.0001026 & 2.62 & 96\,600
              & 0.0000363 & 7.63 & 91\,500 \\
28.842 & 5.49 & 0.001236 & 0.17 & \llap{1}09\,100
              & 0.000439 & 0.56 & \llap{1}04\,000
              & 0.0001113 & 2.41 & 97\,200
              & 0.0000394 & 7.02 & 92\,100 \\
28.858 & 5.51 & 0.001330 & 0.16 & \llap{1}09\,200
              & 0.000472 & 0.52 & \llap{1}04\,100
              & 0.0001198 & 2.23 & 97\,300
              & 0.0000424 & 6.51 & 92\,200 \\
28.884 & 5.62 & 0.001505 & 0.14 & \llap{1}08\,400
              & 0.000535 & 0.45 & \llap{1}03\,300
              & 0.0001356 & 1.96 & 96\,700
              & 0.0000480 & 5.73 & 91\,600 \\
28.908 & 5.56 & 0.001627 & 0.13 & \llap{1}09\,500
              & 0.000578 & 0.41 & \llap{1}04\,400
              & 0.0001466 & 1.80 & 97\,700
              & 0.0000519 & 5.29 & 92\,600 \\[0.05cm]
\hline\\[-0.2cm]
\end{tabular}}
%
%
\end{center}
\end{table*}
\noindent
As a check, an approximate solution exists in closed form.  To the extent
that the term $1/z^2$ in Eq.\,(48) can be neg\-lected, \mbox{$[1 -
(1/z^2)]^{-\frac{1}{2}} \approx 1$}, the integral becomes an incomplete
Gamma function $\Gamma(h,x)$,
\begin{eqnarray}
\int_{\sqrt{r_{\rm subl}/q}}^{\infty} z^{\frac{5}{2}} \! \left( \! 1 \!-\!
  \frac{1}{z^2} \! \right)^{\! -\frac{1}{2}} \! \exp \, (-Bz) \, dz \, \approx
  {\hspace{2.1cm}} & & \nonumber \\
\int_{\sqrt{r_{\rm subl}/q}}^{\infty} z^{\frac{5}{2}} \exp\, (-Bz) \, dz =
  B^{-\frac{7}{2}} \, \Gamma \! \left( \! {\textstyle \frac{7}{2}}, B
  \sqrt{\frac{r_{\rm subl}}{q}} \right) \! .
  {\hspace{0.5cm}} & &
\end{eqnarray}
Since for $h > 3$
\begin{eqnarray}
\Gamma(h,x) & = & (h\!-\!3)(h\!-\!2)(h\!-\!1) \Gamma(h\!-\!3,x)
  \nonumber \\[0.1cm]
 & & + \, x^{h-1} e^{-x} \! \left[ 1 \!+\! {\displaystyle
     \frac{h\!-\!1}{x}} \!+\! {\displaystyle \frac{(h\!-\!2)(h\!-\!1)}{x^2}}
     \right]
\end{eqnarray}
and
\begin{equation}
\Gamma({\textstyle \frac{1}{2}},x) = \sqrt{\pi} \left[ 1 \!-\! {\rm
erf}(\sqrt{x}) \right],
\end{equation}
the final form of an approximate solution to Eq.\,(48) in close form is
\begin{eqnarray}
\Delta a_{\rm subl} & = & \frac{\Lambda \sigma}{k_{\rm G} \rho}
  \sqrt{\frac{\mu}{\pi \epsilon\:\!{\Re} T_0}} \left(\!\frac{L}{{\Re}
  T_0}\!\right)^{\!\!-\frac{7}{2}} \!\! r_0^{\frac{3}{2}} \! \left\{
  \raisebox{0ex}[3ex][2ex]{} \! {\textstyle \frac{15}{8}} \sqrt{\pi}
  \left[ 1 \!-\! {\rm erf} \! \left( \! \sqrt{b} \right) \right] \right.
  \nonumber \\[0.15cm]
  & & \left. + \, b^{\frac{5}{2}} \exp (-b) \left[ 1 \!+\! \frac{5}{2b}
  \!+\! \frac{15}{4 b^2} \right] \! \right\},
\end{eqnarray}
where
\begin{equation}
b = \frac{L}{{\Re} T_0} \sqrt{ \frac{r_{\rm subl}}{r_0}} = B
  \sqrt{\frac{r_{\rm subl}}{q}}.\\[0.25cm]
\end{equation}
Equations (51) and (55) represent a constraint on the relevant properties
of grains:\ $\Delta a_{\rm subl}$, $\rho$, $\mu$, $r_{\rm subl}$, and $L$.
Next, we investigate this constraint for the material that made up the
streamer of comet C/2012~S1.

\subsection{Determination of Heliocentric Distance at Point of
Disconnection and Range of Sublimation Heat for Dust Grains in the Streamer}
In Table 7 we summarize the angular distances of the measured points of
disconnection (complete grain sublimation) in the streamer.  Viewed from
SOHO, the angles 1$^\circ\!$.5 and 1$^\circ\!$.9 correspond, at its distance
from the Sun, to heliocentric distances of $\sim$5.5 and $\sim$7.0~{\Rsun},
respectively.

The discrete distances of the points of disconnection require an
explanation.  The streamer's fainter part, terminating at 5.5~{\Rsun} in
the post-perihelion images, should also show up in the preperihelion
images.  Unfortunately, in projection it overlaps the northern boundary of
the tail from Event~2 and is not recognized as a separate feature.  The
positions of the point of disconnection that terminates at 7.0~{\Rsun}
were determined from the preperihelion images despite the overlap only
thanks to the exceptional brightness (Figure~8).  By contrast, in the
post-perihelion images the faint part of the streamer and the tail from
Event~2 clearly separate from one another.  The point of disconnection at
7.0~{\Rsun} is undetected in these images either because it is too far to
fit in or because it is no longer bright enough (in part because of
backscatter).  This interplay of coincidences suggests that
the streamer consists of at least two components.  It is in fact likely
that there are quite a few more than two overlapping components --- or
separate ``substreamers'' --- whose points of disconnection are located
between 5.5 and 7.0~{\Rsun}, but, except for the two, they are too faint
to detect.

Grains in the streamer are sorted by distance from the head so that, for
a given time of release from the comet and near-zero release velocities,
the smaller the acceleration by solar radiation pressure, $\beta$, the
closer to the head they are.  Thus, a proper approach is to determine~$\beta$
that the grains at the point of disconnection were subjected to in each
measured image by applying the equations of grain motion.  This solution
also pinpoints the position of the grains in space, thus furnishing the
heliocentric distance at complete sublimation, $r_{\rm subl}$.

A complication is that the time of release or the equivalent heliocentric
distance cannot for the streamer be unequivocally determined (Sec.~3.2);
we carried out the computations by assuming the grain release at 10, 20,
50, and 100~AU from the Sun (Table~8).  The knowledge of $\beta$ at a
particular distance of the point of disconnection then provides us with
a good estimate for the grains' size and therefore for $\Delta a_{\rm
subl}$ in Eqs.\,(51) and (55).  With the scattering efficiency for
radiation pressure assumed to be unity (an excellent approximation for
very large grains; e.g., van de Hulst 1957), the acceleration $\beta$ (in
units of the Sun's gravitational acceleration equal to 0.593 cm~s$^{-2}$
at 1~AU), exerted by solar radiation pressure on a grain whose diameter
is $\delta$ (in microns) and bulk density $\rho$ (in g cm$^{-3}$), is
expressed by a well-known relation (e.g., \mbox{Sekanina} et al.\ 2001),
\begin{equation}
\beta(\rho,\delta) = \frac{1.148}{\rho \, \delta},
\end{equation}
which is to be solved together with Eq.\,(21), the adopted relationship
between the grain's bulk density and size.

\begin{figure*}[ht]
\vspace{-1.77cm} 
\hspace{0.4cm} 
\centerline{
\scalebox{0.7}{ 
\includegraphics{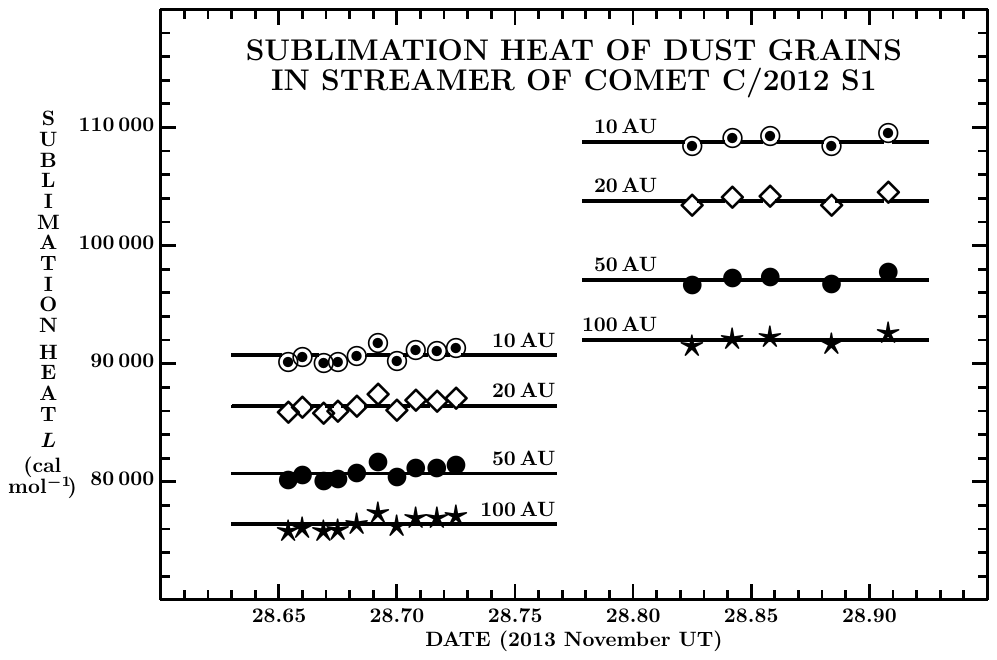}}} 
\vspace{-10.33cm} 
\caption{Sublimation heat $L$ of dust grains in the streamer of comet C/2012~S1
derived from the heliocentric distance $r_{\rm subl}$ of the streamer's
point of disconnecton in 15 images taken with the C2 coronagraph on board
SOHO.  Different symbols are used to plot the results for each of the four
assumed heliocentric distances at grain release from the comet:\ 10, 20, 50,
and 100~AU.  The distance of complete sublimation, $r_{\rm subl}$, averaged
7.02~{\Rsun} for{\vspace{-0.04cm}} the 10 preperihelion images on the left,
but 5.54~{\Rsun} for{\vspace{-0.04cm}} the five post-perihelion images.
This difference is responsible for the systematic shift in the range of
sublimation heat established from the pre- and post-perihelion images.
Note that the sublimation heat for grains released at 10~AU, as derived
from the preperihelion images, is very close to the sublimation heat for
very large grains (pebbles) released at 100~AU, as derived from the
post-perihelion images.{\vspace{0.4cm}}}
\end{figure*}

From the described procedure we have so far been able to gain information
on three of the five grain-dependent quantities that enter Eqs.\,(51) or
(55).  The only two remaining unknowns are the molar weight $\mu$ and the
sublimation heat $L$.  The effect of $\mu$ on $L$ in the range of our
solutions is shown below to be small, on the order of \mbox{$\pm$1--2}~percent
for $\mu$ varying by a factor of two or so.  The sublimation heat $L$ of the
dust in the streamer is by far the most important quantity.  We are now ready
to use Eq.\,(51) [with Eq.\,(55) as a check] to determine $L$ from the
perceived sublimation effect that causes the streamer's disconnection.

Our nominal runs in Table 8 used a generic molar weight of 100~g~mol$^{-1}$.
Not listed is the bulk density, which was always between 0.4 and
0.6~g~cm$^{-3}$ and can readily be ascertained from Eq.\,(21).  The numbers
are from the numerical integrations of Eq.\,(51); the approximate,
close-form formula gave, as expected, consistently smaller values for
$\Delta a_{\rm subl}$, mostly between 75 and 80~percent of the accurate
value, thus fulfilling its task.

All measured preperihelion images resulted in nearly the same value for the
sublimation heat, which depends significantly on the adopted heliocentric
distance $r_{\rm rls}$ at the time of dust release from the comet and
varies between about 91\,000~cal~mol$^{-1}$ for 10~AU from the Sun to
76\,000~cal~mol$^{-1}$ for 100~AU from the Sun.  The sublimated grain
diameter ialso depends a little on the time of imaging, as the critical
sublimation distance $r_{\rm subl}$ is reached by grains of gradually
decreasing size.  During the 1.7~hours spanned by the preperihleion images,
the diameter of the sublimated grains dropped by a factor of $\sim$2,
amounting to a few millimeters for the assumed release distance of
10~AU but $\sim$10~cm or more for 100~AU.

The smaller set of post-perihelion images yielded similarly consistent
results.  Because of the smaller heliocentric distances $r_{\rm subl}$,
the sublimation heat came out higher, between 109\,000~cal~mol$^{-1}$
for the release distance of 10~AU and 92\,000~cal~mol$^{-1}$ for 100~AU.
During the 2~hours spanned, the diameter of the sublimated grains dropped
by a factor of only 1.5 or so, varying from \mbox{1--2 mm} for the release
distance of 10~AU to as much as several centimeters for 100~AU.

Because the size of the sublimated away grains was decreasing with the
time of imaging, the heliocentric distance at complete grain sublimation,
$r_{\rm subl}$, should be slightly increasing with time in Table~8.  In
spite of some noise, a trend like this is indeed apparent in both the
preperihelion and post-perihelion sets, and is also confirmed by the
correlation coefficients, 0.70 and 0.71, respectively, even though the
average rates, \mbox{$2.6 \pm 0.9$ {\Rsun} day$^{-1}$} from{\vspace{-0.05cm}}
the preperihelion images and \mbox{$1.1 \pm 0.6$ {\Rsun} day$^{-1}$} from
the post-perihelion images, are poorly determined.  The respective average
values of $r_{\rm subl}$ are \mbox{$7.02 \pm 0.09$ {\Rsun}} and \mbox{$5.54
\pm 0.05$ {\Rsun}}, with the errors much smaller than the difference,
\mbox{$1.48 \pm 0.10$ {\Rsun}}.

\begin{figure*}[ht]
\vspace{-2.85cm} 
\hspace{0.2cm} 
\centerline{
\scalebox{0.7}{ 
\includegraphics{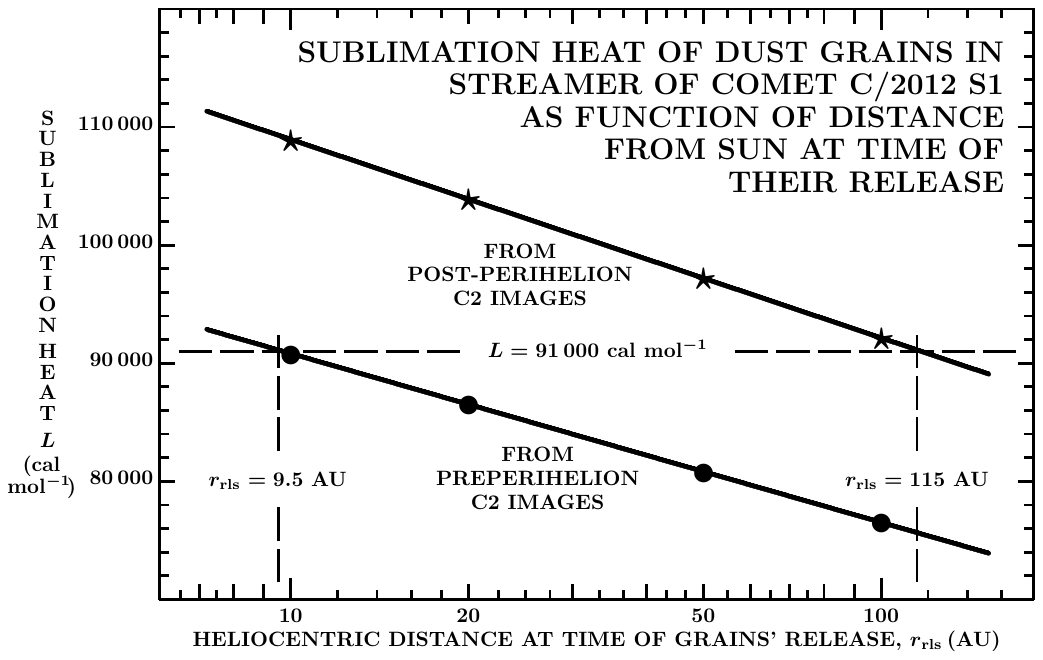}}} 
\vspace{-9.3cm} 
\caption{Sublimation heat $L$ of dust grains in the streamer plotted
against heliocentric distance $r_{\rm rls}$ at the time of their release.
Based on the averaged data points from Figure~16, this plot displays an
exponential dependence of the sublimation heat $L$ on the distance $r_{\rm
rls}$.  Note that the slopes of the lines of sublimated grains in the
preperihelion and post-perihelion images are not the same.  A solution
that satisfies the constraint of a constant sublimation heat is depicted by
a dashed line parallel to the axis of abscissae.  In an intriguing scenario,
the presence of grains with diameters of 1 mm or so and released 9.5~AU
from the Sun is suggested to explain the points of disconnection of the
streamer in the preperihelion images, while the presence of pebbles
several centimeters in diameter and released 115~AU from the Sun is
required to explain the points of disconnection of the streamer in the
post-perihelion images.  The sublimation heat of the streamer's material
is approximately 90\,000 cal mol$^{-1}$, close to that of atomic
iron.{\vspace{0.35cm}}}
\end{figure*}

The values of the sublimation heat derived from the 15 images taken with
the C2 coronagraph are plotted in Figure~16.  To the extent that the range
of assumed heliocentric distances at the times of grain release from the
comet cover all realistic possibilities, one can conclude from the figure
that the sublimation heat of the grains that made up the streamer was
between 75\,000 and 110\,000 cal mol$^{-1}$.  This is a highly refractory
material, comparable, for example, to atomic iron (for which \mbox{$L =
88\,100$ cal mol$^{-1}$} in a temperature range of 1730--3130~K or a
vapor-pressure range of 1--10$^5$\,Pa),\footnote{See, for example, {\tt
http://en.wikipedia.org/wiki/Iron}.} but not as refractory as silicates
(120\,000--130\,000~cal mol$^{-1}$).  One can definitely exclude not only
water ice, but also such substances as atomic sodium (25\,000 cal
mol$^{-1}$) and other similar metals that are much less refractory than
the above lower limit on $L$ suggests.  As remarked, the choice of a molar
weight has almost no effect on the value of the sublimation heat; replacing
100~g mol$^{-1}$ with, for example, 50~g~mol$^{-1}$ leads to $L$ smaller by
1200--1700~cal mol$^{-1}$.  

The results in Table 8 allow one to plot the average sublimation heat of
material in the streamer as a function of heliocentric distance, $r_{\rm
rls}$, at the time of release from the nucleus, separately from the
preperihelion and post-perihelion images (Figure~17).  Since $r_{\rm rls}$
correlates closely with the grain diameter, the difference between the
two exponential relationships can be attributed either to differences in
the dimensions of released grains of essentially the same material, or to
different sublimation heats and, thus, to different materials.  The first
option sounds more credible, because particle dimensions in comets are
known to vary widely; the other option appears factitious and is more
difficult to justify.

Accepting the first option, the sublimation heat of the material that made
up the streamer is the same for both the preperihelion and post-perihelion
images.  This constraint leads to an intriguing result:\ the solutions imply
that a likely sublimation heat of the streamer's material amounts to about
90\,000~cal mol$^{-1}$.  For a sublimation heat significantly lower, the
positions of the point of disconnection in the post-perihelion images would
require boulder-sized or larger fragments to have been released at enormous
heliocentric distances, on the order of many hundreds or even thousands AU;
for a sublimation heat significantly higher, the positions of the point of
disconnection in the preperihelion images would require that the heliocentric
distances at release be too small, inconsistent with the arguments presented
in Sec.~3.2.  In a narrow range of acceptable scenarios, the grains that
completely sublimated away in the preperihelion images were released from the
comet near 10~AU from the Sun and were initially close to 1~mm across, while
the grains that completely sublimated away in the post-perihelion images
were released near 100~AU from the Sun and were initially more than
10~times larger, several centimeters across.  One such solution is
depicted in Figure~17, with two clouds of grains of a sublimation heat
of 91\,000~cal mol$^{-1}$ released at, respectively, 9.5~AU and 115~AU.
Returning to Eq.\,(47), we find that even at a heliocentric distance as
small as 5.5~{\Rsun}, more than 98~percent of the solar radiation absorbed
by a dust particle made up of a material of this heat of sublimation is
spent on thermal reradiation; at larger distances the fraction is greater
still.  This means that Eq.\,(47) approximates the grain's equilibrium
temperature quite satisfactorily and Eq.\,(51) is appropriate for
determining the sublimation effect on dust in the range of relevant
dimensions.

\subsection{Activity of Comet C/2012 S1 Far from the Sun}

Even though the presented solution is the best apparent choice given the
options offered by Figure~17, one should ask whether this dust particle
sublimation model can be defended on physical grounds.  Reviewing briefly
the past work on cometary activity at large heliocentric distances, we
note that, on the observational side, the appearance of dynamically new
comets has provided evidence of preperihelion activity at distances of
more than 10~AU.  One long-known source of this information is the
orientation of narrow dust tails of such comets with perihelia beyond
2--3~AU (Sekanina 1975).\footnote{Strangely, major deviations in the
orientation of these tails (mistaken for plasma tails) from the
antisolar direction were in the 1960s thought to offer evidence, beyond
2~AU from the Sun, for now a long-abandoned hypothesis of a solar ``breeze''
(Chamberlain 1960, 1961).} Later, the same conclusion was reached based,
among others, on orbital evidence (Marsden et al.\ 1978), on an extent
of sources of activity on the nucleus (Rickman et al.\ 1991), and on
the depletion of carbon-chain molecules (A'Hearn et al.\ 1995).  Fairly
recently, dynamically new comets have been directly observed as active
objects displaying their tails at distances of more than 10~AU from
the Sun preperihelion (Meech et al.\ 2009).

On the theoretical side, a consensus has been emerging over the past
four decades (e.g., Patashnick et al.\ 1974, Klinger 1980, Smoluchowski
1981, Herman \& Podolak 1985, Prialnik 1992, 2006, Jenniskens \& Blake 1996,
Enzian et al.\ 1998, Korsun \& Ch\"{o}rny 2003, Gonz\'alez et al.\ 2008)
that a major source of cometary activity far from the Sun are processes in
amorphous water ice that must exist in comets if ice deposition had taken
place under the conditions of very low temperatures and low pressures.

The progress in the understanding of the physical processes involving
amorphous water ice at low temperatures has been accelerated by numerous
laboratory experiments.  A typical procedure is to deposit water vapor in
a vacuum chamber on a receptor cooled usually to 10--20~K; guest gases,
such as CO, CO$_2$, CH$_4$, NH$_3$, N$_2$, etc., are co-deposited either
simultaneously or subsequently (e.g., Bar-Nun et al.\ 1985, 1987, Laufer
et al.\ 1987, 2005, Schmitt \& Klinger 1987, Schmitt et al.\ 1989, Ayotte
et al.\ 2001).  Large amounts of the guest gases get trapped in the pores
of amorphous ice, with an excess, if any, freezing on the surface.  Upon
gradually warming up the receptor, it is observed that once the excess gas
sublimates away from the surface, intense release of gas from within the
ice begins at 35--37~K.  This gas evacuates pore space in the ice by its
slow annealing, a process that proceeds in the laboratory stepwise as
long as the temperature keeps increasing, and still can continue at
temperatures as high as 110~K.  Release of icy grains was observed in
laboratory experiments when gas was escaping at high rates.  Meech et
al.\ (2009) proposed that ``gas release during the annealing process
between $\sim$37~K and 120~K can account for the activity \ldots of comets
at distances [at which the] temperature could not reach the $\sim$120~K
phase transition temperature'' of the ice.  Besides, a transformation of
two metastable phases takes place between 38~K and 68~K, from high-density
to low-density amorphous ice (Jenniskens \& Blake 1994), a transition
that requires a higher activation energy than the annealing.

To the extent that the results of these laboratory experiments are applicable
to dynamically new comets, such as C/2012~S1, they imply dust grains' release
as soon as a temperature of \mbox{$T_{\rm anneal} \simeq 37$~K} has been
reached at the nucleus' surface.  This temperature refers to a heliocentric
distance $r_{\rm anneal}$ given by
\begin{equation}
r_{\rm anneal} = \left( \frac{T_{\rm ref}}{T_{\rm anneal}} \right)^{\!2} \!\!,
\end{equation}
where a reference temperature $T_{\rm ref}$ (at 1~AU from the Sun) depends on
the insolation regime of the nucleus.  The only relevant piece of information
on C/2012~S1 that we are aware of is Li et al.'s (2014) finding that the
nucleus always faced the Sun with one hemisphere until shortly before
perihelion.  This mimicks a case of nonrotating regime, so that \mbox{$T_{\rm
ref} \simeq 394$ K} and
\begin{equation}
r_{\rm anneal} \simeq 113\;{\rm AU}.
\end{equation}
This heliocentric distance essentially coincides with $r_{\rm rls}$ found
in Figure~17 for a sublimation heat of 91\,000 cal mol$^{-1}$ from the
positions of the points of disconnection of the streamer in the
post-perihelion images.

As the comet continued to approach the Sun and~its surface temperature
eventually exceeded 120~K, an exo\-thermic transition from amorphous to
crystalline (cubic) ice began to affect the balance of energy at the
surface.  Crystallization of the annealed, but still gas-laden, amorphous
ice completed the evacuation of the remaining trapped gases and in the
process of their release dust grains were again lifted along.  Because
the crystallization time varies with the temperature $T$ exponentially,
as $\exp(-{\rm const}/T)$, there is no single temperature that defines
the beginning of this process.  However, the crystallization process was
essentially over by the time the temperature reached $\sim$150~K, while
the rate of trapped-gas release peaked close to 130~K or 135~K.  Schmitt
et al.'s (1989) results show that trapped CO$_2$ --- next to CO of
considerable interest --- has a peak evacuation rate near 125~K.  Writing
the relation between the crystallization temperature $T_{\rm cryst}$ and
the crystallization heliocentric distance $r_{\rm cryst}$ in a form
analogous to Eq.\,(58) and taking \mbox{$T_{\rm cryst} = 125\!-\!130$ K},
we find
\begin{equation}
r_{\rm cryst} = 9.2\!-\!9.9\:{\rm AU},
\end{equation}
comparable to $r_{\rm rls}$ in Figure~17 for a sublimation heat of 91\,000~cal
mol$^{-1}$ derived from the positions of the point of disconnection of the
streamer in the preperihelion images.
%

To summarize, we find a remarkably consistent parallelism between the dust
particle sublimation model for the streamer of comet C/2012~S1, on the one
hand, and the existing consensus on the preperihelion activity of dynamically
new comets at very large heliocentric distances, on the other hand.
Specifically, we correlate a dust release event just beyond $\sim$100~AU
with the initial annealing of amorphous water ice at 37\,K and an event
near 10~AU with the exothermic phase change from amorphous to cubic ice at
125--130\,K; the released refractory material, pebble-sized in the first
case and millimeter-sized in the second case, is found to possess thermal
properties that are consistent with the heat of sublimation of
$\sim$91\,000~cal mol$^{-1}$.  Cursory inspection suggests that the
amount of the debris released during the explosive crystallization
of amorphous ice exceeded that released during the annealing, a tentative
conclusion that is qualitatively consistent with laboratory experiments
(Meech et al.\ 2009).  We do not rule out the existence of overlapping
``substreamers'' as products of additional, followup annealing events
between 100 and 10~AU.

\begin{figure*}[t]
\vspace{-1.62cm} 
\hspace{0cm}
\centerline{
\scalebox{0.57}{
\includegraphics{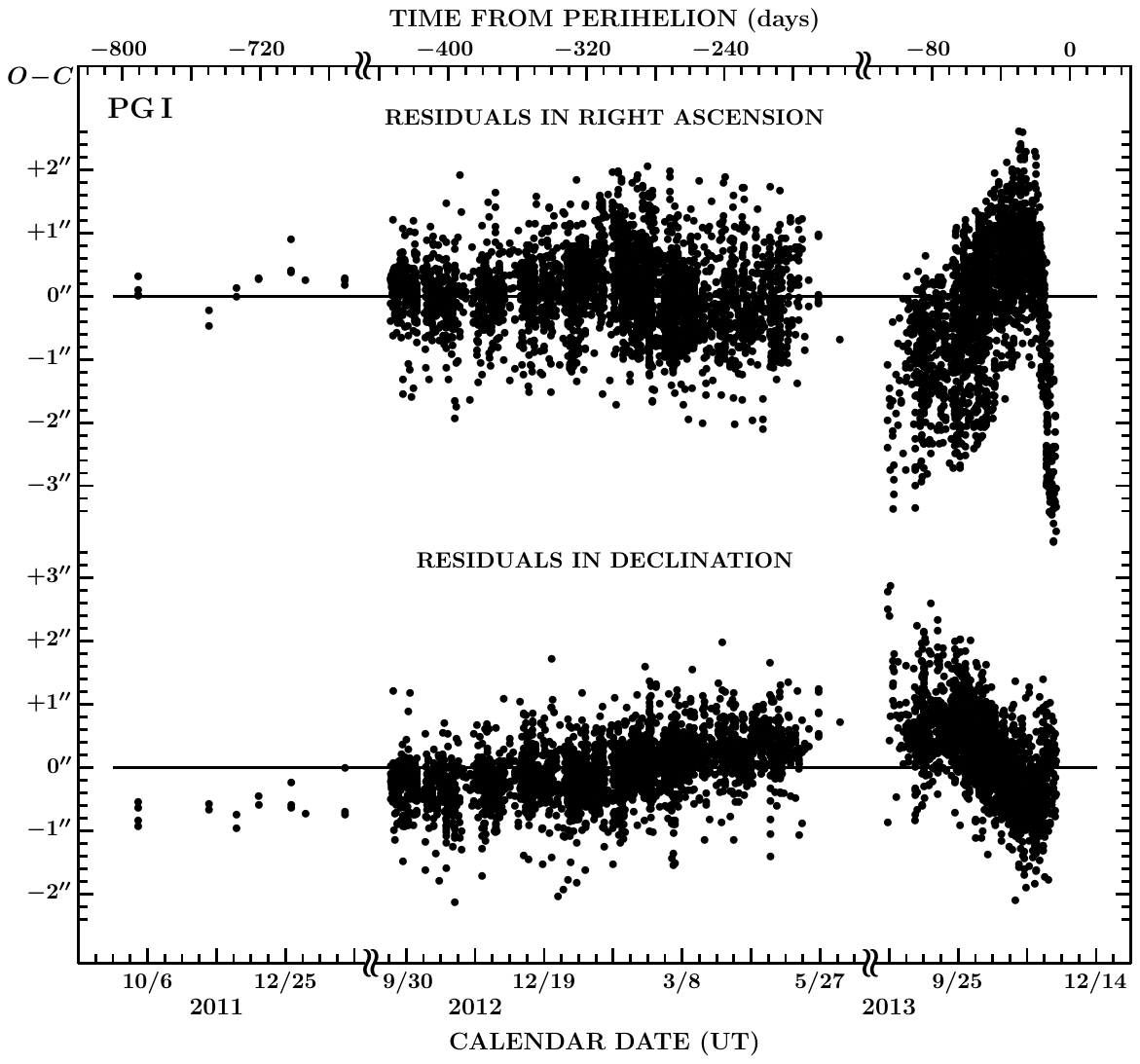}}} 
\vspace{-4.09cm}
\caption{Distribution of residuals for 6166 observations from the gravitational
solution PG\,I.  Very strong trends are apparent in either coordinate after
conjunction with the Sun; a continuous trend is also seen in declination
throughout the pre-conjunction period of time.{\vspace{0.35cm}}}
\end{figure*}

\section{Orbital Modeling and Analysis}

Given the violent events to which the comet was exposed, it is to be
expected that the orbital motion was affected as well and that its analysis
should contribute to the overall understanding of the comet's response.

The {\it Minor Planet Center\/} (MPC) published a number of successively
improved gravitational solutions that eventually covered an orbital arc
from the earliest pre-discovery observations on 2011 September~30 to 2013
November~20 (Williams 2013a), merely two days short of the time of the
comet's last ground-based astrometry.  Only at that point were further
gravitational runs abandoned (Williams 2013b).  As a rule, the positional
residuals from individual observations were not published, and it has been
impossible to make any judgment on the quality of fit and the presence of
systematic trends from a mean residual of $\pm$0$^{\prime\prime}\!$.6, which
depends primarily on an arbitrarily chosen cutoff for rejecting inaccurate
observations.  An exception was the first nongravitational solution (Williams
2013b), for which individual residuals were published, including those of
discarded observations.  Cursory inspection shows that, even though the value
of the resulting nongravitational parameter was one of the largest on record,
systematic trends in the residuals were still present, especially at both
ends of the fitted orbital arc.  This finding suggests that any gravitational
solutions may have been unsatisfactory long before the second half of
November of 2013.

Our objective is to understand effects of the comet's physical state on
its orbital motion.  All computations were carried out by the second author
with an {\it EXORB7\/} orbit-determination code, developed by A.\ Vitagliano.
The perturbations by the eight planets, Pluto, and the three most massive
asteroids~are~\mbox{included} and the standard DE406 library used.  Forced
values of orbital elements and nongravitational parameters can be employed.

First, we considered the ground-based astrometry only (Secs.~6.1--6.4).
After collecting 7770 positional observations of the comet from the MPC's
observations database (see the reference to MPC in footnote 1), we first
eliminated the 1099 that were listed twice, which left us with 6671 entries.
Next, in preliminary runs we discarded all observations that left residuals
greater than $\pm$2$^{\prime\prime}$ from ad hoc osculating solutions that
showed no trends in the residuals.  We eventually ended up with a total of
6177 acceptable ground-based observations between 2011 September~30 and 2013
November~22, spanning 784 days.  All observations were assigned the same
weight.

\begin{table}[t]
\vspace{-0.15cm}
\noindent
\begin{center}
{\footnotesize {\bf Table 9} \\[0.1cm]
{\sc Summary of Purely Gravitational Solutions (PG)\\Derived for Comet
C/2012 S1.}\\[0.1cm]
\begin{tabular}{@{\hspace{0.1cm}}l@{\hspace{0.3cm}}r@{\hspace{0.06cm}}c@{\hspace{0.06cm}}l@{\hspace{-0.04cm}}c@{\hspace{0.18cm}}c@{\hspace{0.3cm}}c@{\hspace{0.05cm}}}
\hline\hline\\[-0.22cm]
 & \multicolumn{3}{@{\hspace{-0.2cm}}c}{End date of used}
 & Number of    & Mean     & Systematic \\[-0.03cm]
Solution
 & \multicolumn{3}{@{\hspace{-0.2cm}}c}{observations$^{\rm a}$}
 & observations & residual & trends \\[0.08cm]
%
%
\hline \\[-0.2cm]
PG\,I  & 2013 & Nov.  &     20 & 6166 & $\pm$0$^{\prime\prime}\!\!$.68
       & enormous \\
PG\,II   &    &       &     10 & 5941 & $\pm$0.59 &  very strong \\
PG\,III  &    & Oct.  &     30 & 5619 & $\pm$0.56 &  strong \\
PG\,IV   &    &       &     15 & 5064 & $\pm$0.54 &  moderate \\
PG\,V    &    & Sept. &     30 & 4579 & $\pm$0.52 &  some in R.A. \\
PG\,VI   &    &       &     15 & 4238 & $\pm$0.50 &  some in R.A. \\
PG\,VII  &    & Aug.  &     31 & 3978 & $\pm$0.48 &  some in R.A. \\
PG\,VIII &    & June & $\;\:$8 & 3923 & $\pm$0.47 &  slight \\
PG\,IX   &    & Jan.  &     31 & 1857 & $\pm$0.43 &  nearly none \\[0.08cm]
\hline\\[-0.2cm]
\end{tabular}}
\parbox{8.32cm}{\scriptsize $^{\rm a}$\,Initial date was always 2011 September
30, the time of the earliest pre-discovery observation.{\vspace{-0.15cm}}}
\end{center}
\end{table}

\subsection{Purely Gravitational Solutions (PG)}
We began with purely gravitational solutions.  To examine the quality of fit
they achieved, we linked all selected astrometric observations, starting with
the earliest pre-discovery entries and ending with different dates in 2013 in
the order of decreasing time span covered.  The most important solutions among
the ones we ran are summarized in Table 9.

\begin{figure*}
\vspace{-1.65cm} 
\hspace{0.5cm}
\centerline{
\scalebox{0.57}{ 
\includegraphics{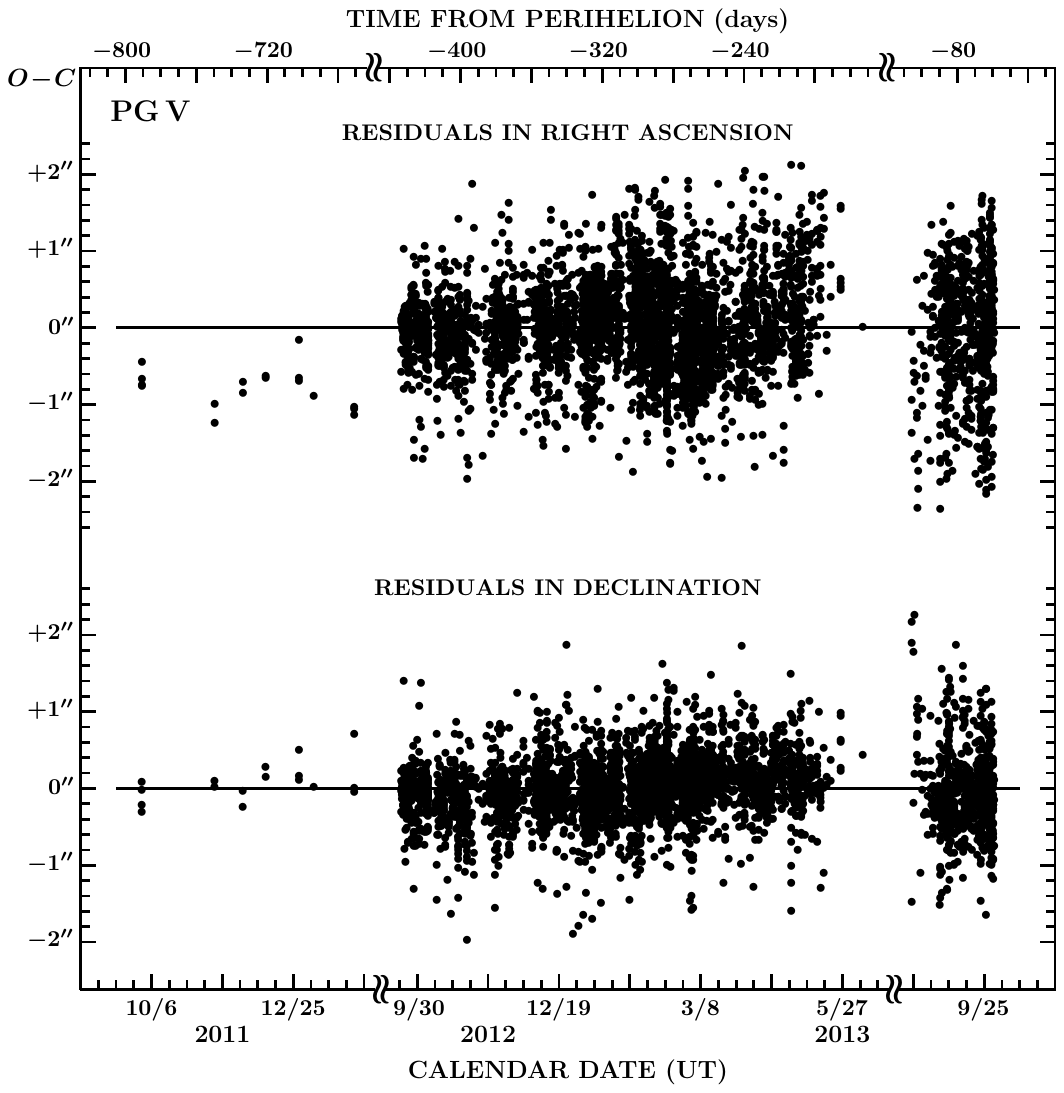}}} 
\vspace{-4.1cm}
\caption{Distribution of residuals for 4579 observations from the gravitational
solution PG\,V.  Clear trends are apparent in right ascension among the
pre-discovery observations and again near 100~days before
perihelion.{\vspace{0.4cm}}}
\end{figure*}

The first gravitational solution, PG\,I, matching the time interval of the
observations fitted by the MPC's last published gravitational solution
(2011 September 30--2013 November 20; Williams 2013a), left a distribution
of residuals presented in Figure 18.  Considerable trends are apparent
especially in right ascension in the period of time following the 2013
conjunction with the Sun, although the fit is poor at other times as well.

The last column of Table~9 indicates that as long as any post-conjunction
observations were included (the solutions up to PG\,VII), the fit was,
primarily in right ascension, never satisfactory.  As an example, the
residuals from the solution PG\,V are displayed in Figure~19.  The
residuals in R.A.\ from the pre-discovery, late pre-conjunction and early
post-conjunction observations are unacceptable.  The trend in the residuals
in R.A.\ from pre-discovery observations was not completely rectified even
by an improved solution PG\,VIII.  To get a nearly-perfect fit by a
gravitational solution, it was necessary to shorten the covered orbital
arc to only about 16~months, terminating it at the end of January 2013,
when the comet was some 4.9~AU from the Sun.  The orbital elements from
this solution, PG\,IX, are presented in Table~10 and the residuals are
plotted in Figure~20.  The residuals from the pre-discovery observations
now show only a marginal trend in declination and none in R.A.

\begin{table}[hb]
\noindent
\vspace{-0.15cm} 
\begin{center}
{\footnotesize {\bf Table 10} \\[0.08cm]
{\sc Orbital Elements of Comet C/2012 S1 for the Period of\\2011 September
30--2013 January 31 (Solution PG\,IX).}\\[0.1cm]
\begin{tabular}{l@{\hspace{0cm}}r@{\hspace{0.08cm}}c@{\hspace{0.08cm}}l}
\hline\hline\\[-0.2cm]
%
%
%
Epoch of osculation (TT)        & 2013$\;\;$ & \llap{N}o\rlap{v} & $\;$24.0 \\
Time of perihelion, $t_\pi$ (TT) & 2013 Nov 28.7843 & $\pm$ & 0.0018 \\
Argument of perihelion, $\omega$ & 345$^\circ\!$.5639 & $\pm$
                                                      & 0$^\circ\!$.0001 \\
Longitude of ascending node, $\Omega$ & 295$^\circ\!$.6547 & $\pm$
                                                      & 0$^\circ\!$.0002 \\
Orbital inclination, $i$         & 62$^\circ\!$.3887 & $\pm$
                                                     & 0$^\circ\!$.0012 \\
Perihelion distance, $q$ (AU)    & 0.01244418 & $\pm$ & 0.00000035 \\
Orbital eccentricity, $e$        & 1.00000158 & $\pm$ & 0.00000008 \\[0.08cm]
\hline \\[-0.65cm] 
\end{tabular}}
%
%
\end{center}
\end{table}

\begin{figure}
\vspace{-1.55cm} 
\hspace{2cm}
\centerline{
\scalebox{0.555}{
\includegraphics{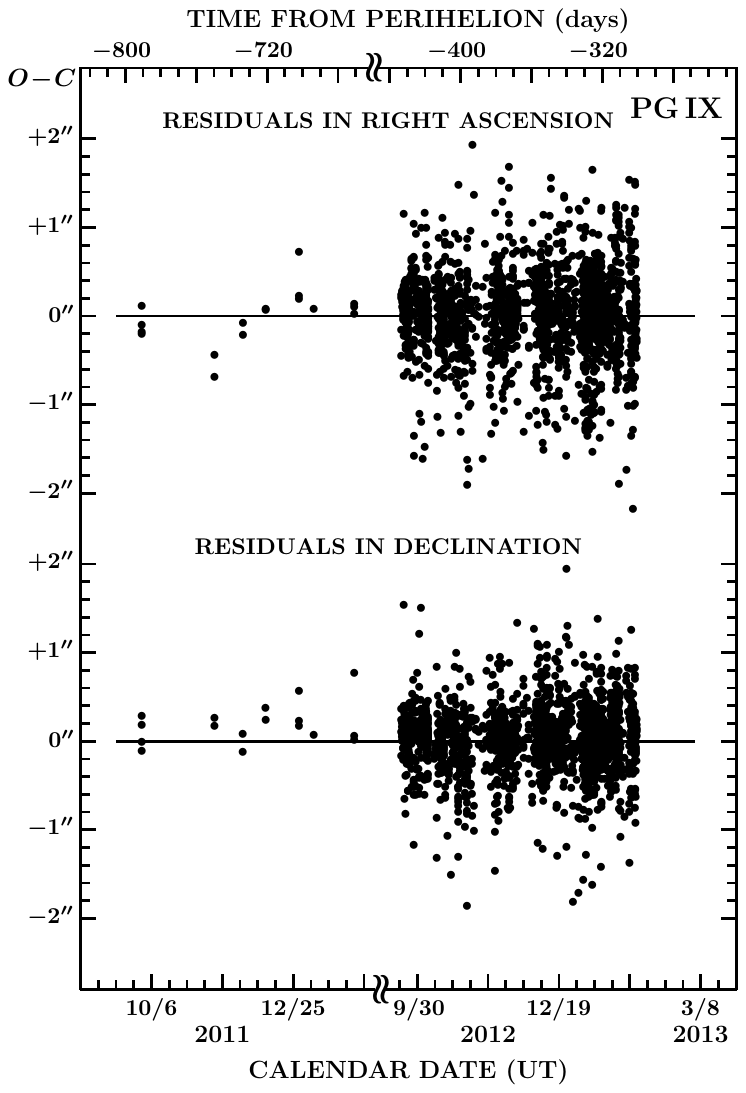}}} 
\vspace{-3.95cm}
\caption{Distribution of residuals for 1857 observations from the gravitational
solution PG\,IX.  Practically no trends are apparent in either coordinate.
This solution covers the longest observed orbital arc that can be fitted
without use of nongravitational terms in the equations of
motion.{\vspace{-0.1cm}}}
\end{figure}

\begin{table*}[t]
\vspace{-0.17cm}
\noindent
\begin{center}
{\footnotesize {\bf Table 11} \\[0.1cm]
{\sc Summary of Standard Nongravitational Solutions (SN) for Comet C/2012 S1
 and\\Parametric Values $A_1$, $A_2$, and $A_3$.}\\[0.1cm]
\begin{tabular}{l@{\hspace{0.2cm}}r@{\hspace{0.07cm}}c@{\hspace{0.07cm}}l@{\hspace{0.5cm}}c@{\hspace{0.5cm}}l@{\hspace{0.5cm}}c@{\hspace{0.4cm}}r@{\hspace{0.07cm}}c@{\hspace{0.07cm}}l@{\hspace{0.4cm}}r@{\hspace{0.06cm}}c@{\hspace{0.06cm}}l@{\hspace{0.4cm}}r@{\hspace{0.06cm}}c@{\hspace{0.06cm}}l}
\hline\hline\\[-0.2cm]
         & \multicolumn{3}{@{\hspace{-0.3cm}}c}{End date of} & Number  & &
 & \multicolumn{9}{@{\hspace{0cm}}c}{Parametric values (10$^{-8}$\,AU
 day$^{-2}$)} \\
         & \multicolumn{3}{@{\hspace{-0.3cm}}c}{observations} & of obser- &
 & Mean & \multicolumn{9}{@{\hspace{0cm}}c}{\rule[0.7ex]{5.9cm}{0.4pt}} \\
Solution & \multicolumn{3}{@{\hspace{-0.3cm}}c}{employed$^{\rm a}$} & vations
 & Option & residual & \multicolumn{3}{@{\hspace{0.25cm}}c}{$A_1$}
 & \multicolumn{3}{@{\hspace{0.1cm}}c}{$A_2$}
 & \multicolumn{3}{@{\hspace{0.35cm}}c}{$A_3$}\\[0.1cm]
%
%
\hline \\[-0.15cm]
SN\,I    & 2013 & Nov. & 22 & 6177 & SN\,I$_1$ & $\pm$0$^{\prime\prime}\!\!$.56
         & +8.69 & $\pm$ & 0.11 & & \lpts & & & \lpts & \\
         & & & & & SN\,I$_2$ & $\pm$0.56
         & +8.06 & $\pm$ & 0.12 & +0.60 & $\pm$ & 0.03 & & \lpts & \\
         & & & & & SN\,I$_3$ & $\pm$0.55
         & +10.82 & $\pm$ & 0.34 & $-$0.54 & $\pm$ & 0.14 & $-$1.89 & $\pm$
         & 0.22 \\[0.08cm]
SN\,II   & 2013 & Nov. & 10 & 5941 & SN\,II$_1$ & $\pm$0.55
         & +10.03 & $\pm$ & 0.24 & & \lpts & & & \lpts & \\
         & & & & & SN\,II$_2$ & $\pm$0.55
         & +9.41 & $\pm$ & 0.24 & +0.82 & $\pm$ & 0.06 & & \lpts & \\
         & & & & & SN\,II$_3$ & $\pm$0.55
         & +9.24 & $\pm$ & 0.39 & +0.97 & $\pm$ & 0.27
         & +0.26 & $\pm$ & 0.46 \\[0.08cm]
SN\,III  & 2013 & Oct. & 30 & 5619 & SN\,III$_1$
         & $\pm$0.55 & +10.25 & $\pm$ & 0.49 & & \lpts & & & \lpts & \\
         & & & & & SN\,III$_2$ & $\pm$0.54
         & +9.54 & $\pm$ & 0.49 & +1.63 & $\pm$ & 0.11 & & \lpts & \\
         & & & & & SN\,III$_3$ & $\pm$0.54
         & +10.01 & $\pm$ & 0.50 & +5.02 & $\pm$ & 0.62
         & +5.9 & $\pm$ & 1.1 \\[0.08cm]
SN\,IV   & 2013 & Oct. & 15 & 5064 & SN\,IV$_1$ & $\pm$0.54
         & +4.4 & $\pm$ & 1.4 & & \lpts & & & \lpts & \\
         & & & & & SN\,IV$_2$ & $\pm$0.53
         & +4.5 & $\pm$ & 1.4 & +4.66 & $\pm$ & 0.36 & & \lpts & \\
         & & & & & SN\,IV$_3$ & $\pm$0.53
         & +11.4 & $\pm$ & 3.9
         & +8.5 & $\pm$ & 2.1
         & +6.7 & $\pm$ & 3.5$\;\:$ \\[0.08cm]
SN\,V    & 2013 & Sept. & 30 & 4579 & SN\,V$_1$ & $\pm$0.52
         & $-$10.0 & $\pm$ & 3.5 & & \lpts & & & \lpts & \\
         & & & & & SN\,V$_2$ & $\pm$0.52
         & $-$6.8 & $\pm$ & 3.5 & +8.94 & $\pm$ & 0.96 & & \lpts & \\
         & & & & & SN\,V$_3$ & $\pm$0.51
         & $-$149 & $\pm$ & 22 & $-$37.8 & $\pm$ & 7.0
         & $-$78 & $\pm$ & 12 \\[0.08cm]
\hline\\[-0.15cm]
\end{tabular}}
\parbox{15cm}{\hspace{0.55cm}\scriptsize $^{\rm a}$\,Initial date was always
 2011 September 30, the time of the earliest pre-discovery
 observation.}\\[0.55cm]
\end{center}
\end{table*}

The comet's original barycentric reciprocal semimajor axis, $(1/a_{\rm b})_{\rm
orig}$, derived {\vspace{-0.08cm}}from this set of orbital elements, is equal
to \mbox{$+0.000\,035 \pm 0.000\,006\;{\rm AU}^{-1}$}, within 2$\sigma$ of
an average perihelion-distance corrected value of \mbox{$+0.000\,046\;{\rm
AU}^{-1}$} (Marsden et al.\,1978), confirming that there is no doubt about
C/2012~S1 being a dynamically new comet, arriving from the Oort Cloud.

In summary, purely gravitational solutions are generally found to fail
fitting the orbital motion of comet C/2012~S1 except at heliocentric
distances larger than $\sim$5~AU.  We conclude that it is inappropriate
to ignore the outgassing-driven nongravitational effects at smaller
heliocentric distances and that their neglect at distances below 1.4~AU
leads to strong systematic trends in the distribution of residuals.

\subsection{Standard Nongravitational Solutions (SN)}
To satisfy the need for incorporating nongravitational terms into the equations
of motion, we turned to the standard `Style II' formalism of Marsden et al.\
(1973).  The applied dimensionless law $g_{\rm ice}(r)$ is based on a premise
that a comet's nongravitational acceleration is driven by momentum transfer
from the outgassing of water ice.  This nongravitational law, a function of
heliocentric distance $r$ only and therefore symmetrical with respect to
perihelion, is in Marsden et al.'s formalism expressed by an empirical
formula,
\begin{equation}
g_{\rm ice}(r) = \alpha \left( \! \frac{r}{r_0} \right)^{\! -m} \! \left[
 1 \!+\!  \left( \! \frac{r}{r_0} \! \right)^{\!n} \right]^{\! -k} \!\!,
\end{equation}
where \mbox{$r_0 = 2.808$} AU is a scaling heliocentric distance, at which
the fraction of the solar energy spent in water-ice sublimation is about
0.023 times the fraction spent in reradiation of the surface.  The values of
the exponents are \mbox{$m = 2.15$}, \mbox{$n = 5.093$}, \mbox{$nk = 23.5$},
and \mbox{$\alpha = 0.1113$} is a normalization coefficient that forces
\mbox{$g_{\rm ice}(1 \, {\rm AU}) = 1$}.  For a low Bond albedo and unit
emissivity, these constants apply to a so-called isothermal model, which
averages the Sun's incident radiation over the surface of a spherical
nucleus by assuming that the temperature does not vary from place to place.
In reality, of course, the temperature does vary over the surface, but
for the orbit-determination purposes the law from Eq.\,(61) has over the
four decades since its inception in 1973 provided excellent service and
still is employed worldwide nowadays.

The magnitude of the nongravitational acceleration at 1~AU from the Sun is
given by the components of a right-handed coordinate system in three cardinal
directions tied to the orbital plane:\ a radial component $A_1$, pointing in
the antisolar direction; a transverse component $A_2$; and a normal component
$A_3$.  They are expressed in units of 10$^{-8}$\,AU~day$^{-2}$, equivalent to
\mbox{$2.004 \times 10^{-5}$}\,cm~s$^{-2}$.  The magnitude of the acceleration
at a distance $r$ from the Sun is therefore given as \mbox{$\sqrt{A_1^2 \!+\!
A_2^2 \!+\!  A_3^2} \; g_{\rm ice}(r)$}.  However, $A_3$ is seldom determined,
as part of an orbital solution, with satisfactory accuracy, while for
single-apparition comets a meaningful value of $A_1$, if sufficiently large
for reliable detection, is always positive (indicating that the acceleration
points away from the Sun)\footnote{It is not always appreciated that the
parameter $A_1$ of this symmetrical nongravitational law has a very different
meaning in orbital solutions that link successive returns of short-period
comets (Sekanina 1993).} and, typically, it exceeds $A_2$ in absolute value,
often by one order of magnitude (Marsden \& Williams 2008).  As a result, $A_1$
is the prime nongravitational parameter to solve for.  Next, it is customary
to solve for $A_1$ and $A_2$, and only quite rarely for $A_1$, $A_2$, and
$A_3$, which usually yields a meaningless $A_3$ with an error exceeding the
parameter's nominal value.  These three options are in the following marked
with subscripts 1, 2, and 3, respectively.

\begin{figure*}
\vspace{-1.65cm} 
\hspace{0cm}
\centerline{
\scalebox{0.57}{ 
\includegraphics{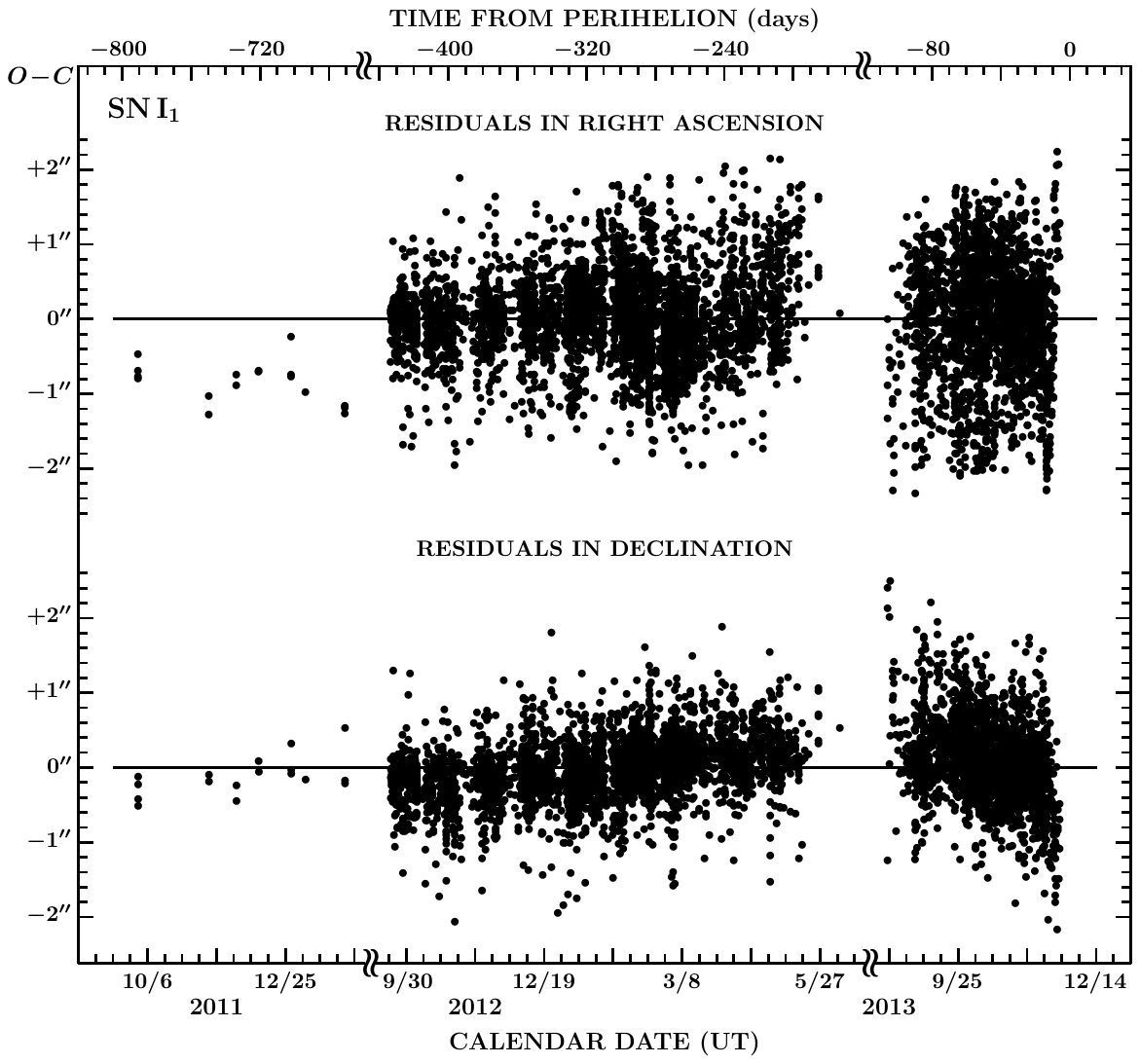}}} 
\vspace{-4.1cm}
\caption{Distribution of residuals for 6177 observations between 2011
September 30 and 2013 November 22 from the standard nongravitational solution
SN\,I$_1$.  Although the quality of fit is better than by an equivalent
gravitational solution, fairly strong systematic trends in both right
ascension and declination still~persist.{\vspace{0.43cm}}}
\end{figure*}
Some of the standard nongravitational (SN) solutions that we ran are listed
in Table~11.  We began testing these solutions by comparing them with the
MPC's orbits.  The first option of our first solution, SN\,I$_1$ in
Table~11, using all accepted observations from 2011 September 30 through
2013 November 22, compares favorably with the MPC's first two published
nongravitational sets of elements (Williams 2013b, 2013c), based,
respectively, on 6120 and 6138 observations from the same period of time
and resulting in \mbox{$A_1 = +8.93 \pm 0.12$} and \mbox{$A_1 = +9.13$} in
units of \mbox{$10^{-8} \, {\rm AU\;day}^{-2}$}.  The second option of the
solution, SN\,I$_2$, similarly compares commendably with the MPC's most
recently published nongravitational orbit (Williams 2014), based on 6217
observations and giving \mbox{$A_1 = +7.84$} and \mbox{$A_2 = +0.64$} in
the same units.  An agreement with the orbital results by Nakano (2013c),
who derived \mbox{$A_1 = +6.09 \pm 0.24$} and \mbox{$A_2 = +0.66 \pm 0.07$}
in the same units, is somewhat less satisfactory in the radial component.

The distribution of residuals from the SN\,I$_1$ solution is plotted in
Figure~21.  Similar distributions resulted from the SN\,I$_2$ and
SN\,I$_3$ runs listed in Table~11.  The quality of fit in Figure~21 is
better than from the equivalent gravitational solutions, but not free
from trends in either coordinate.  We found the match between the standard
nongravitational model and observations disappointing for two reasons:\
(i)~solving in addition to $A_1$ also for $A_2$ and $A_3$ offered virtually
no improvements; and (ii)~comparison of Table~11 with Table~9 suggests that,
except for SN\,I and SN\,II, the standard nongravitational solutions failed
to reduce the mean residual over that given by the equivalent gravitational
solutions.   Playing only minor roles, the transverse and normal components
could be neglected.  And, finally, all SN solutions with the end date before
November 2013 led to poorly defined and/or meaningless values of the
parameters.

\subsection{Modified Nongravitational Solutions (MN)}
The failure of the standard nongravitational law $g_{\rm ice}(r)$ made us
search for improvements.  The most critical parameter in Eq.\,(61) is the
scaling distance $r_0$.  After introducing the law, Marsden et al.\ (1973)
investigated the physical meaning of $r_0$ and found out that on the
assumptions of an isothermal model's constant Bond albedo and emissivity
and constant values of the exponents $m$, $n$, and $k$, the distance $r_0$
measured essentially the heat of sublimation $L$ of the volatile substance
that dominates momentum transfer to the nucleus, varying to a first
approximation inversely as its square,
\begin{equation}
r_0 \simeq \left( \frac{\rm const}{L} \right)^{\!2} \! .
\end{equation}
With the sublimation heat of 11\,400~cal~mol$^{-1}$ for water ice, the constant
equals 19\,100~AU$^{\frac{1}{2}}$~cal~mol$^{-1}$ for the isothermal model
under consideration here.

It should be noted that the only basis for introducing this generic form of
what we call a modified nongravitational law is, as pointed out by Marsden
et al.\ (1973), a similarity in the shapes of normalized sublimation curves
for a variety of species except for major horizontal shifts in a plot of
log\,(sublimation rate) against log\,$(r/r_0)$ that generally fit Eq.\,(62).
In other words, a normalized sublimation rate is incomparably less sensitive
to the values of the exponents $m$ (which always slightly exceeds 2), $n$,
and $k$, than to the scaling distance $r_0$.  This approach is a useful tool
to examine momentum-transfer effects in the orbital motion due to species of
unknown identity.
%

%
\begin{table*}[t]
\vspace{-0.17cm}
\noindent
\begin{center}
{\footnotesize {\bf Table 12} \\[0.1cm]
{\sc Summary of Modified Nongravitational Solutions (MN) and Solution
 Accounting for Nucleus' Dwindling Dimensions (DD)\\for Comet C/2012 S1,
 Constants $r_0$, {\rend}, and $\xi$, and Parametric Values $A_1$, $A_2$,
 and $A_3$.}\\[0.1cm]
\begin{tabular}{l@{\hspace{0.11cm}}c@{\hspace{0.17cm}}c@{\hspace{0.2cm}}l@{\hspace{0.2cm}}c@{\hspace{0.3cm}}c@{\hspace{0.3cm}}c@{\hspace{0.3cm}}r@{\hspace{0.06cm}}c@{\hspace{0.06cm}}l@{\hspace{0.2cm}}r@{\hspace{0.06cm}}c@{\hspace{0.06cm}}l@{\hspace{0.2cm}}r@{\hspace{0.06cm}}c@{\hspace{0.06cm}}l}
\hline\hline\\[-0.2cm]
 & & Number & & & Distance & Mass
 & \multicolumn{9}{@{\hspace{0cm}}c}{Parametric values (10$^{-8}$\,AU
   day$^{-2}$)} \\
 & Start and end dates of & of obser- & & Mean
 & \raisebox{0.06cm}{$\,r_{_{\scriptstyle 0}}$ or {\rend}} & erosion rate
 & \multicolumn{9}{@{\hspace{0cm}}c}{\rule[0.7ex]{5.5cm}{0.4pt}} \\
Solution & observations used & vations & Option & residual & (AU)
 & exponent $\xi$ & \multicolumn{3}{@{\hspace{0.25cm}}c}{$A_1$}
 & \multicolumn{3}{@{\hspace{0.1cm}}c}{$A_2$}
 & \multicolumn{3}{@{\hspace{0.35cm}}c}{$A_3$}\\[0.1cm]
%
%
\hline \\[-0.2cm]
MN\,I    & 2011\,Sept.\,30--2013\,Nov.\,22 & 6177 & MN\,I$_1$
         & $\pm$0$^{\prime\prime}\!\!$.56 & $3.11 \pm 0.16$ & \lpts
         & +8.67 & $\pm$ & 0.11 & & \lpts & & & \lpts & \\
         & & & MN\,I$_2$ & $\pm$0.56
         & $3.50 \pm 0.18$ & \lpts & +7.87 & $\pm$ & 0.12 & +0.53 & $\pm$
         & 0.03 & & \lpts & \\
         & & & MN\,I$_3$ & $\pm$0.55
         & $2.27 \pm 0.13$ & \lpts & +12.81 & $\pm$ & 0.28 & $-$2.15 & $\pm$
         & 0.16 & $-$4.81 & $\pm$ & 0.27 \\[0.08cm]
MN\,II   & 2013\,Feb.\,1--2013\,Oct.\,30 & 3762 & MN\,II$_1$ & $\pm$0.58
         & $1.93 \pm 0.12$ & \lpts & +21.23 & $\pm$ & 1.13 & & \lpts & &
         & \lpts & \\
         & & & MN\,II$_2$ & $\pm$0.58 & $2.10 \pm 0.15$ & \lpts
         & +16.86 & $\pm$ & 0.89 & +2.04 & $\pm$ & 0.30 & & \lpts & \\[0.08cm]
DD\,I    & 2011\,Sept.\,30--2013\,Nov.\,22 & 6177 & DD\,I$_1$
         & $\pm$0.57 & 0.024; fixed & $2.68 \pm 0.04$ & +7.82 & $\pm$ & 0.11
         & & \lpts & & & \lpts & \\
         & & & DD\,I$_2$ & $\pm$0.56 & 0.024; fixed
         & $2.60 \pm 0.04$ & +7.02 & $\pm$ & 0.11 & +0.54 & $\pm$ & 0.02
         & & \lpts & \\
         & & & DD\,I$_3$ & $\pm$0.56 & 0.024; fixed
         & $2.35 \pm 0.06$ & +3.46 & $\pm$ & 0.27 & +1.25 & $\pm$ & 0.06
         & +1.31 & $\pm$ & 0.09 \\[0.08cm]
\hline\\[-0.1cm]
\end{tabular}}
%
%
\end{center}
\end{table*}

Thus, to test the use of modified nongravitational solutions, we retained the
values of $m$, $n$, and $k$ for water ice from Eq.\,(61), but searched for
the best fit by varying $r_0$ in a law $g(r; r_0)$, given by the same formal
expression as the law $g_{\rm ice}(r)$ and normalized to \mbox{$g(1\;{\rm AU};
r_0) = 1$}.

\begin{figure}[b]
\vspace{-1.5cm} 
\hspace{-2.4cm}
\centerline{
\scalebox{0.595}{
\includegraphics{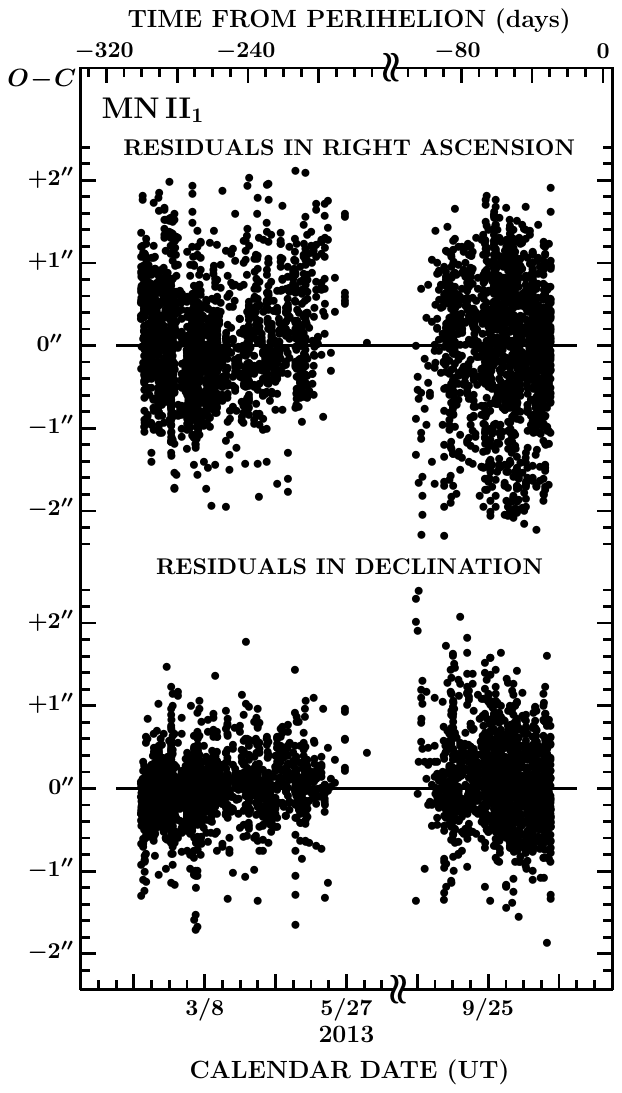}}} 
\vspace{-3.97cm}
\caption{Distribution of residuals for 3762 observations between 2013
February 1 and October 30 from the modified nongravitational solution
MN\,II$_1$. No obvious trends are apparent in either coordinate.  This
solution yields the most representative set of orbital elements (Table 13)
that we are able to offer for the 9-months long period of
time.{\vspace{-0.05cm}}}
\end{figure}

Table 12 lists two examples of the modified solutions.  The first, MN\,I,
covered the entire period of ground-based observations and confirmed that the
value of the scaling distance used in the standard nongravitational law was
essentially valid over such long periods of time.  A weighted mean of $r_0$
from the three options of the MN\,I solution equals 2.82~AU, in a remarkably
good agreement with \mbox{$r_0 = 2.808$ AU} used in the standard law $g_{\rm
ice}(r)$.  The distributions of residuals were practically identical to
those of the equivalent SN solutions and are not displayed.

The second example of the modified solutions, MN\,II, covered only a limited
period of time.  In Sec.~6.1 we found that the gravitational solution PG\,IX
offered a very satisfactory match to all observations from the period of
time ending 2013 January~31.  The period of time covered by the solution
MN\,II started the next day and extended to 2013 October 30, thus avoiding
the stormy period of activity (Sec.~2.1).  The scaling distances $r_0$,
derived as part of the two listed options, MN\,II$_1$ and MN\,II$_2$, agree
with each other within the errors and are only moderately smaller than
2.8~AU.  Both options provide nearly identical distributions of residuals
that show no obvious trends in either coordinate.  The residuals from the
solution MN\,II$_1$ are in Figure~22 and the respective orbital elements,
in Table~13, are judged to be the most representative for the given period
of time that we are able to offer.  We also tried to calculate an option
MN\,II$_3$, but no clearly defined minimum on the curve of the sum of squares
of residuals was found at \mbox{$r_0 < 5$ AU} and for all tested values of
$r_0$ greater than 3.4~AU the radial component $A_1$ came out negative,
suggesting that given the overall degree of accuracy, an excessive number
of parameters was attempted to be solved for in this option.

\begin{table}[b]
\noindent
\vspace{0.1cm}
\begin{center}
{\footnotesize {\bf Table 13} \\[0.08cm]
{\sc Orbital Elements of Comet C/2012 S1 for the Period of\\2013 February
 1--October 30 (Solution MN\,II$_1$).}\\[0.05cm]
\begin{tabular}{@{\hspace{0.1cm}}l@{\hspace{0.35cm}}r@{\hspace{0.05cm}}c@{\hspace{0.05cm}}l@{\hspace{0.1cm}}}
\hline\hline\\[-0.2cm]
%
%
%
Epoch of osculation (TT)        & 2013$\;\;$ & \llap{N}o\rlap{v} & $\;$24.0 \\
Time of perihelion, $t_\pi$ (TT) & 2013 Nov 28.78122 & $\pm$ & 0.00007 \\
Argument of perihelion, $\omega$ & 345$^\circ\!$.57506 & $\pm$
                                                      & 0$^\circ\!$.00007 \\
Longitude of ascending node,\,\rlap{$\Omega$} & 295$^\circ\!$.65252 & $\pm$
                                                      & 0$^\circ\!$.00006 \\
Orbital inclination, $i$         & 62$^\circ\!$.40108 & $\pm$
                                                     & 0$^\circ\!$.00036 \\
Perihelion distance, $q$ (AU)    & 0.01244488 & $\pm$ & 0.00000013 \\
Orbital eccentricity, $e$        & 0.99994358 & $\pm$ & 0.00000003 \\[0.05cm]
Nongravitational parameters \rlap{of law $g(r;r_0)$:} &  &  & \\
\hspace{0.3cm}Scaling distance $r_0$ (AU)      & 1.93 & $\pm$ & 0.12 \\
\hspace{0.26cm}Radial component\,$A_1$\,(\rlap{10$^{-8}$\,AU\,day$^{-2}$)}
 & 21.23 & $\pm$ & 1.13 \\[0.05cm]
\hline \\[-0.58cm]
\end{tabular}}
\end{center}
\end{table}

Even though the modified nongravitational law \mbox{$g(r;r_0)$} proves
a useful tool in instances when the orbital motion is clearly inconsistent
with the standard law, $g_{\rm ice}(r)$, its use is essentially a
last-resort-type of a solution.  The most beneficial byproduct of its
successful application is presented by Eq.\,(62):\ one learns whether
the nongravitational forces affecting the comet's motion were due
primarily to sublimation of a species more volatile or more refractory
than water ice, if $r_0$ deviates substantially from 2.8~AU.

\subsection{Nongravitational Solutions Accounting for\\Nucleus' Dwindling
Dimensions (DD)}
The nongravitational laws under consideration have so far failed
to account for the nucleus' gradual erosion, whose existence has in the
previous sections been demonstrated beyond any doubt.  Effects of cascading
fragmentation on the comet's orbital motion could likewise be approximately
mimicked by numerically simulating the dwindling dimensions of the nucleus
modeled as a single body.  To incorporate an eroding nucleus into the
momentum-transfer considerations, we begin with a basic conservation of
momentum equation.  If at time $t$ the mass of the nucleus is ${\cal M}(t)$,
its mass loss rate by erosion is $\dot{\cal M}(t)$, the outflow velocity of
the eroded mass is $\nu(t)$, the erosion-driven acceleration of the nucleus
is $\eta(t)$, and an average momentum-transfer efficiency, depending
primarily on the erosion rate distribution over the nuclear surface, is
\mbox{$\Psi < 1$}, the differentiated conservation of momentum equation is
\begin{equation}
\dot{\cal M}(t) \, \nu(t) + \Psi {\cal M}(t) \, \eta(t) = 0.
\end{equation}
If at time $t$ a spherical nucleus of radius ${\cal R}(t)$ is shrinking at
a rate of $\dot{\cal R}(t)$, the acceleration equals
\begin{equation}
\eta(t) = -\frac{3 \Psi}{2} \, \frac{\dot{\cal R}(t)}{{\cal R}(t)} \, \nu(t)
 = -\frac{3 \Psi}{2 \rho} \, \frac{\dot{\cal E}(t)}{{\cal R}(t)} \, \nu(t),
\end{equation}
where the minus sign, indicating that the direction of the dynamical impulse
on the nucleus is opposite to that of the outflowing mass, is in the
following considerations inconsequential.  The rate of dwindling nuclear
size, $\dot{\cal R}(t)$, is expressed as a function of the effective mass
erosion (or sublimation) rate $\dot{\cal E}(t)$ per unit surface area and
the bulk density $\rho$.  The relation between the nuclear radius
${\cal R}(t)$ at time $t$ and its radius ${\cal R}(t_0)$ at a reference
time $t_0$, is given by
\begin{equation}
{\cal R}(t) = {\cal R}(t_0) - \frac{1}{\rho} \int_{t_0}^{t} \dot{\cal E}(t)
 \, dt,
\end{equation}
where the rate $\dot{\cal E} > 0$.

It is more convenient to integrate over heliocentric distance $r$ than over
time.  In that case it is necessary to distinguish whether or not the
integration is carried out through perihelion, at time $t_\pi$.  If it
is, one needs to integrate in two parts, from $t_0$ to $t_\pi$ and from
$t_\pi$ to $t$.  Because C/2012~S1 disintegrated before perihelion, we
contemplate from now on only that option.  In order to derive an expression
for the momentum-transfer law that is consistent with the formalism of the
employed orbit determination code, we introduce two approximations.

The first approximation has to do with the conversion of the integration
variable from time to heliocentric distance and is of the same nature as
that used in Sec.~5.2 to derive Eq.\,(55).  In a parabolic case, an
increment $dt$ is before perihelion related to an increment $dr$ by
\begin{equation}
dt = -c_0 \, r^{\frac{1}{2}} \! \left( 1 \!-\! \frac{q}{r}
 \right)^{\!-\frac{1}{2}} \! dr \approx -c_0 \, r^{\frac{1}{2}} dr,
\end{equation}
where $c_0$ is a constant and $q$ is the perihelion distance. It is noted
that this approximation is inadmissible when the integration is carried out
through perihelion.

The other approximation concerns the erosion rate, for which --- based on
our experience with $g_{\rm ice}(r)$ --- we assume a power law of
heliocentric distance,
\begin{equation}
\dot{\cal E}(r) = h_0 \, r^{-\xi},
\end{equation}
where \mbox{$h_0 > 0$} and \mbox{$\xi > 2$} are constants.  This expression
allows one to iteratively adjust $\xi$ to the observations and thus to
learn from the degree of steepness about the nature of a momentum-transfer
effect.

Inserting from Eqs.\,(66) and (67) into Eq.\,(65), we obtain, after
identifying $t_0$ with the time of complete disintegration, {\tend}, when
\mbox{${\cal R}\mbox{(\tend)} = 0$} and $r = \mbox{\rend}$,
\begin{equation}
{\cal R}(t) = \frac{c_0 h_0}{\rho} \!\! \int_{\mbox{\rrend}}^{r} \!\!
 r^{\frac{1}{2} - \xi} \, dr = \frac{c_0 h_0 \mbox{{\rend}}^{\!\!\frac{3}{2}
 - \xi}}{\rho \, (\xi \!-\! \frac{3}{2})} \! \left[ 1 \!-\! \left( \!
 \frac{r}{\,\mbox{\rend}} \! \right)^{\!\! \frac{3}{2} - \xi} \right] \! .
\end{equation}
Inserting Eqs.\,(67) and (68) into Eq.\,(64) and approximating the outflow
velocity with a constant, \mbox{$\nu(t) = \nu_0$}, the expression for the
acceleration $\eta(r)$ is finally
\begin{equation}
\eta(r) = \frac{3 \Psi \nu_0 (\xi \!-\! \frac{3}{2}) \,
 {\mbox{\rend}}^{\!\!-\frac{3}{2}}}{2 c_0 a_0} \, {\cal G}(r;\mbox{\rend},\xi),
\end{equation}
where
\begin{equation}
{\cal G}(r;\mbox{\rend},\xi) = a_0 \, \frac{\left( \! {\displaystyle
 {\frac{r}{\,\mbox{\rend}}}} \! \right)^{\!\! -\xi}}{1 \!-\! \left( \!
 {\displaystyle \frac{r}{\,\mbox{\rend}}} \! \right)^{\!\! \frac{3}{2} -
 \xi}}
\end{equation}
and $a_0$ is a normalization constant,
\begin{equation}
a_0 = \mbox{\rend}^{\!\!-\xi} \!-\! \mbox{\rend}^{\!\!-\frac{3}{2}}.
\end{equation}
The function for ${\cal G}(r;\mbox{\rend},\xi)$ in Eq.\,(70) formally matches
the general expression for the momentum-transfer law $G(r;r_0,b_0)$ in an
upgraded {\it EXORB\/} orbit determination code,\footnote{At our request,
A.\ Vitagliano kindly implemented the necessary modifications to his
orbit determination code.} which reads
\begin{equation}
G(r;r_0,b_0) = \alpha_0 \! \left( \! \frac{r}{\,r_0} \! \right)^{\!\! -m}
 \!\!  \left[1 \!+\! b_0 \! \left( \! \frac{r}{\,r_0} \! \right)^{\!\! n}
 \right]^{\! -k} \!\! ,
\end{equation}
when $r _0 = \mbox{\rend}$, $b_0 = -1$, $m = \xi$, $n = \frac{3}{2} - \xi$,
$k = 1$, and $\alpha_0 = a_0$.  Whereas $r_0$ in $g(r;r_0)$ measures the
degree of volatility of a species [note that \mbox{$G(r;r_0,+1) \equiv
g(r;r_0)$}], the value of {\rend} in ${\cal G}(r;\mbox{\rend},\xi)$ determines
the point in the orbit where the disintegrating object ultimately perishes.
This is a major difference that makes the ${\cal G}(r;\mbox{\rend},\xi)$
law diverge to infinity at \mbox{$r = \mbox{\rend}$} and undefined at
\mbox{$r \le \mbox{\rend}$}, that is, at \mbox{$t \ge \mbox{\tend}$}.  In
practice, of course, no comet perishes at any particular point, but along a
finite arc of the orbit.  Accordingly, {\rend} should be perceived as a
{\nopagebreak}dynamical parameter that describes a general location in the
{\nopagebreak}orbit where the comet's motion ceased to follow any consistent
pattern.

It should be remarked that a more complex function for $\dot{\cal E}$, such
as $g(r;r_0)$, would require another modification of the orbit determination
code.  In principle, however, there is no limit to experimentation along these
lines and a host of functions can in the future be tested in an effort to
further refine the method of accounting for the nongravitational effects in
the orbital motions of comets.

In order to understand{\vspace{-0.05cm}} the meaning of $\xi$ from Eq.\,(67)
in terms of a relationship between $\dot{\cal E}(r)$ and $G(r;r_0,+1)$ within
a limited range of heliocentric distances, we examine this exponent, defined
in that equation for any heliocentric distance by
\begin{equation}
\xi = -\frac{\partial}{\partial \ln r} \, {\ln \dot{\cal E}(r)},
\end{equation}
as a function of the parameters of the $G(r;r_0,+1)$ law by requiring that
at heliocentric distances near $r^\prime$ it also satisfies a
condition
\begin{equation}
\xi(r^\prime) \simeq -\frac{\partial}{\partial \ln r}\,\ln G(r^\prime;r_0,+1).
\end{equation}
From Eq.\,(72) we find that
\begin{equation}
\xi(r^\prime) \simeq \! \left\{ \begin{array}{l@{\hspace{0.5cm}}l}
 \! m & \mbox{if $r^\prime \ll r_0$,} \\
 \! m \!+\! {\textstyle \frac{1}{2}}nk & \mbox{if $r^\prime = r_0$,} \\
 \! m \!+\! nk & \mbox{if $r^\prime \gg r_0$,} \\[-0.05cm]
 \! m \!+\! {\displaystyle \frac{nk}{1\!+\!(r^\prime/r_0)^{-n}}}
 & \!\!\:\mbox{elsewhere.}
 \end{array}  \right.
\end{equation}
With the values of the exponents for $g_{\rm ice}(r)$ from Eq.\,(61),
we find that \mbox{$\lim_{r \rightarrow 0} \xi(r) \approx 2$},
\mbox{$\xi(\frac{2}{3} r_0) \approx 5$}, and \mbox{$\xi(r_0) \approx
14$}, which are useful specific values to keep in mind.

We applied the law from Eq.\,(70) to the whole set of ground-based observations
to derive the solution DD\,I, listed in Table 12.  The computations followed
an approach, in which we first chose an arbitrary value of {\rend} and a
sequence of arbitrary values of $\xi$.  We optimized $\xi$ for the given
{\rend} by searching for the best fit with a minimum sum of squares of
residuals, \mbox{$\Sigma({\rm o}\!-\!{\rm c})^2 = {\rm min}$}.  We repeated
this procedure for a sequence of chosen values of {\rend} and searched for
an optimized pair of values {\rend} and $\xi$ using again \mbox{$\Sigma({\rm
o}\!-\!  {\rm c})^2$}.  We found that the fit continued to improve with
decreasing {\rend} down to 0.01~AU, the least value employed in the
computations.  The quality of fit had a tendency to level off at
\mbox{$\mbox{\rend} \,\lapeq\, 0.05$}~AU.  For example, in the option
DD\,I$_1$ (Table 12), the sum of squares of residuals from an optimized
solution for \mbox{$\mbox{\rend} = 0.01$}~AU amounted to 98.3 percent of
the sum for \mbox{$\mbox{\rend} = 0.3$}~AU, but fully 99.92~percent of the
sum for \mbox{$\mbox{\rend} = 0.05$}~AU.  A formal parabolic fit through
a number of $\Sigma({\rm o}\!-\!{\rm c})^2$ points showed that its minimum
was reached at a slightly negative (and therefore meaningless) {\rend},
but with a standard deviation exceeding its value.  Fixing {\rend} at
0.024~AU, a heliocentric distance at which all activity ceased 3.5~hours
before perihelion (Sec.~3.9), is at 1.7~times the standard deviation, well
within the uncertainties involved.

Table 12 shows that in terms of the mean residual the DD\,I solution is
competitive with MN\,I and also with SN\,I from Table~11.  The distributions
of residuals (not displayed) are also very similar.  It appears that the
proposed nongravitational law that accounts for the nucleus' dwindling
dimensions does not enjoy any advantage {\nopagebreak}over the other
nongravitational laws in applications to observations taken long before the
effect of dwindling dimensions could dominate.  The role of this law in
applications to observations from times that are close to the disintegration
of the nucleus is investigated in Sec.~6.6.
 
\subsection{Astrometric Positions from STEREO-B}
In an effort to extend the comet's astrometric observa\-tions beyond the
period of observability from the ground, which ended on 2013 November 22, we
searched for adequate coronagraphic observations with instruments on board
SOHO, STEREO-A, and STEREO-B.  Because the comet did not enter the field of
view of the C2 coronagraph until November~28.5~UT and the field of view of
the COR-2A coronagraph until November~28.2~UT, by far the best prospects
for acquiring acceptable spaceborne astrometry from the period after
November~22 were offered by the COR-2 coronagraph on board STEREO-B; the
comet entered this instrument's field of view as early as November 26.2~UT
and the image taken at 6:25~UT was the first with the center of the head
separated from the field's edge enough that it could be astrometrically
determined; the second author measured all positions of the comet until
15:00 UT on November~28.  The CCD images from November 26--27 were 2048 by
2048 pixels (equivalent to a pixel size of $\sim$15 arcsec or approximately
12\,000~km at the comet), except between 9:56 and 14:24~UT on the 26th,
when all were 1024 by 1024 pixels (with a pixel size twice as large); none
of these images was measured.  All images on November~28 were 1024 by
1024~pixels and, starting at about 10:54 UT, they showed an elongated
nuclear region, making the measurement increasingly uncertain.  Table~14
lists all derived astrometric positions until 12:00~UT November~28, when
the comet was a little over 8~{\Rsun} from the Sun.

The acquisition and reduction of the comet's images were accomplished with
the use of the {\it Astrometrica\/} software package.\footnote{See the
website {\tt http://www.astrometrica.at.}}  An aperture radius of 2~pixels,
the smallest that {\it Astrometrica\/} allows, was employed.  The brightest
pixel in the comet's head was located first and then the centroid was shifted
to match this position as close as possible, usually within 0.3~pixel.  The
decision to proceed this way was made in response to a recommendation by
Yeomans et al.\ (2004) that the brightest pixel defines the position of the
nucleus more accurately than a best-match two-dimensional Gaussian fit.
\begin{table*}[ht]
\vspace{-0.15cm}
\noindent
\begin{center}
{\footnotesize {\bf Table 14} \\[0.08cm]
{\sc Astrometric Positions of the Head of Comet C/2012 S1 Measured in Images
Taken\\with COR-2 Coronagraph on Board STEREO-B on 2013 November
26--28.}\\[0.13cm] 
\begin{tabular}{r@{\hspace{0.5cm}}l@{\hspace{0.05cm}}r@{\hspace{0.7cm}}c@{\hspace{0.6cm}}c@{\hspace{0.4cm}\vline\hspace{0.35cm}}r@{\hspace{0.5cm}}l@{\hspace{0.05cm}}r@{\hspace{0.7cm}}c@{\hspace{0.6cm}}c}
\hline\hline\\[-0.31cm]
\raisebox{-1.1ex}[2ex][4ex]{No.$\!\!\!$}
 & \multicolumn{2}{@{\hspace{-0.28cm}}c}{\raisebox{-1.1ex}[2ex][4ex]{2013 (UT)}}
 & \raisebox{-1.1ex}[2ex][4ex]{R.A.(2000)}
 & \raisebox{-1.1ex}[2ex][4ex]{Dec.(2000)}
 & \raisebox{-1.1ex}[2ex][4ex]{No.$\!\!$}
 & \multicolumn{2}{@{\hspace{-0.28cm}}c}{\raisebox{-1.1ex}[2ex][4ex]{2013 (UT)}}
 & \raisebox{-1.1ex}[2ex][4ex]{R.A.(2000)}
 & \raisebox{-1.1ex}[2ex][4ex]{Dec.(2000)}\\[-0.13cm]
\hline \\[-0.2cm]
& &
  & \hspace{0.01cm}$^{\rm h}$\hspace{0.22cm}$^{\rm m}$\hspace{0.23cm}$^{\rm s}$
& \hspace{0.43cm}$^\circ$\hspace{0.28cm}$^\prime$\hspace{0.28cm}$^{\prime\prime}$
& & &
& \hspace{-0.01cm}$^{\rm h}$\hspace{0.25cm}$^{\rm m}$\hspace{0.19cm}$^{\rm s}$
& \hspace{0.43cm}$^\circ$\hspace{0.29cm}$^\prime$\hspace{0.28cm}$^{\prime\prime}$ \\[-0.23cm]
  1 & Nov. & 26.26762 & 06 48 56.46 & +21 08 53.1 &  64 & Nov.
    & 27.64263 & 06 48 48.81 & +21 10 55.9 \\
  2 & & 26.28845 & 06 48 57.00 & +21 08 54.2 &
 65 & & 27.65305 & 06 48 48.13 & +21 11 08.4 \\
  3 & & 26.30929 & 06 48 57.37 & +21 08 50.7 &
 66 & & 27.66346 & 06 48 47.40 & +21 11 11.1 \\
  4 & & 26.33012 & 06 48 57.85 & +21 08 46.9 &
 67 & & 27.68430 & 06 48 45.87 & +21 11 36.0 \\
  5 & & 26.35095 & 06 48 58.34 & +21 08 33.3 &
 68 & & 27.69471 & 06 48 45.89 & +21 11 38.1 \\
  6 & & 26.37179 & 06 48 58.94 & +21 08 31.7 &
 69 & & 27.70513 & 06 48 44.37 & +21 11 51.1 \\
  7 & & 26.39262 & 06 48 59.06 & +21 08 28.5 &
 70 & & 27.72596 & 06 48 42.79 & +21 12 08.1 \\
  8 & & 26.41345 & 06 48 59.70 & +21 08 26.3 &
 71 & & 27.73638 & 06 48 42.10 & +21 12 20.9 \\
  9 & & 26.60096 & 06 49 02.88 & +21 07 50.8 &
 72 & & 27.74680 & 06 48 42.10 & +21 12 22.9 \\
 10 & & 26.62180 & 06 49 03.44 & +21 07 40.9 &
 73 & & 27.76763 & 06 48 39.70 & +21 12 50.2 \\
 11 & & 26.64263 & 06 49 03.96 & +21 07 38.0 &
 74 & & 27.77805 & 06 48 38.94 & +21 13 03.8 \\
 12 & & 26.66346 & 06 49 04.31 & +21 07 36.8 &
 75 & & 27.78846 & 06 48 38.92 & +21 13 06.3 \\
 13 & & 26.68430 & 06 49 04.44 & +21 07 34.8 &
 76 & & 27.80930 & 06 48 36.61 & +21 13 33.8 \\
 14 & & 26.70513 & 06 49 04.83 & +21 07 31.6 &
 77 & & 27.81971 & 06 48 35.79 & +21 13 48.4 \\
 15 & & 26.72596 & 06 49 04.55 & +21 07 31.5 &
 78 & & 27.83013 & 06 48 35.02 & +21 13 52.4 \\
 16 & & 26.74680 & 06 49 05.06 & +21 07 30.4 &
 79 & & 27.85096 & 06 48 33.37 & +21 14 19.8 \\
 17 & & 26.76763 & 06 49 05.23 & +21 07 28.7 &
 80 & & 27.86138 & 06 48 32.61 & +21 14 34.1 \\
 18 & & 26.78846 & 06 49 04.97 & +21 07 26.0 &
 81 & & 27.87180 & 06 48 31.02 & +21 14 49.1 \\
 19 & & 26.80930 & 06 49 05.32 & +21 07 22.3 &
 82 & & 27.89263 & 06 48 29.40 & +21 15 07.8 \\
 20 & & 26.83013 & 06 49 05.67 & +21 07 21.4 &
 83 & & 27.90305 & 06 48 28.50 & +21 15 21.5 \\
 21 & & 26.85096 & 06 49 05.40 & +21 07 21.7 &
 84 & & 27.91346 & 06 48 27.65 & +21 15 35.9 \\
 22 & & 26.87180 & 06 49 05.75 & +21 07 19.7 &
 85 & & 27.93430 & 06 48 25.17 & +21 16 06.2 \\
 23 & & 26.89263 & 06 49 05.97 & +21 07 17.6 &
 86 & & 27.94471 & 06 48 24.33 & +21 16 21.3 \\
 24 & & 26.91346 & 06 49 05.54 & +21 07 18.0 &
 87 & & 27.95513 & 06 48 22.74 & +21 16 38.0 \\
 25 & & 26.93430 & 06 49 06.27 & +21 07 24.7 &
 88 & & 27.97596 & 06 48 20.28 & +21 17 18.6 \\
 26 & & 26.95513 & 06 49 05.77 & +21 07 24.3 &
 89 & & 27.98638 & 06 48 19.41 & +21 17 33.2 \\
 27 & & 26.97596 & 06 49 06.03 & +21 07 22.3 &
 90 & & 27.99680 & 06 48 17.69 & +21 17 50.7 \\
 28 & & 26.99680 & 06 49 05.39 & +21 07 24.0 &
 91 & & 28.01763 & 06 48 16.58 & +21 18 13.2 \\
 29 & & 27.01763 & 06 49 06.70 & +21 07 17.9 &
 92 & & 28.02805 & 06 48 14.85 & +21 18 38.6 \\
 30 & & 27.03846 & 06 49 05.50 & +21 07 19.5 &
 93 & & 28.03846 & 06 48 12.50 & +21 18 53.1 \\
 31 & & 27.05930 & 06 49 05.86 & +21 07 24.6 &
 94 & & 28.05930 & 06 48 09.38 & +21 19 41.9 \\
 32 & & 27.08013 & 06 49 04.98 & +21 07 35.1 &
 95 & & 28.06971 & 06 48 09.65 & +21 19 50.2 \\
 33 & & 27.10096 & 06 49 04.66 & +21 07 32.2 &
 96 & & 28.08013 & 06 48 07.90 & +21 20 14.0 \\
 34 & & 27.12180 & 06 49 04.97 & +21 07 33.3 &
 97 & & 28.10096 & 06 48 04.42 & +21 21 14.6 \\
 35 & & 27.14263 & 06 49 04.57 & +21 07 34.4 &
 98 & & 28.11138 & 06 48 02.14 & +21 21 22.9 \\
 36 & & 27.16346 & 06 49 04.08 & +21 07 45.1 &
 99 & & 28.12180 & 06 48 00.39 & +21 21 49.1 \\
 37 & & 27.18430 & 06 49 04.22 & +21 07 43.6 &
100 & & 28.14263 & 06 47 57.06 & +21 22 42.2 \\
 38 & & 27.20513 & 06 49 03.68 & +21 07 44.6 &
101 & & 28.15305 & 06 47 57.30 & +21 22 50.2 \\
 39 & & 27.22596 & 06 49 04.10 & +21 07 53.9 &
102 & & 28.16346 & 06 47 53.64 & +21 23 25.0 \\
 40 & & 27.24680 & 06 49 02.89 & +21 07 56.8 &
103 & & 28.18430 & 06 47 49.97 & +21 24 23.2 \\
 41 & & 27.26763 & 06 49 02.85 & +21 08 06.9 &
104 & & 28.19471 & 06 47 48.16 & +21 24 50.4 \\
 42 & & 27.28846 & 06 49 02.32 & +21 08 08.4 &
105 & & 28.20513 & 06 47 46.31 & +21 25 20.7 \\
 43 & & 27.30930 & 06 49 01.79 & +21 08 18.3 &
106 & & 28.22596 & 06 47 43.14 & +21 26 25.0 \\
 44 & & 27.33013 & 06 49 01.28 & +21 08 20.6 &
107 & & 28.23638 & 06 47 41.25 & +21 26 53.8 \\
 45 & & 27.35096 & 06 49 00.79 & +21 08 31.8 &
108 & & 28.24680 & 06 47 37.97 & +21 27 24.9 \\
 46 & & 27.37180 & 06 49 00.08 & +21 08 33.0 &
109 & & 28.26763 & 06 47 33.93 & +21 28 26.1 \\
 47 & & 27.39263 & 06 48 59.34 & +21 08 45.2 &
110 & & 28.27805 & 06 47 32.81 & +21 28 52.9 \\
 48 & & 27.41346 & 06 48 58.97 & +21 08 47.0 &
111 & & 28.28846 & 06 47 30.86 & +21 29 26.0 \\
 49 & & 27.43430 & 06 48 57.64 & +21 09 01.3 &
112 & & 28.30930 & 06 47 25.12 & +21 30 57.3 \\
 50 & & 27.44471 & 06 48 57.69 & +21 09 02.0 &
113 & & 28.31971 & 06 47 23.26 & +21 31 32.8 \\
 51 & & 27.45513 & 06 48 57.05 & +21 09 13.3 &
114 & & 28.33013 & 06 47 19.68 & +21 32 06.7 \\
 52 & & 27.47596 & 06 48 56.37 & +21 09 15.5 &
115 & & 28.35096 & 06 47 14.34 & +21 33 35.2 \\
 53 & & 27.48638 & 06 48 55.76 & +21 09 27.5 &
116 & & 28.36138 & 06 47 12.94 & +21 34 06.0 \\
 54 & & 27.49680 & 06 48 55.76 & +21 09 27.6 &
117 & & 28.37180 & 06 47 08.74 & +21 35 05.3 \\
 55 & & 27.51763 & 06 48 54.37 & +21 09 42.9 &
118 & & 28.39263 & 06 47 03.06 & +21 36 37.6 \\
 56 & & 27.52805 & 06 48 54.49 & +21 09 44.2 &
119 & & 28.40305 & 06 47 01.81 & +21 37 08.5 \\
 57 & & 27.53846 & 06 48 53.81 & +21 09 55.1 &
120 & & 28.41346 & 06 46 58.19 & +21 38 07.1 \\
 58 & & 27.55930 & 06 48 53.08 & +21 10 08.9 &
121 & & 28.43430 & 06 46 52.46 & +21 39 43.4 \\
 59 & & 27.56971 & 06 48 52.38 & +21 10 10.3 &
122 & & 28.44471 & 06 46 48.87 & +21 40 44.0 \\
 60 & & 27.58013 & 06 48 51.75 & +21 10 22.4 &
123 & & 28.45513 & 06 46 45.24 & +21 41 44.4 \\
 61 & & 27.60096 & 06 48 50.96 & +21 10 26.0 &
124 & & 28.47596 & 06 46 37.87 & +21 43 50.8 \\
 62 & & 27.61138 & 06 48 50.22 & +21 10 38.3 &
125 & & 28.48638 & 06 46 34.23 & +21 44 54.9 \\
 63 & & 27.62180 & 06 48 49.56 & +21 10 50.8 &
126 & & 28.49680 & 06 46 27.99 & +21 46 28.7 \\[0.1cm]
\hline\\[-0.2cm]
\end{tabular}}
\end{center}
\end{table*}

\subsection{Orbital Solutions Linking Ground-Based and Spaceborne
 Astrometric Observations}
Although the quality of the astrometric observations derived from
the STEREO-B images is, because of the COR-2 detector's large pixel size,
inferior compared to the ground-based astrometry, the extension of the
orbital arc more than offsets this drawback.  Indeed, in terms of true
anomaly, the whole period of time covered by the ground-based data
amounts to a range of less than 18$^\circ$, while the 2.2-day long
period of time covered by the data in Table~14 is equivalent to a
range of 39$^\circ$, more than twice as much!  In addition, analysis
of the orbital motion close to the Sun {\nopagebreak}should offer unique
complementary information on the comet's physical state during this
critical time.

\begin{table*}[t]
\vspace{-0.17cm}
\noindent
\begin{center}
{\footnotesize {\bf Table 15} \\[0.1cm]
{\sc Summary of Gravitational and  Nongravitational Solutions for Comet C/2012
 S1 That Employ\\Ground-Based and Spaceborne Astrometry, and Their Parameters
 $A_1$, $A_2$, and $A_3$.}\\[0.1cm]
\begin{tabular}{l@{\hspace{0.15cm}}r@{\hspace{0.07cm}}c@{\hspace{0.07cm}}l@{\hspace{0.25cm}}c@{\hspace{0.35cm}}l@{\hspace{0.35cm}}c@{\hspace{0.45cm}}c@{\hspace{0.4cm}}c@{\hspace{0.35cm}}r@{\hspace{0.06cm}}c@{\hspace{0.06cm}}l@{\hspace{0.35cm}}r@{\hspace{0.06cm}}c@{\hspace{0.06cm}}l@{\hspace{0.35cm}}r@{\hspace{0.06cm}}c@{\hspace{0.06cm}}l}
\hline\hline\\[-0.2cm]
 & \multicolumn{3}{@{\hspace{-0.3cm}}c}{Start date of} & Number & &
 & Distance & Mass & \multicolumn{9}{@{\hspace{0cm}}c}{Parametric values
   (10$^{-8}$\,AU day$^{-2}$)$^{\rm c}$} \\
 & \multicolumn{3}{@{\hspace{-0.3cm}}c}{observations} & of obser- & & Mean
 & \raisebox{0.04cm}{$r_0$ or {\rend}} & erosion rate
 & \multicolumn{9}{@{\hspace{0cm}}c}{\rule[0.7ex]{5.9cm}{0.4pt}} \\
Solution & \multicolumn{3}{@{\hspace{-0.3cm}}c}{employed$^{\rm a}$} & vations
 & Option & residual\rlap{$^{\rm b}$} & (AU) & exponent $\xi$
 & \multicolumn{3}{@{\hspace{0.25cm}}c}{$A_1$}
 & \multicolumn{3}{@{\hspace{0.1cm}}c}{$A_2$}
 & \multicolumn{3}{@{\hspace{0.2cm}}c}{$A_3$}\\[0.1cm]
%
%
\hline \\[-0.2cm]
PGs\,I   & 2011 & Sept. & 30 & 6303 & \,\ldots\ldots
         & $\pm$3$^{\prime\prime}\!\!$.55 & \lpts & \lpts & & \lpts & & & \lpts
         & & & \lpts & \\[0.08cm]
PGs\,II  & 2013 & Oct.  & 31 & $\;\:$684 & \,\ldots\ldots & $\pm$7.74
         & \lpts & \lpts & & \lpts & & & \lpts & & & \lpts & \\[0.08cm]
SNs\,I   & 2011 & Sept. & 30 & 6303 & SNs\,I$_1$ & $\pm$2.84 & \lpts & \lpts
         & +10.45 & $\pm$ & 0.19 & & \lpts & & & \lpts & \\
         & & & & & SNs\,I$_2$ & $\pm$2.74 & \lpts & \lpts
         & +9.34 & $\pm$ & 0.20 & +0.97 & $\pm$ & 0.06 & & \lpts & \\
         & & & & & SNs\,I$_3$ & $\pm$2.04 & \lpts & \lpts
         & $-$0.62 & $\pm$ & 0.36 & +4.73 & $\pm$ & 0.13
         & +6.45 & $\pm$ & 0.19 \\[0.08cm]
SNs\,II  & 2013 & Oct. & 31 & $\;\:$684 & SNs\,II$_1$ & $\pm$4.72
         & \lpts & \lpts & +52.2 & $\pm$ & 2.3 & & \lpts & & & \lpts & \\
         & & & & & SNs\,II$_2$ & $\pm$3.75 & \lpts & \lpts
         & +113.6 & $\pm$ & 5.7 & $-$12.6 & $\pm$ & 1.1 & & \lpts & \\
         & & & & & SNs\,II$_3$ & $\pm$3.61 & \lpts & \lpts
         & +65.1 & $\pm$ & 9.5 & $-$0.2 & $\pm$ & 2.2
         & +8.5 & $\pm$ & 1 3 \\[0.08cm]
MNs\,I   & 2013 & Oct. & 31 & $\;\:$684 & MNs\,I$_1$ & $\pm$2.77
         & 0.438 $\pm$ 0.024 & \lpts & +4.61 & $\pm$ & 0.15 & & \lpts
         & & & \lpts & \\
         & & & & & MNs\,I$_2$ & $\pm$2.77 & 0.442 $\pm$ 0.025 & \lpts
         & +5.24 & $\pm$ & 0.27 & +0.25 & $\pm$ & 0.14 & & \lpts & \\
         & & & & & MNs\,I$_3$ & $\pm$2.76 & 0.444 $\pm$ 0.025 & \lpts
         & +6.09 & $\pm$ & 0.83
         & +0.11 & $\pm$ & 0.44
         & $-$0.15 & $\pm$ & 0.35 \\[0.08cm]
NAs\,I   & 2013 & Oct. & 31 & $\;\:$684 & NAs\,I$_1$ & $\pm$2.78 & \lpts
         & \lpts & +17.75 & $\pm$ & 0.56 & & \lpts & & & \lpts & \\
         & & & & & NAs\,I$_2$ & $\pm$2.79 & \lpts & \lpts
         & +16.36 & $\pm$ & 0.94 & +0.99 & $\pm$ & 0.54 & & \lpts & \\
         & & & & & NAs\,I$_3$ & $\pm$2.80 & \lpts & \lpts
         & +13.1 & $\pm$ & 2.5 & +2.8 & $\pm$ & 1.4
         & +1.4 & $\pm$ & 1.0 \\[0.08cm]
DDs\,I   & 2013 & Oct. & 31 & $\;\:$684 & DDs\,I$_1$ & $\pm$2.77
         & 0.003 $\pm$ 0.010 & 3.90 $\pm$ 0.10 & +4.37 & $\pm$ & 0.14 &
         & \lpts & & & \lpts & \\[0.08cm]
DDs\,II  & 2013 & Oct. & 31 & $\;\:$684 & DDs\,II$_1$ & $\pm$2.77
         & 0.024; fixed & 3.86 $\pm$ 0.11 & +4.56 & $\pm$ & 0.15 & & \lpts
         & & & \lpts & \\
         & & & & & DDs\,II$_2$ & $\pm$2.79 & 0.024; fixed & 3.94 $\pm$ 0.11
         & +3.58 & $\pm$ & 0.19 & +0.23 & $\pm$ & 0.11 & & \lpts & \\
         & & & & & DDs\,II$_3$ & $\pm$2.69 & 0.024; fixed & 4.49 $\pm$ 0.18
         & +0.08 & $\pm$ & 0.14 & +0.81 & $\pm$ & 0.10
         & +0.44 & $\pm$ & 0.06 \\[0.08cm]
\hline\\[-0.22cm]
\end{tabular}}
\parbox{17.4cm}{\scriptsize $^{\rm a}$\,End date was always 2013 November
 28, the date of perihelion passage.}\\[-0.04cm]
\parbox{17.4cm}{\scriptsize $^{\rm b}$\,This is an unweighted mean
 residual.}\\[-0.02cm]
\parbox{17.4cm}{\scriptsize $^{\rm c}$\,Except that for all MNs\,I and
 NAs\,I options the unit is 10$^{-15}$ AU day$^{-2}$.}\\[0.4cm]
\end{center}
\end{table*}

As with the ground-based observations, preliminary runs were made to obtain
ad hoc orbital solutions to check the residuals from the 126 STEREO astrometric
observations in Table~14.  No entry was discarded, as none of the residuals
exceeded the pixel size.  The 90~data points from November~26--27 displayed
a scatter within $\pm$10~arcsec in either coordinate and each was assigned
a weight of 0.2 the weight of the ground-based observations, while the 36~data
points from November~28 had residuals well within $\pm$20~arcsec in either
coordinate and each was assigned a weight of 0.1.  The resulting set of all
ground-based and spaceborne astrometric observations totaled 6303.

\subsubsection{Runs linking all 6303 observations.}
To test the linkage of the ground-based and spaceborne data in
orbital computations, we began with a gravitational solution, PGs\,I, and
with three options of the nongravitational solution SNs\,I, each of them
based on all 6303 observations (a suffix ``s'' was added to every solution
that included spaceborne observations to distinguish it from solutions
based on the ground-based observations only).

The results are listed at the top of Table~15 and the distributions of
residuals are plotted in Figure~23.  The residuals show that neither the
gravitational solution nor any nongravitational solution based on the
standard law $g_{\rm ice}(r)$ could successfully link the comet's ground-based
observations with its STEREO ones.  Very strong systematic trends in the
residuals of up to 5~arcsec in the ground-based data and in excess of
1~arcmin(!) in the STEREO data are seen in Figure~23.  The fact that the
fit based on the standard nongravitational law is only marginally better
than the gravitational fit is particularly disappointing:\ the $g_{\rm ice}$
law overcorrected the residuals from the last ground-based observations and
reduced the residuals from the STEREO observations by only some 10--20~arcsec,
but did not remove the exponentially diverging trends.  A physically
meaningless negative value of $A_1$ resulted from the run SNs\,I$_3$ in
Table~15.

What is the meaning of the trends in the residuals from the gravitational
solution in Figure~23?  From the strongly negative trend in right ascension
and a moderately positive trend in declination in the last week of
ground-based observations, the comet's observed geocentric motion relative
to the computed one was increasingly toward the west-northwest.  With
respect to the Earth, the comet's motion was at that time in position
angle 116$^\circ$, that is, to the east-southeast, in exactly the opposite
direction.  During the two days of STEREO-B observations, the trend in the
residuals was strongly positive in right ascension and almost equally
strongly negative in declination, so that in reference to the spacecraft
the comet's observed motion relative to the computed one was increasingly
toward the southeast.  The comet's motion with respect to STEREO-B was in
position angle 318$^\circ$, that is, to the northwest, again in exactly the
opposite direction.  In summary, the distribution of residuals during the
last two weeks before perihelion show consistently that the comet's
nucleus --- or what was measured in its place --- was increasingly lagging
in the orbit behind the position expected from the gravitational law:\ {\it
the comet was rapidly decelerating\/}.  The deceleration did not however
follow the standard nongravitational law, indicating apparently that
{\it outgassing of water was not the primary trigger\/} of the
nongravitational effects, the enormous production of water during
Event~2 (Secs.~2.2--2.3) notwithstanding.

\begin{figure*}[ht]
\vspace{-1.2cm} 
\hspace{0.25cm}
\centerline{
\scalebox{0.72}{
\includegraphics{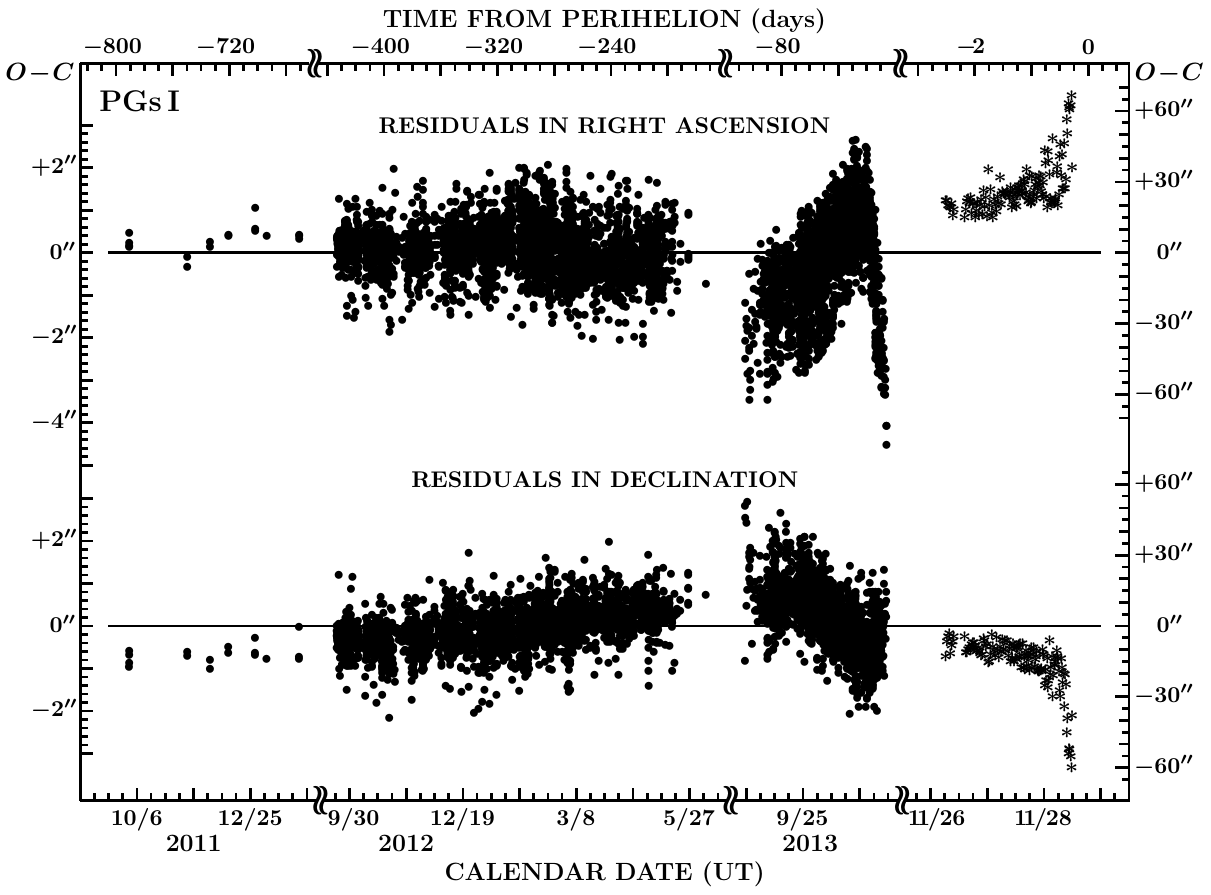}}} 

\vspace{-9.15cm} 
\hspace{0.25cm}
\centerline{
\scalebox{0.72}{
\includegraphics{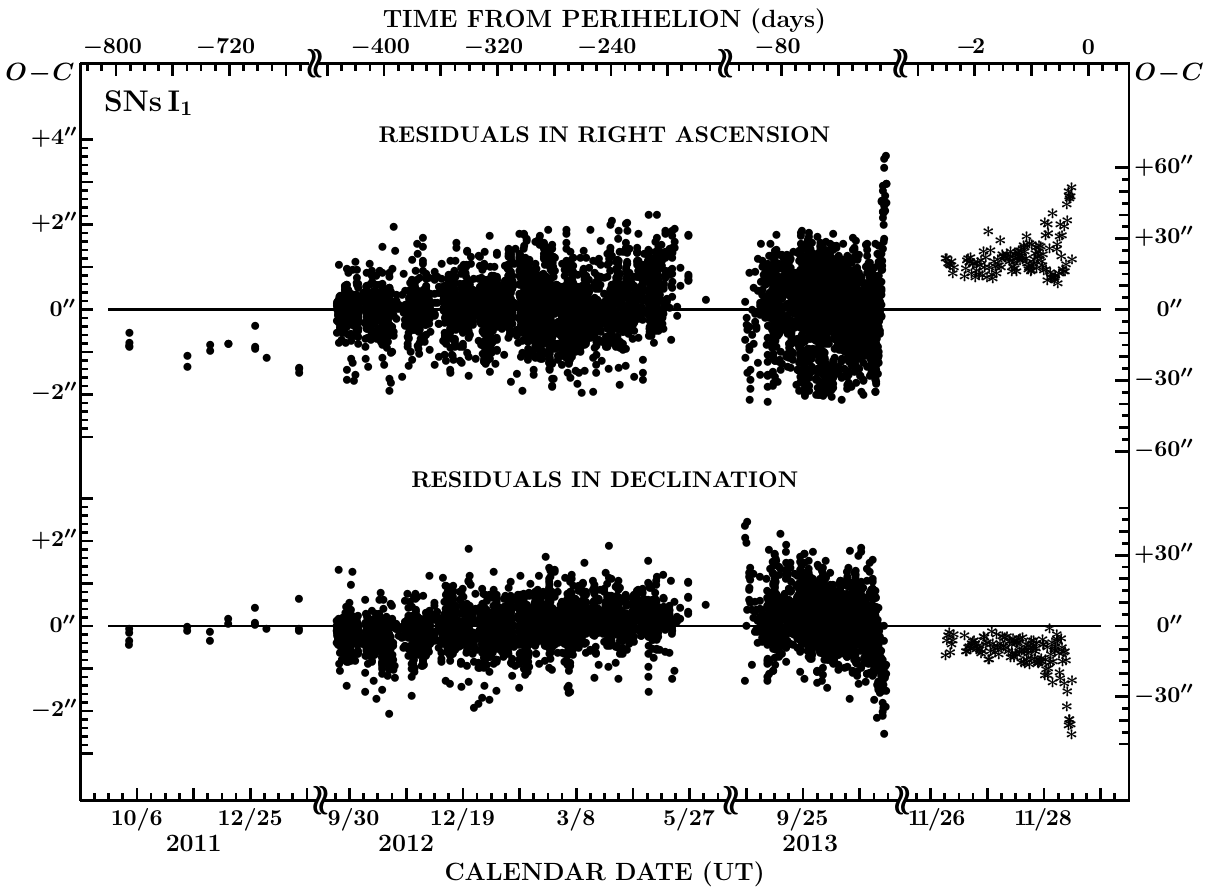}}} 
\vspace{-8.65cm}
\caption{Distributions of residuals for 6303 observations made between 2011
September 30 and 2013 November 28, from the solutions PGs\,I (top) and
SNs\,I$_1$ (bottom).  The ground-based data are depicted by bullets and
their scales are on the left ordinate axis, the spaceborne data (the
clusters on the far right) by asterisks and their scales (with steps a factor
of 18 smaller) are on the right ordinate axis.  The time scale is expanded
for the spaceborne data by a factor of 40.  The ground-based portion
of the top plot resembles the distribution in Figure~18, while the quality
of fit to the ground-based data at the bottom has somewhat deteriorated
compared to that in Figure~21.  The spaceborne data display an astonishingly
large discrepancy with both solutions, showing residuals of up to nearly
70~arcsec.}
\end{figure*}

\subsubsection{Runs PGs, SNs, and MNs linking observations from\\2013
 October 31--November 28.}
To examine whether the trends in the residuals from the STEREO-B data
that dominate the appearance of Figure~23, could be due to an effect
of an overwhelming preponderance of ground-based data on the fitting
algorithm, we repeated the PGs and SNs runs over a much shorter period
of time, with the number of ground-based observations reduced by about
one order of magnitude.  Since the MN\,II$_1$ solution provided a very
satisfactory fit to the ground-based observations up to 2013 October~30,
we now chose for the limited period of time a span from 2013 October~31 to
November~28.

The results of these alternative runs, PGs\,II and SNs\,II, are listed in
Table~15, while their distributions of residuals are in Figure~24.  Cursory
inspection reveals clear trends in the residuals from the ground-based
observations and the diverging trends in the residuals from the
spaceborne observations that are equally prominent as in Figure~23
(in the case of PGs\,II) or only slightly reduced (in the case of
SNs\,II$_1$).  Similar trends are also exhibited in the distributions
of residuals from SNs\,II$_2$ and SNs\,II$_3$, which are not shown.  We
are satisfied that the exponentially diverging residuals from the STEREO
observations were not a byproduct of the data selection process.  It is
noted that the unweighted mean residuals for these runs came out to be
higher than in the solutions PGs\,I and SNs\,I.  This was due to the
fact that the fraction of less accurate spaceborne data increased from
2~percent to more than 18~percent of the total data used.

The next step was an application of the modified nongravitational law
$g(r; r_0)$, keeping the exponents $m$, $n$, and $k$ constant, as in
Eq.\,(61), and optimizing the fit by varying only the scaling distance
$r_0$.  The results, referred to in Table~15 as the MNs\,I solution, were
astonishing:\ $r_0$ came out to be a mere 0.44~AU and the fit was excellent,
with no trends in the distribution of residuals, as illustrated in Figure~25.
From Eq.\,(62) it follows that the comet's motion during this period of time
was affected by one or more sublimating species whose sublimation heat was
near \mbox{$L \approx 29\,000$ cal mol$^{-1}$}.  The obvious candidate is
atomic sodium,\footnote{See, e.g., {\tt http://en.wikipedia.org/wiki/Sodium.}}
known to sublimate profusely near perihelion from sungrazing comets and whose
heat of sublimation is near 25\,000 cal mol$^{-1}$.  

We conclude from the runs of the modified nongravitational law that there
is a contradiction between the nature of the outgassing-driven effects in
the comet's motion at larger heliocentric distances (as described in
Secs.~6.2--6.4, summarized in Tables~11--12, and resulting in the scaling
distances of 2--4~AU) and near the Sun.  For this reason we felt that there
was no chance of successfully linking all 6303 observations from 2011--2013
with a single modified law and we made no effort to do so.

\begin{figure}[ht]
\vspace{0.2cm} 
\hspace{-2.4cm}
\centerline{
\scalebox{0.66}{
\includegraphics{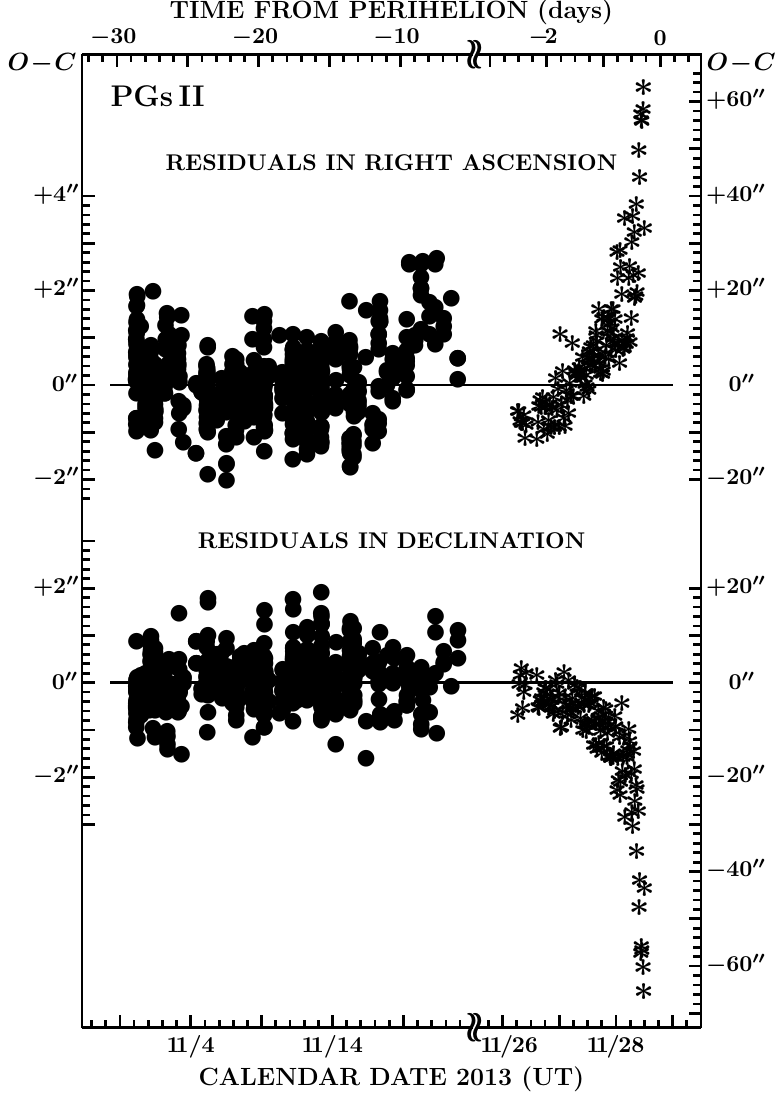}}} 

\vspace{-7.05cm} 
\hspace{-2.4cm}
\centerline{
\scalebox{0.66}{
\includegraphics{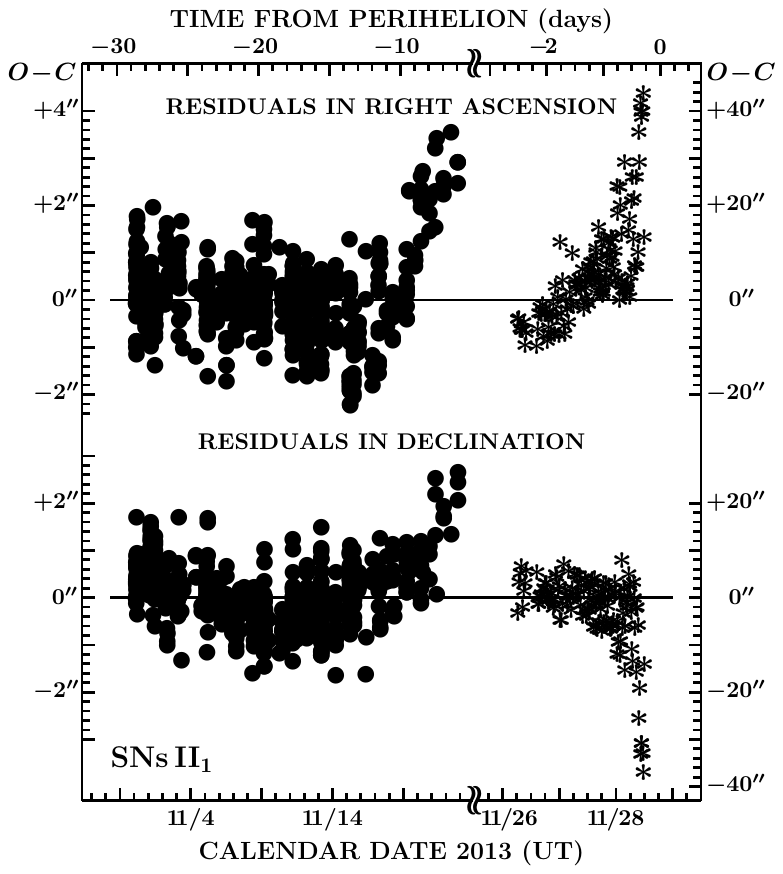}}} 
\vspace{-7.9cm}
\caption{Distributions of residuals for 684 observations made between 2013
October 31 and November 28 from the solutions PGs\,II (top) and SNs\,II$_1$
(bottom).  The spaceborne $O \!-\! C$ data are compressed by a factor of
10 compared to the scales of the ground-based data.  See the caption to
Figure~23 for more details.{\vspace{-0.45cm}}}
\end{figure}

\subsubsection{Nongravitational solutions based on sublimation\\of sodium
(NAs)}

\begin{figure}[ht]
\vspace{-2.83cm} 
\hspace{-2.2cm} 
\centerline{
\scalebox{0.66}{
\includegraphics{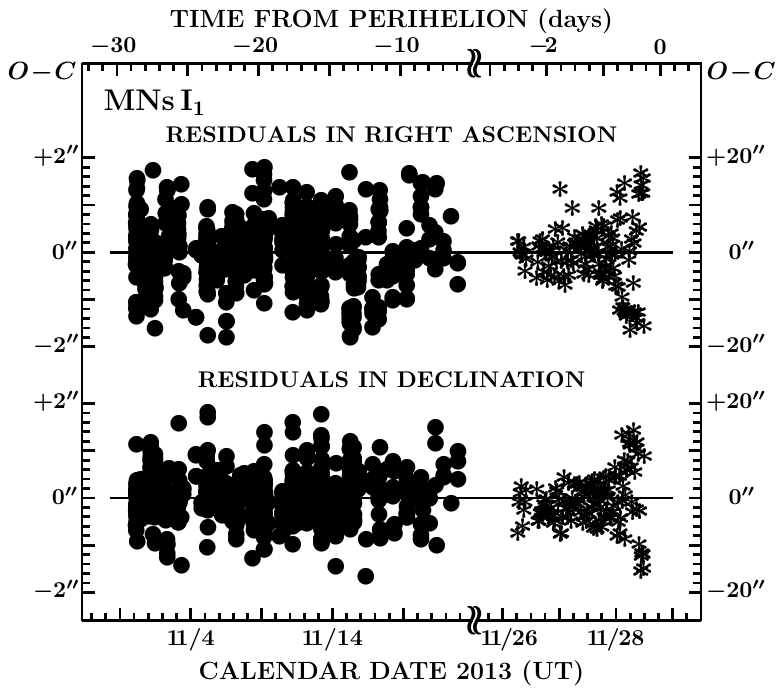}}} 
\vspace{-7.88cm}
\caption{Distribution of residuals for 684 observations made between 2013
October 31 and November 28 from the run MNs\,I$_1$.  Bullets are the
ground-based data, asterisks the STEREO-B data.  The $O \!-\! C$ scales of
the latter are compressed by a factor of 10 compared to those of the former;
recall that the November 28 data are twice as inaccurate as the November 26--27
data.  Contrary to Figures~23 and 24, no trends in the residuals are apparent.
Nearly identical distributions of residuals resulted from other two
MNs\,I runs as well as the three NAs\,I runs.  Only slightly different
distributions, at the very end of the data set, resulted from the DDs\,I
and DDs\,II solutions; see Table~15 for a summary of all these
runs.{\vspace{0.4cm}}}
\end{figure}

Evidence that near the Sun the comet's orbital motion may have been subjected
to strong outgassing of atomic sodium motivated us to formulate a new
$g(r; r_0)$ law, equivalent to $G(r; r_0, +1)$ from Eq.\,(71), with constants
that fit specifically the heliocentric-distance {\nopagebreak}variations in
the sublimation rate per unit surface area of atomic sodium derived for the
isothermic model directly from the dependence of the saturated sodium-vapor
pressure on the temperature.  The constants that define this empirical
sodium-sublimation law we refer to as $g_{\rm Na}(r)$ were found by least
squares as follows:
\begin{eqnarray}
r_0 & = & 0.3458\;{\rm AU}, \nonumber \\
m   & = & 2.089, \nonumber \\
n   & = & 3.603, \\
k   & = & 4.896, \nonumber \\
\alpha_0 & = & 1.3948 \times 10^9. \nonumber
\end{eqnarray}
Eleven values of the sodium sublimation rate between 0.01~AU and 0.8~AU were
fitted by this law with a mean residual of $\pm$5~percent.

The results, identified in Table~15 as a solution NAs\,I, show that the
$g_{\rm Na}(r)$ law fitted the astrometric observations between 2013 October
31 and November 28 just as well as the modified nongravitational law with
\mbox{$r_0 \simeq 0.44$ AU}.  This is demonstrated by comparing the NAs\,I$_1$
and MNs\,I$_1$ runs.  The distribution of residuals is at first sight
indistinguishable from that in Figure~25, and we do not display it.  The
transverse and normal components of the nongravitational accelerations are
not well defined, as in previous cases.  In summary, we submit that the high
probability of a major influence of sodium outgassing on the comet's orbital
motion at heliocentric distances smaller than $\sim$1~AU is hereby confirmed.
The issues as to whether the outgassing of sodium was the only major trigger
of the observed nongravitational motion and what are the implications for
the comet's nucleus are addressed in Sec.~6.7.

\subsubsection{Nongravitational solutions based on\\sublimation of silicates}
Sublimation of silicate grains, such as olivine or pyroxene, depends
significantly on the grains' orbits, specifically on their perihelion
distances (Sec.~3.9), which, in turn, are a function of the Sun's radiation
pressure acceleration and the time of the grains' ejection from the nucleus.
To investigate potential effects of silicate sublimation directly from the
surface of the comet's nucleus, we followed the procedure described in
Sec.~6.6.3 for atomic sodium and employed the dependence of the saturated
vapor pressure on temperature for forsterite,\footnote{The relationship
between the vapor pressure and temperature is followed closely also above
2163\,K, the melting point of forsterite.} the magnesium-rich end-member
of the olivine solid solution series (Mg$_2$SiO$_4$), measured by Hashimoto
(1990).  For $g_{\rm for}(r)$, the same $g(r; r_0)$ type of empirical law,
we derived the constants as follows:
\begin{eqnarray}
r_0 & = & 0.01486\;{\rm AU}, \nonumber \\ 
m   & = & 2.634, \nonumber \\
n   & = & 5.155, \\
k   & = & 3.320, \nonumber \\
\alpha_0 & = & 1.2591 \times 10^{36}. \nonumber
\end{eqnarray}
Ten values of the forsterite sublimation rate between 0.0055~AU and 0.03~AU
were fitted with a mean residual of $\pm$2.5~percent.

Application of this law to the observations from 2013 October~31 to
November~28, produced no positive results.  A solution that included only
$A_1$ failed to converge, and solutions that also included $A_2$ and $A_3$
crashed.  Given the value of $r_0$ in Eqs.\,(77), which is equivalent to
about 3.2~{\Rsun}, and the heliocentric distance of the measured image
closest to the Sun (Table~14), which exceeds 8~{\Rsun}, any
forsterite-sublimation based solution forces a variation in the
nongravitational acceleration close to $r^{-20}$, utterly incompatible
with any realistic fit.

\subsubsection{Nongravitational solutions accounting for nucleus' dwindling
 dimensions near the Sun (DDs)}
In Sec.~6.4 we concluded that this type of nongravitational law offered
competitive, but not superior, nongravitational solutions over extended
periods of time, when the observations close to perihelion were not
included.

We now tested this law, ${\cal G}(r; \mbox{\rend}, \xi)$, on the same data
set that was so satisfactorily fitted by the modified nongravitational law
and the sodium-sublimation based law.  Table~15 shows that the incorporation
of the dwindling-dimensions effect (DDs) offers equivalent solutions.

We first applied the approach, described extensively in Sec.~6.4, that allowed
us to optimize, successively, both {\rend} and $\xi$.  This procedure led
for the DDs\,I$_1$ run ($A_1$ only) to \mbox{$\mbox{\rend} = 0.003 \pm
0.010$ AU} and \mbox{$\xi = 3.90 \pm 0.10$}.  We noted that the optimized
value of {\rend}, though nominally inside the Sun, was only about 2$\sigma$
away from the heliocentric distance of 0.024~AU, a point of termination of
all activity, which already was substituted for {\rend} in Sec.~6.4 and
which now defined the solution that is referred to as DDs\,II in Table~15.
We then ran the three options with an individually optimized erosion-rate
exponent $\xi$.

\begin{table*}[t]
\noindent
\vspace{-0.28cm}
\begin{center}
{\footnotesize {\bf Table 16} \\[0.08cm]
{\sc Orbital Elements of Comet C/2012 S1 for the Period of 2013 October
 31--November 28\\(Solutions MNs\,I$_1$, NAs\,I$_1$, and DDs\,II$_1$).}\\[0.1cm]
\begin{tabular}{l@{\hspace{0.4cm}}r@{\hspace{0.08cm}}c@{\hspace{0.08cm}}l@{\hspace{0.4cm}}r@{\hspace{0.08cm}}c@{\hspace{0.08cm}}l@{\hspace{0.4cm}}r@{\hspace{0.08cm}}c@{\hspace{0.08cm}}l}
\hline\hline\\[-0.2cm]
Orbital parameter/element
  & \multicolumn{3}{@{\hspace{0.95cm}}c}{Solution MNs\,I$_1$}
  & \multicolumn{3}{@{\hspace{0.95cm}}c}{Solution NAs\,I$_1$}
  & \multicolumn{3}{@{\hspace{1.1cm}}c}{Solution DDs\,II$_1$} \\[0.08cm]
%
%
\hline \\[-0.2cm]
Epoch of osculation (TT)        & 2013$\;\;$ & \llap{N}o\rlap{v} & $\;$24.0
                                & 2013$\;\;$ & \llap{N}o\rlap{v} & $\;$24.0
                                & 2013$\;\;$ & \llap{N}o\rlap{v} & $\;$24.0 \\
Time of perihelion, $t_\pi$ (TT) & 2013 Nov 28.77687 & $\pm$ & 0.00008
                                 & 2013 Nov 28.77570 & $\pm$ & 0.00008
                                 & 2013 Nov 28.77633 & $\pm$ & 0.00008 \\
Argument of perihelion, $\omega$
                         & 345$^\circ\!$.5501 & $\pm$ & 0$^\circ\!$.0010
                         & 345$^\circ\!$.5402 & $\pm$ & 0$^\circ\!$.0010
                         & 345$^\circ\!$.5430 & $\pm$ & 0$^\circ\!$.0010 \\
Longitude of ascending node, $\Omega$
                         & 295$^\circ\!$.6684 & $\pm$ & 0$^\circ\!$.0010
                         & 295$^\circ\!$.6724 & $\pm$ & 0$^\circ\!$.0009
                         & 295$^\circ\!$.6785 & $\pm$ & 0$^\circ\!$.0009 \\
Orbital inclination, $i$
                         &  62$^\circ\!$.3300 & $\pm$ & 0$^\circ\!$.0032
                         &  62$^\circ\!$.3144 & $\pm$ & 0$^\circ\!$.0030
                         &  62$^\circ\!$.3231 & $\pm$ & 0$^\circ\!$.0032 \\
Perihelion distance, $q$ (AU)    & 0.0124714 & $\pm$ & 0.0000012
                                 & 0.0124769 & $\pm$ & 0.0000012
                                 & 0.0124781 & $\pm$ & 0.0000012 \\
Orbital eccentricity, $e$        & 0.9999886 & $\pm$ & 0.0000018
                                 & 1.0000327 & $\pm$ & 0.0000018
                                 & 1.0000070 & $\pm$ & 0.0000017 \\[0.05cm]
Nongravitational law \rlap{and parameters:} & & & & & & & & & \\
\hspace{0.3cm}Type of law & & \llap{$g(r$};\,\rlap{$r_0)$}
                    & & & \llap{$g_{\rm N}$}$_{\rm a}$\rlap{$(r)$}
                    & & & \llap{${\cal G}(r;\:$}\rlap{$\mbox{\rend}\!,\xi)$} \\
\hspace{0.3cm}Distance $r_0$ or {\rend} (AU)
       & 0.438 & $\pm$ & 0.024 & & \lpts & & & \llap{0.024};\rlap{\,fixed} & \\
\hspace{0.3cm}Mass erosion rate exponent $\xi$ & & \lpts & & & \lpts &
       & 3.86 & $\pm$ & 0.11 \\
\hspace{0.3cm}Radial component
       $A_1$\,\rlap{(10$^{-15}\,$AU\,day$^{-2})\,^{\rm a}$}
& +4.61 & $\pm$ & 0.15 & +17.75 & $\pm$ & 0.56 & +4.56 & $\pm$ & 0.15 \\[0.08cm]
\hline \\[-0.2cm]
\end{tabular}}
\parbox{17.65cm}{\scriptsize $^{\rm a}$\,Except for the solution DDs\,II$_1$,
 for which the unit is 10$^{-8}$\,AU day$^{-2}$.{\vspace{0.35cm}}}
\end{center}
\end{table*}

The distributions of residuals from all DDs\,II runs were nearly identical to
the distribution in Figure~25.  The only readily detectable differences were in
the residuals from the very last STEREO-B observation, on November~28.4968~UT
(Table~14), which, from the DDs\,II$_1$ run, amounted to $-$24~arcsec in right
ascension and +25~arcsec in declination.  Since this measurement was admittedly
uncertain because of the elongation of the comet's image (Sec.~6.5), these
large residuals (but still smaller than a pixel size), are inconsequential.

Nevertheless, we found consistently that the effect of dwindling dimensions
of the nucleus did not play a major role in fitting the astrometric data,
even though the law based on it offered a match that was competitive with
those by some of the other employed laws and superior to the standard
nongravitational law (the SN and SNs runs).  It is possible that better
results could be achieved with more complex expressions for the DD law, but
such options cannot be exercised without first modifying the respective
function in the orbit-determination software, such as the EXORB code.

Since we judge the three nongravitational laws about equally successful in
fitting the astrometric data in the period of 2013 October~31--November~28,
we present in Table~16 the sets of orbital elements from each of the
solutions MNs\,I$_1$, NAs\,I$_1$, and DDs\,II$_1$.  The sets differ from
one another much more than a few times the formal mean errors due to the
fact that the influence of the different nature of the nongravitational
laws affecting the orbital elements is not included in the formal errors.

\subsection{Implications for C/2012 S1 from\\Results of Orbital Analysis}
The motion of C/2012 S1 was too complex to fit with a single, all-inclusive
orbital solution.  We found that the entire arc of the observed orbit could
be divided into three parts:

(1) From the first observation at 9.4~AU down to 4.9~AU, that is, from
790~days to about 300~days before perihelion, the comet's motion was
satisfactorily fitted by a gravitational solution, which implied an original
barycentric reciprocal semimajor axis, $(1/a_{\rm b})_{\rm orig}$, equal
to \mbox{$+0.000\,035 \pm 0.000\,006$~AU$^{-1}$}, equivalent to an
initial aphelion distance of \mbox{$57\,000  \pm 10\,000$~AU} and an
orbital period of \mbox{$4.8 \pm 1.2$~million} years, thus confirming
that the comet arrived from the Oort Cloud.

(2) Between $\sim$300 days and $\sim$30 days before perihelion, at a range
of heliocentric distances from $\sim$4.9~AU down to $\sim$1~AU, the comet's
motion was subjected to a nongravitational acceleration, due to the momentum
transferred primarily, but perhaps not solely, from outgassing of water; the
best orbital solution was achieved by a modified nongravitational law with
a scaling distance of \mbox{$r_0 \simeq 2$ AU}, fairly close to that of the
standard law in the Style~II formalism of Marsden et al.\ (1973).

(3) At heliocentric distances smaller than 1~AU, within $\sim$30~days
of perihelion, the comet was moving in an orbit affected by strong
nongravitational forces; the scaling distance of the modified law dropped
dramatically to about 0.44~AU, suggesting that sublimation of sodium,
rather than water ice, dominated the effect.  An acceleration due to the
Sun's radiation pressure could possibly have been a contributing factor in
a late stage.

The sets of orbital elements that best fit the astrometric observations
are presented in Table~10 for the first of the three periods of time, in
Table~13 for the second period, and in Table~16 for the third period.  This
last table contains three orbits, each of which fitted the observations
about equally well.  We already commented on the differences among the
individual elements from the three solutions being much greater than the
formal errors.  The inherent diversity of the laws in Table~16 manifests
itself also in an extrapolated prediction of the true time of perihelion
passage, which for a given set of elements is achieved by integrating it to
the osculation epoch at perihelion.

For the set of elements from the solution DDs\,II$_1$ this integration
could not in fact be performed because, by definition, the nongravitational
acceleration reached a singularity at \mbox{$\mbox{\rend} = 0.024$ AU},
3.5~hours before perihelion.  The diverse nature of the other two laws is
reflected in a difference of 10.4~minutes between the obtained results:\
November~28.7829~TT from the MNs\,I$_1$ run and November~28.7757~TT from
the NAs\,I$_1$ run.  Since there was no reason to prefer either solution,
we adopted their average, which was used throughout this paper when counting
times from perihelion passage:
\begin{eqnarray}
t_\pi & = & 2013\;{\rm November}\;28.7793 \pm 0.0036\;{\rm TT}
            \nonumber \\
      & = & \hspace{0.52cm}2013/11/28,{\rm 18\!:\!42.2} \pm 5.2\;{\rm TT}.
\end{eqnarray}

\begin{table}[ht]
\begin{center}
{\footnotesize {\bf Table 17} \\[0.1cm]
{\sc Nongravitational Accelerations of Comet C/2012 S1 Computed from
 Orbital Solutions in Table 16.}\\[0.2cm]
\begin{tabular}{c@{\hspace{0.95cm}}c@{\hspace{1.07cm}}c@{\hspace{1cm}}c}
\hline\hline\\[-0.2cm]
 & \multicolumn{3}{@{\hspace{-0.05cm}}c}{Computed nongravitational
 acceleration} \\
 & \multicolumn{3}{@{\hspace{-0.05cm}}c}{(units of Sun's gravitational
 acceleration)} \\
Distance & \multicolumn{3}{@{\hspace{-0.05cm}}c}{\rule[0.7ex]{5.5cm}{0.4pt}} \\
from Sun & Solution   & Solution   & Solution \\
(AU)     & MNs\,I$_1$ & NAs\,I$_1$ & DDs\,II$_1$ \\[0.08cm]
\hline \\[-0.2cm]
0.04     &  0.0072$\;\:$    & 0.0121$\;\:$    & 0.0876$\;\:$    \\
0.20     &  0.0052$\;\:$    & 0.0056$\;\:$    & 0.0031$\;\:$    \\
0.50     &  0.00003\rlap{3} & 0.00000\rlap{5} & 0.00056\rlap{0} \\[0.08cm]
\hline\\[-0.3cm]
\end{tabular}}
%
%
\end{center}
\end{table}

The preponderant component of the nongravitational acceleration in the most
satisfactory solutions was the radial component, $A_1$, which was in all such
cases positive, pointing away from the Sun and confirming that the comet's
orbital motion was decelerated.  This obviously means that the momentum
transferred to the nucleus was from the material sublimating preferentially
in the sunward direction, as expected.  The transverse and normal components,
$A_2$ and $A_3$, were, as a rule, one order of magnitude smaller and
their inclusion in the solution usually failed to improve the quality
of fit.  The runs, in which only $A_1$ was solved for, were consistently
preferred as the most dependable ones.

Because the nucleus disintegrated into pebbles and dust before it reached a
heliocentric distance of 5~{\Rsun}, we see no role whatsoever for the Sun's
tidal forces.  Similarly, we do not see any effect on the orbit from
sublimation of silicates, apparently because the distance from the Sun was
still too large even at the time of the last astrometric observation.  We
suggest that thermal stress due to increasing temperature gradients triggering
explosions of pressurized water vapor and other volatile species from heated
and/or thermally damaged reservoirs in the nucleus' interior was most probably
responsible for episodic primary fragmentation of the nucleus that subsequently
turned into cascading fragmentation of its debris.  Although thermal stresses
in the nuclei of sungrazing comets are substantially smaller before perihelion
than afterward (Sekanina \& Chodas 2012), the cohesion of the nucleus of
C/2012~S1 was obviously much too low to withstand them.  The integrated effect
on the orbital motion by continuous sublimation of sodium atoms (and, in
smaller amounts, other materials) was at heliocentric distances of less
than 1~AU apparently greater than by episodic outbursts of water molecules.

It is highly enlightening to compute the magnitude of the comet's
nongravitational acceleration as a function of heliocentric distance from
the three laws in Table~16.  For three distances these accelerations (in
units of the Sun's gravitational acceleration at the same distance) are
listed in Table~17.  It is known that for the relatively few one-apparition
comets, for which the nongravitational parameters could be determined, the
acceleration is usually on the order of 10$^{-5}$, seldom 10$^{-4}$, the Sun's
gravitational acceleration (Marsden \& Williams 2008), as indeed it was for
C/2012~S1 at larger heliocentric distances (Tables~12 and 13; about \mbox{$7
\times 10^{-4}$} from the solution MN\,II$_1$).  Nongravitational accelerations
on the order of several times 10$^{-3}$ the Sun's gravitational acceleration or
higher, seen in Table~17, are unheard of.  If, for example, the accelerations
in the first row of Table~17 are due to sublimation of sodium atoms, the
parent bodies from which they were released had typical dimensions in a
centimeter to submeter size range at most, depending in part on the velocities
of release (assumed hundreads of meters per second), bulk density (assumed
a fraction of 1~g~cm$^{-3}$), and the degree of sublimation anisotropy (assumed
on the order of 0.1).  This is the same range of dimensions that we estimated
from independent evidence in Sec.~4.3.  We should add that it is possible that
a fraction of the nongravitational acceleration could be due to solar radiation
pressure, although probably only a small fraction because centimeter-sized
grains are subject to radiation-pressure accelerations not exceeding 10$^{-4}$
the Sun's gravitational acceleration.

\section{Comparison of C/2012 S1 with Kreutz Sungrazer C/2011 W3 (Lovejoy).}
This last section before the conclusions is dedicated to comparison, especially
in close proximity to the Sun, of the physical behavior of C/2012~S1 with
that of the most recent bright Kreutz sungrazer, C/2011~W3.  To a considerable
extent, the traits of either object are related to the different perihelion
distances and to the very different origin (Oort Cloud vs Kreutz system).
We address this issue because the two comets have sometimes been judged to
possess similar properties.  We limit our comparison primarily to images in the
C2 and C3 coronagraphs on board SOHO, but similar studies can also be based
on other instruments.  We find major differences between the two objects,
as follows:

(i) {\it Saturation of the coronagraphs' CCD detectors before perihelion\/}.
The head of C/2012~S1 stopped saturating the detectors just hours before it
disappeared behind the occulting disks of the two instruments; {\it whereas\/}
the head of C/2011~W3 continued to saturate them until its disappearance.

(ii) {\it Morphology of the tail in preperihelion images\/}.  The tail of
C/2012~S1 was in these images bifurcated, consisting of a main feature and
a streamer extending along the orbit; {\it whereas\/} the narrow preperihelion
tail of C/2011~W3 showed no such morphology.

(iii) {\it Origin of the preperihelion tail\/}.  The ejecta in the two
components of the preperihelion tail of C/2012~S1 were very different in
origin, released at vastly different heliocentric distances; {\it whereas\/}
the ejecta in the preperihelion tail of C/2011~W3 were of uniform origin.

(iv) {\it Dynamical nature of the preperihelion tail in post-perihelion
images\/}.  The preperihelion tail of C/2012~S1 in these images was
invariably dominated by the ejecta released during an event around 6~days
before perihelion and was closely approximated by a {\it synchrone\/} fitting
this event; {\it whereas\/} the preperihelion tail of C/2011~W3 consisted of
a stream of continuously emitted submicron-sized particles, described
by a {\it syndyname\/}.

(v) {\it Survival of preperihelion particulate ejecta\/}.  Some of the dust
ejected from C/2012~S1 along the inbound branch of the orbit survived
perihelion passage; {\it whereas\/} none of the solid material released
from C/2011~W3 before perihelion was observed to survive.

(vi) {\it Nucleus condensation in the comet's head after its reappearance
from behind the occulting disk\/}.  C/2012~S1 displayed {\it no nuclear
condensation\/} in the images; {\it where\-as\/} C/2011~W3 was starlike,
saturating the detectors.

(vii) {\it Orientation and outlines of the tail extension near the head in
post-perihelion images\/}. The head of C/2012~S1 exhibited a fan-shaped
feature pointing sideways of the Sun and no extension whatsoever in the
antisolar direction; {\it whereas\/} the head of C/2011~W3 was first tailless,
but soon was developing a new tail that rapidly grew in length and was directed
approximately away from the Sun; later, a second, narrow tail began to show
up exactly along the prolonged radius vector.

(viii) {\it Activity pattern in the light curve\/}. Between 9.4 and 0.7~AU
from the Sun, the light curve of C/2012~S1 showed its preperihelion activity
to have evolved in five progressively contracting cycles; {\it whereas\/} 
the light curve of C/2011~W3 displayed no such pattern; however, this
object was under observation for only a very limited period of time before
perihelion.

(ix) {\it Preperihelion vs post-perihelion brightness variations\/}.  C/2012~S1
was intrinsically bright long before perihelion; on the average, it was
brightening at a very slow rate upon its approach to perihelion, yet appearing
vastly brighter shortly before perihelion than its debris afterwards; {\it
whereas\/} C/2011~W3 was first extremely faint with a steep pace of
brightening before perihelion; it reached its peak brightness shortly after
perihelion.

(x) {\it Nucleus' disintegration\/}.  The nuclei of both comets fell apart
and dissipated, but the demise of the nucleus of C/2012~S1 occurred shortly
before perihelion; {\it whereas\/} the collapse of C/2011~W3 took place
after perihelion.

Most of these differences are due to a major distinction between the two
objects in material strength, with the nucleus of C/2012~S1 being much less
cohesive than that of C/2011~W3.  We notice no obvious similarities between
the two objects in physical behavior beyond the basic features of cometary
appearance and activity.

\section{Conclusions}
The present investigation of comet C/2012~S1 offers a comprehensive analysis
of its light curve, water production curve, morphology of the head and tail,
and orbital motion, with the aim to learn about the comet's evolution and
physical behavior in general, and about the developments that led to the
comet's preperihelion disintegration in particular.  Although emphasis is
on features and events at small heliocentric distances, some information
is acquired on the comet's activity far from the Sun as well.  The results
allow us to arrive at the following conclusions:\\[-0.16cm]

(1) The comet's intrinsic brightness appears to have been evolving in cycles,
five of which, A--E, took place between the heliocentric distances of 9.4~AU
and 0.7~AU, each consisting of an expansion stage and a depletion stage.  The
expansion stage began with an ignition (or activation) point and ended with
a stagnation point, while the depletion stage began with the stagnation point
and ended at the ignition point of the next cycle.

(2) The duration of the cycles grew progressively shorter with decreasing
distance $r$ from the Sun, from $>$290 days for the cycle A to 15 days for
E.  Their extent was much less uneven in terms of $\log r$.  The cycles
were probably caused by activation of limited, discrete reservoirs of ices
on and just beneath the nucleus' surface.

(3) At heliocentric distances smaller than about 2 AU, the available water
production curve and the light curve correlated in a qualitative sense, the
correlation being especially high in November; the peaks on the two curves
either coincided or the light curve trailed Combi et al.'s curve of daily
averages of the water production rate.

(4) Extrapolation of the daily averages of the water production rate and the
{\it Af}$\rho$ data on dust production suggests that between 2013~October~1
and November~25 the comet's loss of mass amounted to nearly \mbox{$6 \times
10^{13}$}\,grams, equivalent to a sphere 0.65~km in diameter at an assumed bulk
density of 0.4~g~cm$^{-3}$.  Given that the nucleus disintegrated completely
before reaching perihelion, this was its diameter at the time of close
approach to Mars on 2013 October~1; it is consistent with the result
based on the HiRISE photometric measurements, according to which the
nucleus was not more than 1~km across.

(5) Examination of the contamination of the HiRISE photometry of the nucleus
by dust ejecta leads to a conclusion that the cross-sectional area of the
nucleus accounted for less than 50~percent of the signal in the brightest
pixel of the comet's image.

(6) The cycle E terminated 16 days before perihelion, at the onset of a
precursor to the first major outburst called Event~1.  A drop in the
intrinsic brightness, which followed, stabilized after a few days, and a
little more than 9~days before perihelion a new flare-up occurred, a
precursor to Event~2.  Although not exceeding Event~1 in peak intrinsic
brightness, Event~2 displayed multiple maxima, lasted for at least 3~days,
and followed an enormous temporary increase in the production of water.

(7) During Events~1 and 2, the sublimating area needed to explain the
water production rate exceeded --- in the case of Event~2 steadily over
a period of several days --- the surface area of the nucleus' sunlit
hemisphere by a factor of, respectively, $\sim$20 and $\sim$40.  Both
the persistence of, and erratic changes in, the greatly elevated water
production rule out an effect of icy grains as a source and, instead,
imply fragmentation of the nucleus  --- due probably to rapidly
increasing thermal stress in its interior.

(8) Events 1 and 2 differ from each other in that most of the comet's
water ice supplies were retained, presumably in the largest fragments
comparable in size to the initial nucleus, in Event~1, whereas the
ice retention was close to nil in Event~2, indicating that the nucleus
was then shattered into much smaller fragments than before.  The activity
of all fragments ceased before perihelion.

(9) A preliminary report of a major drop in gas emission (by a factor of
20 or more) 3~days before perihelion confirms that by then the supplies
of ice in what remained of the comet's nucleus were practically exhausted.

(10) As a result of continuing cascading fragmentation, all boulders dated
from Events~1 and 2 grew progressively smaller with time, and extensive
crumbling of particulates may in part account for a steep intrinsic
brightening in the course of Event~3, which began some 2.4~days before
perihelion, about 40~{\Rsun} from the Sun, and peaked $\sim$16~hours
before perihelion.  The other, probably dominant, contribution to the
brightness was due to sublimation of sodium from the fragmented comet,
which, as it was approaching the Sun, increasingly resembled a cloud of
dust, much of it microscopic.

(11) At the time of maximum light, some 16 hours before perihelion, the
comet's apparent visual magnitude of $-$2 was first thought to be the
onset of the long-anticipated high level of activity.  The misleading
nature of this development is illustrated on an elementary case that the
same brightness is achieved, without any contribution from sublimating
sodium, by a cloud of less than 10$^{12}$\,g of dust particles, each
0.5~micron in diameter and of moderate bulk density.

(12) A major part of our investigation was devoted to a study of the
comet's evolution over a period of 33~hours, based on eight preperihelion
and post-perihelion images taken with the coronagraphs on board the SOHO
and STEREO-A and B spacecraft; our objectives were to fully exploit the
powerful stereoscopic capability provided by the spatial configuration of
the spacecraft for examining the comet's principal morphological features.

(13) Modeling of the preperihelion images showed that the dominant
contributor to a prominent, slightly curved tail, seen in the SOHO and
STEREO images alike, was the microscopic dust that was released during
Event~2, centered on 6~days before perihelion; an unresolved contribution
from Event~1 was probably detected in one of the three studied
preperihelion images.

(14) In the last modeled preperihelion image, the tail made of the
Event~2 dust ejecta displayed a deformation emanating from the head in
the form of a short ``beard.''  Its leading boundary fitted the release
of microscopic dust some 15~hours before perihelion, closely correlating
with the sharp peak on the intrinsic-brightness curve.

(15) The disappearance of the comet's head over a period of 4~hours, from
about 5~hours before perihelion on, has been recorded in real time in images
taken with the C2 coronagraph on board SOHO.  The images show a pointed
extension protruding from an originally rounded head in the direction
of motion, and a truncated leading boundary of the tail.  The forward
tip of the extension was the site of the most massive fragments of the
disintegrating nucleus.  The tail's truncated boundary was another sign
of the termination of Event~3 and was related to the ``beard'' detected
in a set of images taken shortly before perihelion from on board STEREO-A.

(16) The post-perihelion appearance of C/2012~S1 differed from that before
perihelion and resembled a winged object.  A narrow tail, trailing far behind,
was also visible in most images, representing an essentially detached feature
on the other side of the Sun from the location of the wings.  The tail was
again dominated by the dust ejecta from Event~2, as was one of the wings,
which could also be described as a trailing boundary of a dust emission fan.
Although some post-perihelion images, especially those taken with the C3
coronagraph, displayed a pseudo-condensation at the point where the wings
joined together, none of these images showed a true {\it nuclear\/}
condensation.  The object remained headless and the local brightening was
a result of forward scattering of sunlight by porous dust at phase angles
of almost 130$^\circ$.

(17) The leading boundary of the dust emission fan in several images taken
with the coronagraphs on board all three spacecraft after perihelion was
positively identified by modeling as containing the final dust ejecta ever
released from the comet and led to the conclusions that \mbox{$3.5 \pm
0.3$}~hours before perihelion, at 5.2~{\Rsun} from the Sun, (i)~all
activity terminated, never again to be resuscitated; (ii)~the nucleus'
disintegration was completed; and, as a result, (iii)~C/2012~S1 ceased to
exist.

(18) Our modeling of Combi et al.'s daily averages of the water production
rate suggests that, before fragmentation, the nucleus was in fact very
active given its small size, and that some water ice was apparently
sublimating from icy-dust grains in the comet's atmosphere.  We further
find that the conclusions based on our examination of the comet's morphology
are consistent with the water production history

(19) Submicron-sized dust particles were present in the Event~2 tail and
were detected in preperihelion as well as post-perihelion images.  These
particles moved in strongly hyperbolic orbits very different from the orbit
of the comet (and its sizable fragments), some with much larger perihelion
distances and passing perihelion only weeks after the largest debris.  Many
particles survived, as their orbits shielded them from the most hostile
environment near the Sun.

(20) By contrast, we find no submicron-sized grains along the leading
boundary of the emission fan, which is interpreted to mean that they were
absent from the final dust release around 3.5~hours before perihelion.

(21) A separate category is solid material that populated the comet's
streamer, clearly apparent in many preperihelion C2 and C3 images and in
several post-perihelion C2 images; these include two of the modeled images.
The streamer followed closely the projected orbit behind the comet, consisted
of very large pieces of dust released from the nucleus at considerable
heliocentric distances, and was of a similar nature as dust trails of
periodic comets.  From the streamer's orientation and the curvature the
actual time (or distance) of release cannot be determined.

(22) An important property of the streamer was its disconnection from the
head, suggesting dust sublimation at a nearly constant distance from the
Sun, but different in images taken before and after perihelion.  Accounting
for this effect in terms of a self-consistent model shows that if the
sublimation heat of the streamer's material was constant, pebble-sized
gravel should have been released at heliocentric distances up to $\sim$100~AU,
presumably by the annealing of amorphous water ice (at temperatures
as low as 37~K) and millimeter-sized grains should have been released in
large quantities at about 10~AU from the Sun during the heights of the
crystallization of amorphous ice into cubic ice{\vspace{-0.04cm}} (at
temperatures close to 130~K), with the heat of sublimation near 90\,000
cal mol$^{-1}$.  No part of the streamer survived perihelion passage.

(23) The brightness of the dissipating cloud of solid debris was rapidly
diminishing with time after perihelion.  A total cross-sectional area of
dust in the fan-shaped cloud amounted to not more than 50\,000~km$^2$
in a 27-arcmin aperture, used by Nakano to derive the magnitude from a C3
image taken on November 29.38 UT, 0.6~day post-perihelion.  From a model
that employs this cross-sectional area and includes assumptions about a
size-distribution law of the debris, we estimate that the largest surviving
{\it inert\/} fragments of the nucleus were almost certainly smaller than
$\sim$0.25~meter and may even have been sub-centimeter in diameter.  It
follows from conclusion (8) that no {\it active\/} fragments survived.

(24) Comprehensive orbital analysis of the comet shows that its motion could
successfully be fitted by a gravitational solution only at heliocentric
distances greater than 4.9~AU.  The solution confirms that the object
originated in the Oort Cloud and began its journey to perihelion at some
50\,000 to 60\,000~AU from the Sun.

(25) Between 4.9~AU and 1~AU from the Sun, the comet's orbital motion was
affected by a nongravitational acceleration driven by the momentum transferred
from the sublimating species, primarily but probably not exclusively water ice.
A satisfactory match to astrometric observations in this range of heliocentric
distances was achieved with a nongravitational law similar to the standard
law used in the Style~II formalism of Marsden et al.\ (1973), but with a
somewhat smaller scaling distance.

(26) At distances from the Sun smaller than 1~AU, the motion of the comet was
subjected to major nongravitational forces, whose variations with time could
not be fitted by the standard law of water production; one of three
nongravitational laws that did match the observed perturbations was based on
the sublimation of sodium.  The nongravitational accelerations near the
Sun were as high as 10$^{-2}$ the Sun's gravitational acceleration and could
represent an effect due to sodium sublimation from the nucleus' debris in
a centimeter to submeter size range; a minor contribution from the Sun's
radiation pressure could not be ruled out.

(27) The radial component, which dominated the nongravitational acceleration,
was always directed away from the Sun, indicating that the comet's orbital
motion was systematically decelerated, obviously because of the prevailing
sunward direction of the flow of sublimating material from the disintegrating
nucleus.

(28) Comparison of C/2012 S1 with the recent bright Kreutz sungrazer C/2011~W3
revealed major differences between the two objects; the nucleus of the latter
held together much more strongly.

(29) We found no contribution from the Sun's tidal forces to the disintegration
of C/2012~S1 and detected no evidence for sublimation of silicates from the
nucleus, apparently because even the last astrometric observation was still
made much too far from perihelion.
\vspace{0.5cm}

We thank M.-T.\ Hui, Guangzhou, China, for sending us an extended version
of the curve of apparent-brightness variations of comet C/2012~S1 he derived
from HI1-A images.  We also thank A.\ Vitagliano, Universit\'a di Napoli
`Federico II', for his positive response to our request for modifications
of his orbit determination code.  This research was carried out in part at
the Jet Propulsion Laboratory, California Institute of Technology, under
contract with the National Aeronautics and Space Administration.\\
\begin{center}
{\footnotesize REFERENCES}
\end{center}
\vspace{-0.3cm}
\begin{description}
{\footnotesize
\item[\hspace{-0.3cm}]
Ag\'undez, M., Biver, N., Santos-Sanz, P., et al. 2014, A\&A, 564, L2
\\[-0.55cm]
\item[\hspace{-0.3cm}]
A'Hearn, M. F., Millis, R. L., Schleicher, D. G., et al. 1995,~Icarus,
  {\hspace*{-0.6cm}}118, 223
\\[-0.55cm]
\item[\hspace{-0.3cm}]
Ayotte, P., Smith, R. S., Stevenson, K. P., et al. 2001, JGR, 106,
  {\hspace*{-0.6cm}}33387
\\[-0.55cm]
\item[\hspace{-0.3cm}]
Bar-Nun, A., Herman, G., Laufer, D., \& Rappaport, M. L. 1985,
  {\hspace*{-0.6cm}}Icarus, 63, 317
\\[-0.55cm]
\item[\hspace{-0.3cm}]
Bar-Nun, A., Dror, J., Kochavi, E., \& Laufer, D. 1987, Phys. Rev.\\
  {\hspace*{-0.6cm}}B, 35, 2427
\\[-0.55cm]
\item[\hspace{-0.3cm}]
Biver, N., Agundez, M., Santos-Sanz, P., et al. 2013, CBET 3711
\\[-0.55cm]
\item[\hspace{-0.3cm}]
Bodewits, D., Farnham, T., \& A'Hearn, M. F. 2013, CBET 3718
\\[-0.55cm]
\item[\hspace{-0.3cm}]
Boehnhardt, H. 2004, in Comets II, ed. M. C. Festou, H. U. Kel-\\
  {\hspace*{-0.6cm}}ler, \& H. A. Weaver (Tucson, AZ: Univ.
  Arizona Press), 301
\\[-0.55cm]
\item[\hspace{-0.3cm}]
Boehnhardt, H., Tubiana, C., Oklay, N., et al. 2013, CBET 3715
\\[-0.55cm]
\item[\hspace{-0.3cm}]
Boehnhardt, H., Vincent, J. B., Chifu, C., et al. 2013, CBET 3731
\\[-0.55cm]
\item[\hspace{-0.3cm}]
Bonev, B. P., DiSanti, M. A., Gibb, E. L., et al. 2013, CBET 3720
\\[-0.55cm]
\item[\hspace{-0.3cm}]
Brueckner, G. E., Howard, R. A., Koomen, M. J., et al. 1995, Sol.
  {\hspace*{-0.6cm}}Phys., 162, 357
\\[-0.55cm]
\item[\hspace{-0.3cm}]
Chamberlain, J. W. 1960, ApJ, 131, 47
\\[-0.55cm]
\item[\hspace{-0.3cm}]
Chamberlain, J. W. 1961, ApJ, 133, 675
\\[-0.55cm]
\item[\hspace{-0.3cm}]
Combi, M. R., Bertaux, J.-L., Quemerais, E., et al. 2013, IAUC{\linebreak}
  {\hspace*{-0.6cm}}9266
\\[-0.55cm]
\item[\hspace{-0.3cm}]
Combi, M. R., Fougere, N., M\"{a}kinen, J. T. T., et al. 2014, ApJ,\linebreak
  {\hspace*{-0.6cm}}788, L7 (5pp)
\\[-0.55cm]
\item[\hspace{-0.3cm}]
Coulson, I. M., Milam, S. N., Villanueva, G. L., et al. 2013, CBET{\linebreak}
  {\hspace*{-0.6cm}}3693
\\[-0.55cm]
\item[\hspace{-0.3cm}]
Crovisier, J., Colom, P., Biver, N., \& Bockel\'ee-Morvan, D. 2013,
  {\hspace*{-0.6cm}}CBET 3711
\\[-0.55cm]
\item[\hspace{-0.3cm}]
Curdt, W., Boehnhardt, H., Vincent, J.-B., et al. 2014, A\&A,\linebreak
  {\hspace*{-0.6cm}}567, L1
\\[-0.55cm]
\item[\hspace{-0.3cm}]
Delamere, W. A., McEwen, A. S., Mattson, S., et al. 2013, CBET
  {\hspace*{-0.6cm}}3720
\\[-0.55cm]
\item[\hspace{-0.3cm}]
Dello Ruso, N., Vervack, Jr., R. J., Kawakita, H., et al. 2013,{\linebreak}
  {\hspace*{-0.6cm}}CBET 3686
\\[-0.55cm]
\item[\hspace{-0.3cm}]
Delsemme, A. H., \& Miller, D. C. 1971, Planet. Space Sci, 19, 1229
\\[-0.55cm]
\item[\hspace{-0.3cm}]
Dohnanyi, J. S. 1969, JGR, 74, 2531
\\[-0.55cm]
\item[\hspace{-0.3cm}]
Enzian, A., Cabot, H., \& Klinger, J. 1998, Planet. Space Sci., 46,{\linebreak}
  {\hspace*{-0.6cm}}851
\\[-0.55cm]
\item[\hspace{-0.3cm}]
Ferr\'{\i}n, I. 2013, eprint arXiv:1310.0552
\\[-0.55cm]
\item[\hspace{-0.3cm}]
Ferr\'{\i}n, I. 2014, Planet. Space Sci., 96, 114
\\[-0.55cm]
\item[\hspace{-0.3cm}]
Fitzsimmons, A., Lacerda, P., Lowry, S., et al. 2013, IAUC 9261
\\[-0.55cm]
\item[\hspace{-0.3cm}]
Fulle, M. 1999, Adv. Space Res., 24 (9), 1087
\\[-0.55cm]
\item[\hspace{-0.3cm}]
Golubev, A. V., Bryukhanov, I. S., Tabolich, A., et al. 2014,\linebreak
  {\hspace*{-0.6cm}}Astron.  Tsirk. 1611
\\[-0.55cm]
\item[\hspace{-0.3cm}]
Gonz\'alez, M., Guti\'errez, P. J., Lara, L. M., \& Rodrigo, R. 2008,
  {\hspace*{-0.6cm}}A\&A, 486, 331
\\[-0.55cm]
\item[\hspace{-0.3cm}]
Hashimoto, A. 1990, Nature 347, 53
\\[-0.55cm]
\item[\hspace{-0.3cm}]
Herman, G., \& Podolak, M. 1985, Icarus, 61, 252
\\[-0.55cm]
\item[\hspace{-0.3cm}]
Hines, D. C., Videen, G., Zubko, E., et al. 2013, ApJ, 780L, 32
\\[-0.55cm]
\item[\hspace{-0.3cm}]
Howard, R. A., Moses, J. D., Vourlidas, A., et al. 2008, Space Sci.
  {\hspace*{-0.6cm}}Rev., 136, 67
\\[-0.55cm]
\item[\hspace{-0.3cm}]
Huebner, W. F., Keady, J. J., \& Lyon, S. P. 1992, Astrophys.{\linebreak}
  {\hspace*{-0.6cm}}Space Sci., 195, 1
\\[-0.55cm]
\item[\hspace{-0.3cm}]
Jenniskens, P., \& Blake, D. F. 1994, Science, 265, 753
\\[-0.55cm]
\item[\hspace{-0.3cm}]
Jenniskens, P., \& Blake, D. F. 1996, Planet. Space Sci., 44, 711
\\[-0.55cm]
\item[\hspace{-0.3cm}]
Keane, J. V., Meech, K. J., Mumma, M. J., et al. 2013, IAUC 9261
\\[-0.55cm]
\item[\hspace{-0.3cm}]
Klinger, J. 1980, Science, 209, 271
\\[-0.55cm]
\item[\hspace{-0.3cm}]
Knight, M. M., \& Battams, K. 2014, ApJ, 782L, 37
\\[-0.55cm]
\item[\hspace{-0.3cm}]
Knight, M. M., \& Walsh, K. J. 2013, ApJ, 776, L5
\\[-0.55cm]
\item[\hspace{-0.3cm}]
Korsun, P. P., \& Ch\"{o}rny, G. F. 2003, A\&A, 410, 1029
\\[-0.55cm]
\item[\hspace{-0.3cm}]
Laufer, D., Kochavi, E., \& Bar-Nun, A. 1987, Phys. Rev. B, 36,{\linebreak}
  {\hspace*{-0.6cm}}9219
\\[-0.55cm]
\item[\hspace{-0.3cm}]
Laufer, D., Pat-El, I., \& Bar-Nun, A. 2005, Icarus, 178, 248
\\[-0.55cm]
\item[\hspace{-0.3cm}]
Li, J.-Y., Weaver, H. A., Kelley, M. S., et al. 2013, CBET 3496
\\[-0.55cm]
\item[\hspace{-0.3cm}]
Li, J.-Y., Kelley, M. S., Knight, M. M., et al. 2014, Am. Astron.{\linebreak}
  {\hspace*{-0.6cm}}Soc., AAS Meeting \#223, \#218.06
\\[-0.55cm]
\item[\hspace{-0.3cm}]
Marcus, J. N. 2007, Int. Comet Quart., 29, 39
\\[-0.55cm]
\item[\hspace{-0.3cm}]
Marsden, B. G., \& Williams. G. V. 2008, Catalogue of Cometary{\linebreak}
  {\hspace*{-0.6cm}}Orbits 2008 (17th ed.; Cambridge, MA: Smithsonian
  Astrophys-{\linebreak}
  {\hspace*{-0.6cm}}ical Observatory), 195\,pp
\\[-0.55cm]
\item[\hspace{-0.3cm}]
Marsden, B. G., Sekanina, Z., \& Yeomans, D. K. 1973, AJ, 78,~211
\\[-0.55cm]
\item[\hspace{-0.3cm}]
Marsden, B. G., Sekanina, Z., \& Everhart, E. 1978, AJ, 83, 64
\\[-0.55cm]
\item[\hspace{-0.3cm}]
McDonnell, J. A. M., Alexander, W. M., Burton, W. M., et~al.{\linebreak}
  {\hspace*{-0.6cm}}1987, in Exploration of Halley's Comet, ESA SP-250,
  ed.~M.~Gre-
  {\hspace*{-0.6cm}}wing, F. Praderie, \& R. Reinhard (Berlin:
  Springer-Verlag), 719
\\[-0.55cm]
\item[\hspace{-0.3cm}]
Meech, K. J., Pittichov\'a, J., Bar-Nun, A., et al. 2009, Icarus,
  201,{\linebreak}
  {\hspace*{-0.6cm}}719
\\[-0.55cm]
\item[\hspace{-0.3cm}]
Meech, K., Keane, J., Yang, B., et al. 2013a, CBET 3693
\\[-0.55cm]
\item[\hspace{-0.3cm}]
Meech, K. J., Yang, B, Kleyna, J., et al. 2013b, ApJ, 776, 20
\\[-0.55cm]
\item[\hspace{-0.3cm}]
Mumma, M. J., DiSanti, M. A., Paganini. L., et al. 2013b, IAUC{\linebreak}
  {\hspace*{-0.6cm}}9263
\\[-0.55cm]
\item[\hspace{-0.3cm}]
Nakano, S. 2013a, CBET 3731
\\[-0.55cm]
\item[\hspace{-0.3cm}]
Nakano, S. 2013b, CBET 3767
\\[-0.55cm]
\item[\hspace{-0.3cm}]
Nakano, S. 2013c, Nakano Note 2587
\\[-0.55cm]
\item[\hspace{-0.3cm}]
Nevski, V., \& Novichonok, A. 2012, CBET 3238
\\[-0.55cm]
\item[\hspace{-0.3cm}]
Opitom, C., Jehin, E., Manfroid, J., et al. 2013a, CBET 3693
\\[-0.55cm]
\item[\hspace{-0.3cm}]
Opitom, C., Jehin, E., Manfroid, J., et al. 2013b, CBET 3711
\\[-0.55cm]
\item[\hspace{-0.3cm}]
Opitom, C., Jehin, E., Manfroid, J., \& Gillon, M. 2013c, CBET{\linebreak}
  {\hspace*{-0.6cm}}3719
\\[-0.55cm]
\item[\hspace{-0.3cm}]
Paganini, L., Blake, G. A., Villanueva, G. L., et al. 2013, IAUC{\linebreak}
  {\hspace*{-0.6cm}}9263
\\[-0.55cm]
\item[\hspace{-0.3cm}]
Patashnick, H., Rupprecht, G., \& Schuerman, D. W. 1974, Nature,
  {\hspace*{-0.6cm}}250, 313
\\[-0.55cm]
\item[\hspace{-0.3cm}]
Prialnik, D. 1992, ApJ, 388, 196
\\[-0.55cm]
\item[\hspace{-0.3cm}]
Prialnik, D. 2006, in Asteroids, Comets, Meteors, Proc.\,229th~IAU
  {\hspace*{-0.6cm}}Symp., ed.\ L.\ Daniela, M.\ Sylvio Ferraz,
  \& F.\ J.\ Angel (Cam-{\linebreak}
  {\hspace*{-0.6cm}}bridge, UK: Cambridge Univ. Press), 153
\\[-0.55cm]
\item[\hspace{-0.3cm}]
Reach, W. T., Sykes, M. V., Lien, D., \& Davies, J. K. 2000,{\linebreak}
  {\hspace*{-0.6cm}}Icarus, 148, 80
\\[-0.55cm]
\item[\hspace{-0.3cm}]
Remijan, A. J., Gicquel, A., Milam, S. N., et al. 2013, CBET 3693
\\[-0.55cm]
\item[\hspace{-0.3cm}]
Rickman, H., Kam\'el, L., Froeschl\'e, C., \& Festou, M. C. 1991,
  AJ,{\linebreak}
  {\hspace*{-0.6cm}}102, 1446
\\[-0.55cm]
\item[\hspace{-0.3cm}]
Schleicher, D. 2013a, IAUC 9260
\\[-0.55cm]
\item[\hspace{-0.3cm}]
Schleicher, D. 2013b, IAUC 9254
\\[-0.55cm]
\item[\hspace{-0.3cm}]
Schmitt, B., \& Klinger, J.\ 1987, in Diversity and Similarity~of{\linebreak}
  {\hspace*{-0.6cm}}Comets, ESA SP-278, ed.\ E.\,J.\,Rolfe \& B.\,Battrick
  (Noordwijk,
  {\hspace*{-0.6cm}}Netherlands:\ ESTEC), 613
\\[-0.55cm]
\item[\hspace{-0.3cm}]
Schmitt,~B.,~Espinasse,~S.,~Grim,~R.\,J.\,A.,~et~al.\,1989,~in~Physics~and{\linebreak}
  {\hspace*{-0.6cm}}Mechanics\, of\, Cometary\, Materials,\, ESA SP-302,
  ed.\ J.\ Hunt \& \\[0.02cm]
  {\hspace*{-0.6cm}}T.\ D.\ Guyenne (Noordwijk, Netherlands: ESTEC), 65
\\[-0.55cm]
\item[\hspace{-0.3cm}]
Sekanina, Z. 1975, Icarus, 25, 218
\\[-0.55cm]
\item[\hspace{-0.3cm}]
Sekanina, Z. 1981, AJ, 86, 1741
\\[-0.55cm]
\item[\hspace{-0.3cm}]
Sekanina, Z. 1988, AJ, 95, 911
\\[-0.55cm]
\item[\hspace{-0.3cm}]
Sekanina, Z. 1992, in Asteroids, Comets, Meteors 1991, ed. A. W.
  {\hspace*{-0.6cm}}Harris \& E. Bowell (Houston, TX: Lunar Planet.\ Inst.), 545
\\[-0.55cm]
\item[\hspace{-0.3cm}]
Sekanina, Z. 1993, AJ, 105, 702
\\[-0.58cm]
\item[\hspace{-0.3cm}]
Sekanina, Z. 2013a, eprint arXiv:1310.1980
\\[-0.58cm]
\item[\hspace{-0.3cm}]
Sekanina, Z. 2013b, CBET 3731{\pagebreak}
\\[-0.55cm]
\item[\hspace{-0.3cm}]
Sekanina, Z. 2013c, CBET 3723
\\[-0.55cm]
\item[\hspace{-0.3cm}]
Sekanina, Z., \& Chodas, P. W. 2012, ApJ, 757, 127 
\\[-0.58cm]
\item[\hspace{-0.3cm}]
Sekanina, Z., Hanner, M.\,S., Jessberger, E.\,K., \& Fomenkova,~M.\,N.\\[-0.04cm]
 {\hspace*{-0.6cm}}2001, in Interplanetary Dust, ed.\ E.\ Gr\"{u}n,
 B.\ {\AA}.\ S.\ Gustafson,{\linebreak}
 {\hspace*{-0.6cm}}S.\ F.\ Dermott, \& H.\ Fechtig (Heidelberg: Springer), 95
\\[-0.58cm]
\item[\hspace{-0.3cm}]
Sekhar, A., \& Asher, D. J. 2014, MNRAS, 437, L71
\\[-0.58cm]
\item[\hspace{-0.3cm}]
Sitko, M. L., Russell, R. W., Kim, D. L., et al. 2013, IAUC 9264
\\[-0.58cm]
\item[\hspace{-0.3cm}]
Smoluchowski, R. 1981, ApJ, 244, 31
\\[-0.58cm]
\item[\hspace{-0.3cm}]
Sykes, M. V., \& Walker, R. G. 1992, Icarus, 95, 180
\\[-0.58cm]
\item[\hspace{-0.3cm}]
Sykes, M. V., Lebofsky, L. A., Hunten, D. M., \& Low, F. 1986,
  {\hspace*{-0.6cm}}Science, 232, 1115
\\[-0.58cm]
\item[\hspace{-0.3cm}]
van de Hulst, H. C. 1957, Light Scattering by Small Particles.\linebreak
 {\hspace*{-0.6cm}}(New York: Wiley \& Sons, 470pp)
\\[-0.58cm]
\item[\hspace{-0.3cm}]
Weaver, H.\,A., Sekanina, Z., Toth, I., et al.\,2001, Science, 292,~1329
\\[-0.58cm]
\item[\hspace{-0.3cm}]
Weaver, H., Feldman, P., McCandliss, S., et al. 2013, CBET 3680
\\[-0.58cm]
\item[\hspace{-0.3cm}]
Williams, G. V. 2013a, MPEC 2013-W13
\\[-0.56cm]
\item[\hspace{-0.3cm}]
Williams, G. V. 2013b, MPEC 2013-W16
\\[-0.56cm]
\item[\hspace{-0.3cm}]
Williams, G. V. 2013c, MPC 86228
\\[-0.56cm]
\item[\hspace{-0.3cm}]
Williams, G. V. 2014, MPC 87064
\\[-0.58cm]
\item[\hspace{-0.3cm}]
Ye, Q., Hui, M.-T., \& Gao, X. 2013, CBET 3718
\\[-0.68cm]
\item[\hspace{-0.3cm}]
Yeomans, D.\,K., Chodas, P.\,W., Sitarski, G., et
  al.\,2004,~in~Comets\\[-0.08cm]
  {\hspace*{-0.6cm}}II, ed.\ M.\ C.\ Festou, H.\ U.\ Keller,
  \& H.\ A.\ Weaver (Tucson, AZ:\\[-0.08cm]
  {\hspace*{-0.6cm}}Univ.\ Arizona Press), 137{\vspace*{-0.1cm}}}
%
\end{description}
\end{document}